\documentclass[12pt,a4paper]{article}
\usepackage{amsmath,amssymb,epsfig}
\usepackage{dcolumn}

\newcolumntype{d}[1]{D{.}{.}{#1}}
\renewcommand{\theequation}{\thesection.\arabic{equation}}
\renewcommand{\thefootnote}{\fnsymbol{footnote}}
\newlength{\extraspace}
\newlength{\extraspaces}
\newcommand{\be}{\begin{equation}
\addtolength{\abovedisplayskip}{\extraspaces}
\addtolength{\belowdisplayskip}{\extraspaces}
\addtolength{\abovedisplayshortskip}{\extraspace}
\addtolength{\belowdisplayshortskip}{\extraspace}}
\newcommand{\ee}{\end{equation}}
 
\newcommand{\ba}{\begin{eqnarray}
\addtolength{\abovedisplayskip}{\extraspaces}
\addtolength{\belowdisplayskip}{\extraspaces}
\addtolength{\abovedisplayshortskip}{\extraspace}
\addtolength{\belowdisplayshortskip}{\extraspace}}
\newcommand{\ea}{\end{eqnarray}}

\newcommand{\bas}{\begin{eqnarray*}
\addtolength{\abovedisplayskip}{\extraspaces}
\addtolength{\belowdisplayskip}{\extraspaces}
\addtolength{\abovedisplayshortskip}{\extraspace}
\addtolength{\belowdisplayshortskip}{\extraspace}}
\newcommand{\eas}{\end{eqnarray*}}
 
\newcounter{subequation}[equation]
\makeatletter

\expandafter\let\expandafter
\reset@font\csname reset@font\endcsname

\def\subeqnarray{\arraycolsep1pt
    \def\@eqnnum\stepcounter##1{\stepcounter{subequation}%
        {\reset@font\rm(\theequation\alph{subequation})}}
\jot5mm     \eqnarray}

\makeatother

\newcommand{\newappendix}[1]{
\vspace{15mm}
\pagebreak[3]
\addtocounter{section}{1}
\setcounter{equation}{0}
\setcounter{subsection}{0}
\setcounter{footnote}{0}
\renewcommand{\theequation}{\Alph{section}.\arabic{equation}}
\begin{flushleft}
{\large\bf \Alph{section}. #1}
\end{flushleft}
\nopagebreak
\medskip
\nopagebreak}

\newcommand{\newsection}[1]{
\vspace{15mm}
\pagebreak[3]
\addtocounter{section}{1}
\setcounter{equation}{0}
\setcounter{subsection}{0}
\setcounter{footnote}{0}
 
\begin{flushleft}
{\large\bf \thesection. #1}
\end{flushleft}
\nopagebreak
\medskip
\nopagebreak}
 
\newcommand{\newsubsection}[1]{
\vspace{1cm}
\pagebreak[3]
 
\addtocounter{subsection}{1}
\noindent{ \bf \thesection.\arabic{subsection} #1}
\nopagebreak
\vspace{2mm}
\nopagebreak}

\newcommand{\NP}[1]{Nucl.\ Phys.\ {\bf #1}}

\newcommand{\1}{\mbox{1\hspace{-.8ex}1}}
\newcommand{\bra}{\langle}
\newcommand{\ket}{\rangle}
\newcommand{\ra}{\rightarrow}

\newcommand{\rra}{\ \longrightarrow \ }

\newcommand{\is}{ &\!=\!& }
\newcommand{\nonum}{\nonumber \\[1.5mm]}
\newcommand{\sspace}{\makebox[1cm]{ }}

\newcommand{\nspace}{\!\!\!\!\!\!\!\!\!\!}

\newcommand{\inv}{^{-1}}

\renewcommand{\th}{{\theta}}
\newcommand{\eps}{{\epsilon}}
\newcommand{\lb}{\lambda}
\newcommand{\sh}{{\rm sh}}
\newcommand{\ch}{{\rm ch}}
\newcommand{\dd}{{\partial}}

\newcommand{\cO}{{\cal O}}

\newcommand{\cU}{{\cal U}}
 
\renewcommand{\d}{\delta}

\newcommand{\gr}{g_{\rm R}}
\newcommand{\Mr}{M_{\rm R}}
\newcommand{\Zr}{Z_{\rm R}}
\newcommand{\so}{\phantom{-}1} 
\newcommand{\sz}{\phantom{-}0} 

\begin{document}
%
\begin{titlepage}
%
\renewcommand{\thefootnote}{\fnsymbol{footnote}}
\begin{flushright}
MPI-PhT/01-18\\
\end{flushright}
\vspace{1cm}

\begin{center}
{\LARGE \bf Does the XY Model have an integrable}\\[4mm]
{\LARGE \bf continuum limit?}
\vspace{2cm}

{\large J. Balog$^1$, M. Niedermaier$^4$, F. Niedermayer$^2$
\footnote{On leave from E\"otv\"os University, Budapest}},\\[2mm]
{\large  A. Patrascioiu$^3$,  E. Seiler$^4$, P. Weisz$^4$}\\[7mm]

{\small\sl $^1$Research Institute for Particle and Nuclear Physics}\\
{\small\sl 1525 Budapest 114, Hungary}
\\[2mm]
{\small\sl $^2$Institute for Theoretical Physics, University of Bern}\\
{\small\sl CH-3012 Bern, Switzerland}
\\[2mm]
{\small\sl $^3$Physics Department, University of Arizona}\\
{\small\sl Tucson, AZ 85721, U.S.A.}
\\[2mm]
{\small\sl $^4$Max-Planck-Institut f\"{u}r Physik}\\
{\small\sl 80805 Munich, Germany}
\vspace{1cm}

{\bf Abstract}
\end{center}
\vspace{-5mm}

\begin{quote}
The quantum field theory describing the massive O(2) nonlinear 
sigma-model is investigated through two non-perturbative 
constructions: The form factor bootstrap based on integrability 
and the lattice formulation as the XY model. The S-matrix, the spin 
and current two-point functions, as well as the 4-point coupling are 
computed and critically compared in both constructions. On the bootstrap
side a new parafermionic super selection sector is found; in 
the lattice theory a recent prediction for the (logarithmic) 
decay of lattice artifacts is probed. 
\end{quote}
\vfill
\setcounter{footnote}{0}
\end{titlepage}


\newsection{Introduction}

The XY-model in two dimensions is of prime interest in the field 
of statistical mechanics, intriguing in particular because of its 
unusual phase transition. On general grounds one expects that a 
suitable scaling limit in the high temperature phase gives rise to 
a massive relativistic quantum field theory (QFT). Though an enormous 
literature exists on the statistical mechanics aspects of the system, 
to the best of our knowledge the nature of this QFT has 
never been systematically explored. As part of a long term 
project on quantum non-linear sigma models we thus address 
here the question:

\noindent {\it What are the qualitative and quantitative 
           features of the QFT obtained from the XY-model by taking 
           the massive continuum limit?}

The question is of interest, first in that it highlights the important 
problem of controlling the approach to the continuum in the lattice 
formulation of QFTs, and second because the proposed continuum QFT seems 
to possess a rich, hitherto unknown superselection structure. 
The problem of controlling the continuum limit of a lattice system 
based on numerical simulations alone is notoriously difficult for 
several reasons. (i) First because one often lacks rigorous knowledge 
of the phase structure and the position(s) of the critical points 
where the correlation length becomes infinite. (ii) Secondly there 
are in general no rigorous results on how a quantity approaches
its continuum limit as a function of the correlation length, even 
if the existence of the limit is taken for granted. 

Knowledge of (i) and (ii) is crucial not only for quantitative 
aspects but also for matters of principle, like the status of 
asymptotic freedom beyond perturbation theory in non-abelian models, 
or more generally the physical differences or similarities of 
the continuum limits of (spin) systems with abelian and non-abelian 
symmetries. So far the application of lattice techniques to the 
extraction of physical 
quantities has concentrated on the non-abelian models.
This suffers unfortunately from both of the before-mentioned 
problems (i) and (ii). Concerning (ii), the usual working hypothesis, 
attributed to Symanzik, is that in (perturbatively) asymptotically 
free theories, physical quantities approach their continuum limit 
rather rapidly with power-like corrections $1/\xi^p$, with 
$p$ a positive integer (up to multiplicative logarithmic corrections).

On the other hand for the XY-model we do have rigorous information 
on point (i). In particular the model with standard action 
is known to have two phases, one massless and the other massive 
\cite{FrohSpen}.
The order of the phase transition has been argued by Kosterlitz
and Thouless (KT) \cite{KT} to be infinite and this
picture has been supported by various numerical studies \cite{XYMC}.
Moreover concerning point (ii), one of the present authors (J.B.) 
\cite{logs}, has argued that for a certain class of lattice actions 
and for certain observables (like the S-matrix or a current 
two-point function) leading lattice artifacts do not depend on 
the choice of lattice action and are {\it calculable}. However, they
vanish extremely slowly, 
generically as inverse powers of the {\it logarithm} of the correlation 
length $\xi$ (e.g. $\sim 1/\ln^2\xi$). One of the goals of the 
present paper is to test this proposal through extensive numerical 
simulations. Its derivation takes advantage of the Sine-Gordon 
description of the KT transition introduced by Amit et.~al.~\cite{Amit}. 
Applying a series of (not entirely rigorous) steps 
invoking universality and making a sequence of mappings their 
analysis also entails the tentative identification: 

{\it The massive continuum limit of the XY model is related to the 
Sine-Gordon (SG) model at its extremal fixed point 
$\beta_{SG}=\sqrt{8\pi}$, in that both systems share subsets of 
fields with identical correlation functions.}

Examples of shared fields are the Noether current (proportional to 
the dual of the gradient of the SG scalar) and the energy momentum tensor.
Similar to Coleman's SG--Thirring correspondence \cite{Coleman,KlaMelz} 
the mapping between the fields does not preserve locality in 
general and is likely not to be strictly one-to-one. Nevertheless 
the correspondence is very useful because the SG model is integrable 
and for such systems a direct non-perturbative continuum approach 
exists to construct the QFT, referred to as the form factor bootstrap. 
In fact, the SG model is the prototype integrable model and
it has played an important role in the development of the
form factor bootstrap method.
Its bootstrap S-matrix was proposed in \cite{ZZ} and a large part of
Smirnov's book \cite{Smir1} is devoted to the study of its soliton
form factors, where also the form factors of the SG scalar and the
Noether current are given.
The second purpose of the present paper is to initiate a 
{\it bootstrap construction} of the O(2) model along similar lines.
Since the form factor approach is largely blind with respect to
the local structure we can borrow many mathematical
techniques from the SG model, but the interpretation   
as form factors of certain local O(2) quantum fields   
requires careful justification, one strategy being 
the comparison with lattice simulations. To have a handy terminology 
we shall refer to the QFT defined through the massive continuum limit
of the lattice XY model as the ``XY QFT'', and to the QFT defined
via the form factor bootstrap as the ``bootstrap O(2) model''. 
The basic proposition to be tested is that both QFTs coincide.

Generally the problem of operator identification and classification 
in the bootstrap framework rests on conserved quantum numbers. 
By definition the O(2) model has a manifest O(2) symmetry.
Remarkably on the level of the bootstrap S-matrix and the scattering 
states a {\it symmetry
enhancement} takes place: They are covariant with respect to 
a larger non-abelian quantum group symmetry. It turns out that 
depending on the nature of certain statistics phases the 
functional equations characterizing the form factors only 
have the manifest O(2) symmetry or are covariant with respect
to the hidden quantum group symmetry. This implies that the 
field operators of the bootstrap O(2) QFT fall into two 
classes: those that are (trivial) O(2) multiplets and those
that are members of a nontrivial quantum group multiplet. 
For example a complete set of scattering states seems to be generated 
both by the spin field and a {\em local} parafermion field of 
Lorentz spin $1/4$. Both are relatively nonlocal and the latter 
is quantum group covariant while the former is not. 
Also the (one--component) Noether current of the O(2) model is a member 
of a hidden isospin 1 quantum group triplet, where however the charge 
$\pm 2$ partners can (already classically) not be expressed as 
local functions of the spin field. The energy-momentum tensor 
is a quantum group singlet. In view of the parafermion the O(2) model 
possesses at least two super selection sectors; the full super selection 
structure and its relation to the quantum group multiplets remains to 
be explored. 

The lay-out of the paper is as follows. We start by briefly 
recording the quantities considered and introduce the 
bootstrap and the lattice formulation. In Section 3 we describe in more 
detail the bootstrap O(2) model and discuss its quantum group invariance. 
We proceed with formulating the form factor equations 
and determine the statistics phases for which they exhibit the 
enhanced quantum group symmetry. Next the $n \leq 4$ particle candidate 
form factors of the spin field and the Noether current are obtained. 
Some details on the quantum group structure 
and the form factor computation are relegated to appendices.
 
The subsequent sections all involve lattice simulations, 
concerning which some general information is 
first collected in Section 4. 
Next, we report on measurements of the XY phase shifts 
using a refinement of the finite size technique first employed 
by L\"{u}scher and Wolff \cite{LuWo} for the O(3) model. 
In Section 6 our measurements of the two-point functions of the spin field
and the Noether current, and of the renormalized zero momentum
coupling $\gr$ are presented. There we also compare the results 
with those obtained by the form factor computations, improving on our 
earlier estimate \cite{SigmacollabII}. Finally in Section 7 we 
attempt some  conclusions.
\vfill
\eject

\newsection{Bootstrap and Lattice O(2) model}

One way to approach the O$(2)$ model is as the $n=2$ member
of the family of O$(n)$ nonlinear $\sigma$-models with the classical 
Lagrangean
\be
{\cal L}^{O(n)}=\frac{1}{2g^2}\,\partial_\mu S^a\partial_\mu S^a\;,
\quad\quad S^aS^a=1,\qquad a=1,2,\dots n.
\label{LagOn}
\ee
From the viewpoint of classical field theory it may be surprising
that this Lagrangean should correspond to a nontrivial QFT, since by
substituting $S_1=\cos\phi,\ S_2=\sin\phi$ it becomes quadratic, 
corresponding to a free theory. Also 
perturbatively the beta-function of the coupling $g^2$ vanishes 
identically for $n=2$, suggesting a trivial scale invariant theory. 
The situation changes, however, when one is trying to  
nonperturbatively construct a QFT corresponding to (2.1). 
In this paper two such approaches are studied, both of which lead to 
a nontrivial massive QFT: the lattice approach which allows the 
construction of a massive continuum limit and the form factor bootstrap 
construction based on the indicated relation to the Sine-Gordon theory. 
These two constructions as well as the comparison of the resulting 
theories are the main subject of this investigation.

\newsubsection{Quantities to be investigated}

Clearly only physical quantities should be considered in this 
comparison. We shall study the S-matrix, the spin and current
two-point functions, and the intrinsic coupling. The S-matrix 
will be discussed in sections 2.2 and 5; for the other quantities 
we collect here the key definitions. They apply to both formulations,
and in fact to any O(2) invariant scalar relativistic QFT with a mass gap. 
Let thus $S^a(x)\,,a=1,2$, denote a two-component (renormalized) scalar 
field (the ``spin field''). For the Fourier transform of its Euclidean 
two-point function we write 
\begin{equation}
G(k)\delta^{a_1a_2}=\int \! d^2 x \,e^{ikx}
\langle S^{a_1}(x)S^{a_2}(0) \rangle\,. 
\end{equation}
Its inverse is supposed to have the usual small momentum expansion 
\begin{equation}
G(k)^{-1}=\Zr^{-1}\Big( \Mr^2 + k^2 + O(k^4)\Bigr)\,.
\label{nxi1}
\end{equation}
The coefficients can be expressed in terms of moments of the spectral 
density $\rho(\mu)$ via 
\be
\Zr=Z \frac{\gamma_2^2}{\delta_2}\,,
\sspace 
\frac{\Mr^2}{M^2}=\frac{\gamma_2}{\delta_2}\,,  
\label{nxi2}
\ee
where $M$ is the mass gap and $\gamma_2$ and $\delta_2$ are the moments
\be
\gamma_2=M^2\int \!d \mu \,\frac{\rho(\mu)}{\mu^2},
\sspace
\delta_2=M^4\int \!d \mu \,\frac{\rho(\mu)}{\mu^4}\,.
\ee
Our normalization for the spectral density is such that 
\be
\label{specrep}
G(k)=Z\int_0^{\infty} \! d\mu \frac{\rho(\mu)}{\mu^2 +k^2}\,,
\ee
with the one-particle contribution given by $\rho^{(1)}(\mu)=
\delta(\mu-M)$. (To avoid irrelevant complications we assume 
that the spectrum of the theory contains a doublet of stable 
particles of mass $M$.)

The intrinsic or renormalized 4-point coupling is an important measure
for the interaction strength of a QFT. A conventional definition 
is 
\be
\gr= - \frac{\Mr^2}{4G(0)^2} \sum_{a,b}G^{aabb}(0,0,0,0)\,,
\label{grdef1}
\ee
where $G^{abcd}$ is defined through the Fourier transform of the
connected 4-point function:
\ba
&& \int\prod_{j=1}^4
\left[ d^2 x_j e^{ik_jx_j}\right]
\langle S^{a_1}(x_1)S^{a_2}(x_2)S^{a_3}(x_3)S^{a_4}(x_4)\rangle_{conn}
\nonum
&& \quad 
=(2\pi)^2\delta^{(2)}(k_1+k_2+k_3+k_4)G^{a_1a_2a_3a_4}(k_1,k_2,k_3,k_4)\,.
\ea
The coupling $\gr$ can then be written as
\be
\gr=- \frac{2\gamma_4}{\gamma_2\delta_2}
\label{grdef2}
\ee
where $\gamma_4$ is defined through
\be
G^{a_1a_2a_3a_4}(0,0,0,0)=\frac{Z^2\gamma_4}{M^6}
\left( 
 \delta^{a_1a_2}\delta^{a_3a_4}
+\delta^{a_1a_3}\delta^{a_2a_4}
+\delta^{a_1a_4}\delta^{a_2a_3}\right)\,.
\ee
In ref.~\cite{SigmacollabII} we computed the moments and coupling 
within the form factor approach in a certain truncation, and in 
Appendix D we present an improved approximation. 

We also consider the two-point function of the Noether current $J_\mu$:
\be
\int\! d^2 x \, e^{ikx}\langle J_\mu(x)J_\nu(0) \rangle
=C\delta_{\mu\nu}+
\frac{I(k)}{k^2}\left(k_\mu k_\nu -k^2\delta_{\mu\nu}\right)\,.
\label{contCurr}
\ee
Here $C$ is the (regularization dependent) coefficient of a 
possible contact term. Only the coefficient $I(k)$ of the transversal 
part is physical. It vanishes at zero momentum due to the assumed 
mass gap. The infinite momentum limit on the other hand is model 
dependent and can be finite or infinite. For the O$(2)$ model it 
is determined in Section~2.2.

\newsubsection{Integrability and bootstrap S-matrix}

A first hint why the O(2) model (with classical Lagrangian (\ref{LagOn}) 
for $n=2$) might be quantum integrable stems from the observation that 
the known bootstrap S-matrix of the O$(n)$, $n\geq 3$, models has a smooth 
$n \ra 2$ limit \cite{Woo}. Here we record this limit and also outline 
the relation to other integrable models. 

Assuming the spectrum of the model consisted of an O$(n)$ vector 
multiplet of massive particles the exact S-matrix of the $n\geq3$ models 
was found by bootstrap methods \cite{ZZ}. For later use we adopt
the projector decomposition 
\be
S_{ab}^{cd}(\th) = S_0(\theta)\,(P_0)_{ab}^{cd} + 
S_1(\theta)\,(P_1)_{ab}^{cd} + S_2(\theta)\,(P_2)_{ab}^{cd}\,,
\label{SOn}
\ee
where
\ba
&& S_0(\th) = \frac{\th +i\pi}{\th -i\pi}\,S_1(\th)\;,\sspace 
S_1(\th) = \frac{(n-2)\th + 2\pi i}{(n-2)\th - 2\pi i} \,S_2(\th)\;,
\nonum
&& S_2(\theta)= -\exp\left\{2i\int_0^\infty\frac{d\omega}{\omega}
\sin\omega\theta  
\left[\frac{e^{-\pi\omega}+e^{-2\pi\frac{\omega}{n-2}}}
{1+e^{-\pi\omega}}\right]
\right\}\;.
\label{SIOn}
\ea
The projectors are those on the O$(n)$ singlet, vector, and symmetric
traceless tensors, i.e.  
\ba
(P_0)_{ab}^{cd}&=&\frac{1}{n}\delta^{ab}\delta^{cd}\,,
\nonum
(P_1)_{ab}^{cd}&=&\frac{1}{2}\Bigl(\delta^{ac}\delta^{bd}
-\delta^{bc}\delta^{ad}\Bigr)\,,
\nonum
(P_2)_{ab}^{cd}&=&\frac{1}{2}\Bigl(\delta^{ac}\delta^{bd}
+\delta^{bc}\delta^{ad}\Bigr) -\frac{1}{n} \delta^{ab}\delta^{cd}\,.
\label{proj}
\ea
Contact to the Lagrangian (\ref{LagOn}) can be made through 
quantum conserved charges of higher spin that prevent particle production. 
Under mild extra assumptions Polyakov \cite{Poly} and L\"uscher 
\cite{Lusch} have shown the existence of such respectively local and 
nonlocal higher spin conserved charges. The latter in particular 
anticipate \cite{Lusch} a Yangian structure and entail the 
factorization equations that dictate the S-matrix.

Much less is known about the O$(2)$ model. A simple observation
is that the amplitudes (\ref{SIOn}) have a smooth
$n\to2$ limit. This suggests that the O$(2)$ model might likewise be 
integrable and that its spectrum consists of a single O$(2)$ doublet
of massive particles whose scattering is described by the
$n\to2$ limit of the S-matrix (\ref{SOn}). Although taking this 
formal $n\to2$ limit is not convincing in itself, the conclusion is
corroborated by the KT theory \cite{KT} of the XY model
and its reformulation in the context of the Sine-Gordon theory \cite{Amit}.
Before turning to the KT theory we thus briefly digress on the Sine-Gordon 
(SG) model. Its Lagrangean can be written as
\begin{equation}
{\cal L}^{\rm SG}=\frac12\partial_\mu\phi\partial_\mu\phi
+\frac{\alpha}{\beta^2}\left[1-\cos(\beta\phi)\right],
\label{LagSG}
\end{equation}
where $\alpha$ has mass dimension $2$ and $\beta$ is the dimensionless
SG coupling. It is also integrable and its spectrum and S-matrix
was also found in \cite{ZZ}. The spectrum depends on $\beta$ 
in a complicated way but it becomes simple for the range 
$8\pi>\beta^2>4\pi$ when it is free of bound states
and consists of a single O$(2)$ vector of massive particles.
In terms of $\nu=\frac{8\pi}{\beta^2}-1$ this corresponds to 
$0 < \nu < 1$, and the S-matrix in this range
can be written in the projector form (\ref{SOn}), with $n=2$, 
where now 
\ba
&& S_0(\th|\nu) = \frac{\sh \frac{\nu}{2}(i\pi + \th)}%
{\sh \frac{\nu}{2}(i\pi - \th)}\,S_2(\th|\nu)\;,
\sspace 
S_1(\th|\nu) =  -\frac{\ch \frac{\nu}{2}(i\pi + \th)}%
{\ch \frac{\nu}{2}(i\pi - \th)}\,S_2(\th|\nu)\;,
\nonum
&& S_2(\th|\nu) = -\exp\left\{2i\int_0^\infty\frac{d\omega}{\omega}
\sin\omega\theta\,\frac{\sinh\frac{\pi\omega(1-\nu)}{2\nu}}
{2\cosh\frac{\pi\omega}{2}\sinh\frac{\pi\omega}{2\nu}}
\right\}.
\label{SSG}
\ea
Note that in the $\beta^2\to 8\pi$ ($\nu\to0$) limit the SG S-matrix
coincides with the $n\to2$ limit of the O$(n)$ S-matrix.

Finally, in the vicinity of $\beta^2 = 8\pi$ the SG model can also be 
related to a fermionic model \cite{Banksetal} formulated in terms 
of a two-component Dirac fermion. It has a manifest SU$(2)$ 
symmetry and is a variant of the chiral Gross-Neveu model with four-fermion 
interaction. The existence of this fermionic model sheds some light 
on the symmetry enhancement in the O(2) model discussed in Section 3.
For the details, however, the difference between $SU_{-1}(2)$ and $SU(2)$
is crucial; c.f.~the discussion at the end of Section 3.2. 

The relation to these other integrable QFTs can in particular 
be used to determine the infinite momentum limit
of the current two-point function $I(k)$ in (\ref{contCurr}). In the 
O(2) model the limit is finite and exactly calculable. 
One way of computing $I(\infty)$ is by noting that it coincides with
the coefficient of the Schwinger term in the current-current commutator. 
This commutator can be evaluated in the SG language using canonical 
quantization, and yields
\be
I(\infty)=\frac{2}{\pi}\,.
\label{2overpi}
\ee
The same result can be obtained using the relation to the before mentioned 
two-fermion model. Here, referring to the asymptotic freedom of the model
in the fermion coupling constant, only a simple free fermion calculation 
has to be done.

\newsubsection{Standard lattice action and KT theory}

The standard lattice action of the XY model is
\begin{equation}
S_{\rm XY}=K\sum_{x,\mu}\big[1-
\cos\big(\varphi(x)-\varphi(x+\hat\mu)\big)\big].
\label{standardS}
\end{equation}
We denote the inverse temperature (inverse of the bare coupling) of the
XY model by $K$ to avoid confusion with the SG coupling $\beta$.

This model has a high temperature phase at small $K$ with exponential
decay of correlations; it has been shown rigorously \cite{FrohSpen} that
at low temperature (large $K$) the correlations decay only like a power
of the distance. The exponential decay disappears therefore at a finite
critical value $K_c$; this is the famous KT transition predicted by 
Kosterlitz and Thouless \cite{KT}. They argued that at not too small
values of $K$ typical configurations of the model can be described
as a combination of `smooth', topologically trivial configurations
(spin waves) and a gas of vortices (of integer topological charge).
The vortices have a logarithmic interaction and therefore form
essentially a two-dimensional Coulomb gas, which has a transition
from a high temperature phase with Debye screening to a low
temperature dipole phase without screening. In the KT picture the
transition is therefore described as `vortex condensation'.
A different perspective of this kind of phase transition was proposed in
\cite{perc} according to which it is driven by the change
from instanton-like defects (vortices) to super-instantons dominating
at low temperatures.

Fr\"ohlich and Spencer \cite{FrohSpen} established a rigorous version of 
the correspondence between the XY model and a type of Coulomb gas 
and used rigorous arguments inspired by the renormalization group 
(grouping of charges into neutral `molecules') to show the absence of 
screening in this gas at low temperature.

Kosterlitz and Thouless employed heuristic energy-entropy considerations 
to show that
in the transition region only vortices of topological charge $\pm 1$
are important and higher vortices can be neglected. It is easy to see 
that this system (spin waves and Coulomb gas with 
unit charge vortices only) is exactly equivalent to the SG model. 
In ref.~\cite{Amit} it was argued that
the extremal SG fixed point $\beta^*=\sqrt{8\pi}\,,\alpha^*=0$ is 
appropriate to describe the KT phase transition. The renormalizability
of the SG model around this point was explicitly demonstrated up to
two-loop order in a double expansion in $\alpha$
and $\delta=\frac{\beta^2-8\pi}{8\pi}$.
\vspace{-4mm}

\newsection{Bootstrap description and symmetry enhancement}

Here we detail on the proposed bootstrap S-matrix, the associated
quantum group structure and its implications for the form factors
and the operator classification.  

\newsubsection{\boldmath{$SU_{-1}(2)$} invariance of the S-matrix}

The candidate S-matrix for the XY QFT can be rewritten as
\be
S_{ab}^{cd}(\th) = S_2(\th)\left[ \d_a^d\d_b^c + 
\frac{\th}{i\pi -\th} \d_{ab}\d^{cd}\right]\;,
\sspace S_2(\th)= 
\frac{\Gamma\left(\frac{1}{2} + \frac{\th}{2\pi i}\right)
\Gamma\left(- \frac{\th}{2\pi i}\right)}%
{\Gamma\left(\frac{1}{2} - \frac{\th}{2\pi i}\right)
\Gamma\left(\frac{\th}{2\pi i}\right)}\;.
\label{S1}
\ee
The S-matrix (\ref{S1}) satisfies the usual S-matrix postulates 
with the charge conjugation matrix $C_{ab} = \delta_{ab}$ and 
normalization $S_{ab}^{cd}(0) = -\d_a^d\d_b^c$. 
Note the nontrivial limit
\be
S_{ab}^{cd}(\pm \infty) = \d_a^d\d_b^c - \d_{ab} \d^{cd}\;,
\quad S(\pm \infty)^2 = \1\;.
\label{Slimit}
\ee 
The symmetry group of the massive ${\rm O}(2)$ model is of course 
${\rm O}(2)$, as far as the Lagrangian and the functional measure 
is concerned. The proposal (\ref{S1}) however entails 
that on the level of the S-matrix and the scattering states a
symmetry enhancement takes place, in that on them a nonabelian 
symmetry operates. In view of the known $\cU_q(su(2))$ 
quantum group symmetry of the Sine-Gordon S-matrix \cite{RS}
and the identification 
$q = - e^{i\pi(-1 + 8\pi/\beta^2)}$, one expects the symmetry to be 
$\cU_{-1}(su(2))$. As a Lie algebra this is the same as 
$\cU_{1}(su(2)) = su(2)$, but the comultiplication in $\cU_{-1}(su(2))$ 
differs from that in $su(2)$. We refer to Appendix A for some basic 
definitions and our conventions on $\cU_q(su(2))$. To simplify the 
notation we shall write $SU_{-1}(2)$ for $\cU_{-1}(su(2))$ from now on. 

The easiest way to see the  $SU_{-1}(2)$ invariance of the 
S-matrix (\ref{S1}) is to perform a projector decomposition. 
Defining $\check{S}_{ab}^{cd}(\th) := 
S_{ab}^{dc}(\th)$ (so that $\check{S}(0) = -\1$) it takes the 
form
\ba
&& \check{S}(\th) = S_2(\th)\left[\frac{i\pi + \th}{i\pi -\th} P_0 + 
P_1\right]\;,\sspace \mbox{with} \nonum
&& 
(P_0)_{ab}^{cd} = \frac{1}{2}\d_{ab}\d^{cd}\;,\sspace 
(P_1)_{ab}^{cd} = \d_a^c\d_b^d - \frac{1}{2}\d_{ab}\d^{cd}\;.
\label{S2}
\ea
Here $P_0 P_1 =0 = P_1 P_0$ and $P_0 + P_1 = \1$. Moreover 
$P_0$ and $P_1$ are the projectors onto the irreducible singlet
and triplet representation of $SU_{-1}(2)$, respectively.
For comparison let us note that the $SU(2)$ 
invariant S-matrix can likewise be written in 
the form (\ref{S2}), but the projectors are given by 
\begin{equation}
SU(2): \sspace 
(P_0)_{ab}^{cd} = \frac{1}{2}\left(\d_a^c\d_b^d - \d_a^d\d_b^c\right)\;,
\quad 
(P_1)_{ab}^{cd} = \frac{1}{2}\left(\d_a^c\d_b^d + \d_a^d\d_b^c\right)\;.
\label{S3}
\end{equation}
The S-matrices (\ref{S1}) and (\ref{S3}) are of course also
invariant under the usual real O(2) transformations. It is 
often advantageous to diagonalize this action by means of 
a unitary basis transformation 
\be
U = \frac{1}{\sqrt{2}}\left( \begin{array}{c c} 1 & i \\
1 & -i \end{array} \right) = U_{\alpha}^{\;\;a} \;,
\label{Udef}
\ee   
where we use Greek letters $\alpha,\beta,\ldots \in \{\pm\}$  
to label the components in the new basis. Explicitly 
$S_{\alpha\beta}^{\gamma \delta}(\th) :=
U_{\alpha}^a U_{\beta}^b S_{ab}^{cd}(\th) U^{\gamma}_c U^{\delta}_d$,
which now has $C_{\alpha\beta} = \delta_{\alpha + \beta,0}$ as its charge 
conjugation matrix. Written in matrix form one finds the familiar 
pattern for (\ref{S2}) 
\be
S(\th) = 
\left( \begin{array}{cccc} 
S_2 & 0   & 0   & 0  \\
0   & S_T & S_R & 0  \\
0   & S_R & S_T & 0  \\
0   &  0  &  0  & S_2
\end{array}\right)\;,
\quad S_T(\th) = \frac{\th}{i\pi -\th}S_2(\th)\;,\;\;\;
S_R(\th) = \frac{i\pi}{i\pi -\th}S_2(\th)\;, 
\label{S4}
\ee
where the rows and columns refer to the $(++,+-,-+,--)$ ordering. 
For the $SU(2)$ invariant S-matrix only the sign of $S_T$ would be 
flipped, which in view of the previous discussion however indicates 
a very different group theoretical structure. Concretely the 
$SU_{-1}(2)$ invariance of (\ref{S4}) amounts to 
\begin{equation}
\Sigma_{\pm} \,\check{S}(\th) = \check{S}(\th) \,\Sigma_{\pm}\,,
\quad \mbox{with} \quad \Sigma_+ = \Sigma_-^T = i 
\left( \begin{array}{c|ccc} 
\sz  & {} & {} & {}\\
-1 & {} & \mbox{\LARGE 0} & {}\\
\so& {} & {} & {}\\
\hline 
\sz  & -1 & 1 & 0 
\end{array} \right)\,,
\label{S5}
\end{equation}
and $\Sigma_{\pm}$ representing the `raising and lowering' operators of 
$SU_{-1}(2)$. 

\newsubsection{Form factors: \boldmath{${\rm O}(2)$} versus 
\boldmath{$SU_{-1}(2)$} covariance}

Form factors in this context are matrix elements of some 
local quantum field between the vacuum and a multi-particle 
scattering state. They can in principle be computed from a recursive
system of functional equations defined largely in terms of 
the given bootstrap S-matrix. Since the S-matrix has the 
enhanced $SU_{-1}(2)$ symmetry it is natural to ask whether 
the associated functional equations are likewise covariant.
Unlike the situation in other models this turns out to be 
{\it not the case} automatically, but it rather hinges on the 
values of certain statistics phases. Since this is a novel
feature we briefly outline the general structure of the 
form factor equations in the O(2) bootstrap theory here. 
Details are relegated to Appendix A. Explicit results for some 
operators of interest are given in the next section and Appendices B
and C.      

The form factors are tensors with respect to the obvious 
(real) action of O(2) rotations. As with the S-matrix it
convenient to diagonalize this action by the unitary
transformation (\ref{Udef}). We write 
$f_{\alpha_n\ldots \alpha_1}(\th_n,\ldots,\th_1)$ 
for the components of some $n$-particle form factor in this
``charged basis''. The terminology is motivated by the fact that 
under a $U(1)$ transformation a form factor picks up a phase
$e^{ie \varphi}$, where $e:= \alpha_n + \ldots + \alpha_1$ 
plays the role of the $U(1)$ charge. Equivalently $e$ is 
the weight with respect to the Cartan subalgebra generator 
of $SU_{-1}(2)$. A form factor (of an operator) of Lorentz 
spin $s$ should also have the homogeneity property
\be
f_{\alpha_n \ldots \alpha_1}(\th_n + u, \ldots ,\th_1 + u) = 
e^{s u} \,f_{\alpha_n \ldots \alpha_1}(\th_n, \ldots ,\th_1)\,.
\label{ffspin}
\ee 

For a fixed particle number $n$ a form factor then has to 
satisfy the functional equations
\begin{subeqnarray}
&& f_{\alpha_n\ldots \alpha_1}(\th_n,\ldots,\th_2,\th_1) = 
S_{\alpha_2\alpha_1}^{\delta\;\,\gamma}(\th_{21})\,
f_{\alpha_n\ldots \alpha_3 \gamma \delta}(\th_n,\ldots,\th_1,\th_2)\,,
\\[2mm]
&& f_{\alpha_n\ldots \alpha_1}(\th_n + 2\pi i,\th_{n-1},\ldots,\th_2,\th_1) = 
\Gamma_{\alpha_n}^\delta \,f_{\alpha_{n-1}\ldots \alpha_1\delta}
(\th_{n-1},\ldots,\th_1,\th_n)\,.
\label{ff3}
\end{subeqnarray}
In the second Eq.~the shift by $2\pi i$ is to be understood in the 
sense of analytical continuation and $\Gamma_{\alpha}^{\beta}$ is a 
constant matrix on whose role we elucidate below. First note that 
the system (\ref{ff3}) decomposes into decoupled sectors with fixed 
$U(1)$ charge $e \in \{n,n-2, \ldots, -n +2,-n\}$ and dimension
$n!/n_-!(n-n_-)!$, where $n_- =(n-e)/2$ is the number of `$-$' 
labels in $(\alpha_n,\ldots,\alpha_1)$. Correspondingly the matrix 
$\Gamma$ is diagonal in this basis but may be different in different 
charge sectors
\be
\Gamma_{\alpha}^{\beta} = \eta_{\alpha}(e) \,
\delta_{\alpha}^{\beta}\;.
\label{ff5}
\ee
The phases $\eta_{\alpha}(e)$ can be thought of as statistics 
phases describing the relative statistics of the (quasilocal) 
operator whose form factors are considered and the field that 
generates the scattering states in a Haag-Ruelle 
type scattering theory \cite{MNcycl}. See e.g.~\cite{Lash} 
for some simple examples. Iterating (\ref{ff3}b) and 
employing the analyticity in $u$ of (\ref{ffspin}) 
one finds the following {\it spin-statistics} relation
\be
\eta_-(e)^{n_-}\, \eta_+(e)^{n-n_-} = e^{2\pi i s}\;.
\label{ffspin-stat}
\ee
A further condition arises if the underlying operator is hermitian. From 
$f_{\alpha_n\ldots \alpha_1}(\th_n, \ldots, \th_1)^*$
\newline 
$= f_{-\alpha_1 \ldots -\alpha_n}(\th_1^* + i\pi, \ldots ,\th_n^* + i\pi)$
one obtains 
\be
\eta_{\alpha}(e) \eta_{-\alpha}(-e)^* =1\;.
\label{ff8}
\ee  

If only O(2) invariance is assumed no further constraints exist
and the phases $\eta_{\alpha}(e)$ are part of the specification of a 
field operator in the bootstrap framework. Collectively they encode
the information about the super selection structure of the theory. 
Since the first Eq. in (\ref{ff3}) is covariant also with respect to
the larger nonabelian $SU_{-1}(2)$ symmetry, it is natural to 
ask whether or not also the second Eq. is covariant for a 
suitable choice of the phases. The covariance requirement links the 
charge $e$ sector with the $e \pm 2$ sectors. It can be 
seen to entail an overdetermined set of relations for the
phases $\eta_{\alpha}(e)$, -- which turn out to be self-consistent. 
The requirement of quantum group covariance thus determines all 
phases $\eta_{\alpha}(e)$ essentially uniquely; c.f.~Appendix A.
For $n \leq 4$ one finds explicitly: 
\ba
n=2: && \eta_+(2) = - \eta_{\pm}(0) = \eta_-(-2)\;,
\nonum
n=3: && \eta_+(3) = \mp \eta_{\pm}(1) = \pm \eta_{\pm}(-1) 
= \eta_-(-3) \;,
\nonum
n=4: && \eta_+(4) = - \eta_{\pm}(2) = \eta_{\pm}(0) =
-\eta_{\pm}(-2) = \eta_-(-4)\;.
\label{cphases}
\ea
Generally, for fixed $n$, the relative signs are given by 
$\eta_{\alpha}(e) \sim \exp \frac{i\pi}{2}(e - n \alpha)$.
Of course the actual phases solving (\ref{cphases}) must 
be chosen $n$-independent. 

Let us illustrate the use of these relations 
in the charge $e=1$ sector (where only the odd particle form factors 
are nonzero). We can take $\eta_{\pm}(1) = e^{\pm 2\pi i s}$ 
as the solution of (\ref{ffspin-stat}). Then (\ref{ff8}) fixes 
$\eta_{\pm}(-1) = e^{\mp 2\pi i s}$. If we now in addition require 
that the field underlying these form factors is quantum group covariant,
the $e = \pm 1$ sectors are linked by (\ref{cphases}), e.g. for 
$n=3$. This yields the condition $e^{\mp 4 \pi i s} = -1$, and we
conclude: $s = 1/4 \,{\rm mod}\, 1/2$. In words, {\it an O(2) doublet
field that is in addition quantum group covariant can only have
Lorentz spin $s = 1/4\, {\rm mod}\,1/2$.} If we had started from
the $SU(2)$ invariant S-matrix (\ref{S3}) instead, no relative 
signs in (\ref{cphases}) would have occurred, and an $SU(2)$ 
doublet of O$(2)$ charge $e = \pm 1$ was forced to have Lorentz
spin $s = 1/2\,{\rm mod}\, 1/2$, as expected.

Next we proceed to the residue equations which link an $n$-particle 
form factor to an $n\!-\!2$ particle form factor. Consistency 
requires that the inverse of the matrix $\Gamma_{\alpha}^{\beta}$ 
appears on the right hand side, irrespective of its concrete form. 
In the charged basis the precise formula is given in (\ref{ffres}). 
For generic phases (\ref{ffres}) will only be O(2) covariant. 
Concretely this means 
that an $n$-particle form factor of $U(1)$ charge $e$ is linked to 
an $n\!-\!2$ particle form factor with the same $U(1)$ charge. 
For the choice (\ref{cphases}) ensuring the $SU_{-1}(2)$ covariance 
of (\ref{ff3}) one expects that also (\ref{ffres}) is quantum group 
covariant, which indeed turns out to be the case. 
Since all form factor Eqs then are covariant a quantum group 
transformation will map one solution into another solution, where 
``solution'' actually means a sequence of functions whose members 
are linked by the (\ref{ffres}).  
Suitable sequences should correspond to (local) quantum fields in the 
O(2) model. We thus find that field operators whose statistics phases 
enjoy the specific relation (\ref{cphases}) form multiplets with 
respect to the quantum group action. Clearly one can concentrate on the 
multiplets transforming irreducibly; the multiplet and the 
associated form factor sequence will then be characterized by an 
isospin quantum number stemming from the representation theory 
of $SU_{-1}(2)$. In appendix A we list the irreducible multiplets 
for $n \leq 4$. 

We can summarize the situation as follows: The functional equations 
characterizing the form factors 
are O(2) invariant and decompose into decoupled sectors with fixed 
$U(1)$ charge $e$. In each sector statistics phases $\eta_{\alpha}(e)$ 
enter that are part of the specification of a field operator (or of 
an O(2) multiplet thereof) in the bootstrap framework. In addition 
operators from {\it different} charge sectors whose statistics phases 
enjoy the particular relation (\ref{cphases}) are members of an 
(irreducible) multiplet with respect to the nonabelian quantum group 
$SU_{-1}(2)$. The existence of these multiplets is a nontrivial
prediction of the bootstrap formulation.     

As remarked earlier there exists a (non-rigorous) transformation 
of the O(2) model into a fermionic model with a manifest SU(2) 
invariance, for which the natural candidate S-matrix is (\ref{S3}),
i.e.~that of the SU(2) chiral Gross-Neveu model. 
As with the O(2) -- Sine-Gordon correspondence we expect that both 
systems share {\em subsets} of fields with identical correlation 
functions. An interesting one-to-one correspondence of the fields 
however is unlikely. To see this let us discuss the relation 
between the bootstrap systems based on the $SU_{-1}(2)$ 
invariant S-matrix (\ref{S1}) and the $SU(2)$ S-matrix (\ref{S3})
in more detail: In the charged basis the mapping 
\be 
|\th_n,\ldots ,\th_1\ket_{\alpha_n \ldots \alpha_1} \rra 
\prod_{j=1}^n (\alpha_j)^j\,
|\th_n,\ldots ,\th_1\ket_{\alpha_n \ldots \alpha_1}\,,
\label{untwist}
\ee
maps states whose exchange relations are governed by the $SU(2)$ S-matrix 
(\ref{S3}) onto those whose exchange relations are governed by 
the $SU_{-1}(2)$ S-matrix (\ref{S1}), and also `untwists' the 
non-trivial comultiplication of $SU_{-1}(2)$ \cite{Smir1}. 
However (\ref{untwist}) does not induce an interesting correspondence
of the form factor sequences. For example if the $SU(2)$ statistics
phases are taken to be unity, the mapping (\ref{untwist}) induces
$\eta_{\alpha}^{\rm induced}(e) = e^{i\pi(e - n \alpha)/2}$ for the  
$SU_{-1}(2)$ bootstrap system. This flips sign under $n \ra n-2$
while the statistics phases of a sequence acceptable for describing 
a $SU_{-1}(2)$ covariant field operator of course must be 
$n$-independent. Mathematically one can set up a correspondence between 
solutions of the form factor equations based on the $SU(2)$ and the 
$SU_{-1}(2)$ S-matrix. One way of doing this is to substitute the respective
S-matrices into the Bethe Ansatz inspired integral formulae of 
\cite{Smir1, Karowskietal}. However this correspondence will in
general {\it not preserve} the spin, the statistics phases, or
even the covariance under the global symmetry group. One must conclude 
that there is no physically interesting one-to-one correspondence 
between the field content of the bootstrap systems based on the 
$SU(2)$ and on the $SU_{-1}(2)$ S-matrix.

\newsubsection{Spin, Parafermion and Current form factors 
for \boldmath{$n \leq 4$}}

With these preparations at hand we now seek to determine the form 
factors of the Noether current $J_{\mu} = \frac{1}{g^2}( S^1 \dd_{\mu} S^2 - 
S^2 \dd_{\mu} S^1)$ and the basic Spin field $S^a,\, a=1,2$. 
The latter is an O(2) doublet and carries Lorentz spin 
$s=0$. From the discussion following (\ref{cphases}) we
conclude that it cannot be a quantum group doublet as well. 
We thus also search for the form factors 
of an additional {\it local} ``parafermion'' field that is a 
$SU_{-1}(2)$ doublet with Lorentz spin $1/4$. (We shall comment
on the relation to Smirnov's parafermion in the SG model \cite{Smir2} 
below.) Technically the construction of form factors for the Spin 
and the parafermion field is very similar. In order to treat both 
cases simultaneously we write $\Phi_s^a(x)$ for the renormalized 
field operators, with $s=0,1/4$, corresponding to the spin and 
the parafermion field, respectively. The objects of interest then are
\ba
\nspace 
\bra 0 |J_{\mu}(0) |\th_n,\alpha_n;\ldots ; \th_1,\alpha_1\ket 
\is
-i \eps_{\mu\nu} \left(\sum_{j=1}^n p^{\nu}(\th_j)\right) 
f_{\alpha_n\ldots \alpha_1}(\th_n,\ldots,\th_1) \;,\quad 
\mbox{$n$ even}\,,
\label{curr1}\\
\nspace 
\bra 0 |\Phi_s^{\alpha}(0) |\th_n,\alpha_n;\ldots ; \th_1,\alpha_1\ket 
\is
f^{\alpha}_{\alpha_n\ldots \alpha_1}(\th_n,\ldots,\th_1) \;,\quad 
\mbox{$n$ odd}\,. 
\label{spin1}
\ea
Here all components refer to the charged basis. The current form 
factors have charge $e =0$ while that of $\Phi_s^{\pm}(x)$ have
charge $e = \pm 1$. The prefactor in the current form factor 
ensures current conservation; the
on-shell momenta are $p_0(\th) =M\ch \th$, $p_1(\th) = M \sh\th$.
As always in the form factor bootstrap Eqs (\ref{curr1}), (\ref{spin1}) 
must be regarded as a ``statement of intent''. That is, the 
right hand is computed through the functional equation while  
the interpretation as the matrix elements aimed at on the 
left hand side has to be justified by additional considerations. 

As input for the recursive functional equations the normalization of 
the starting members has to be fixed. For the current a preferred 
normalization stems from the fact that the associated Noether 
charge $Q$ should induce O(2) rotations on the spins, i.e. 
$[Q, S^a] = i \eps_{ab}S^b$. For the 2-particle form factor 
this converts into 
\be
f_{+-}(\th_2,\th_1) = - f_{-+}(\th_2,\th_1) = 
\frac{2i}{\th_2 - \th_1 - i\pi}
+ \ldots \;, \quad \th_2 \ra \th_1 + i\pi\,,
\label{curr2}
\ee  
where the dots denote regular terms. Writing $f_{+-}(\th_2,\th_1) 
= f(\th_2 -\th_1)$ the function $f(\th)$ has to 
satisfy the functional eqs $f(\th) = S_2(\th)f(-\th)$ and 
$f(\th + 2\pi i) = - f(-\th)$, with $S_2(\th)$ from (\ref{S1}). 
They can be solved in terms of the function
\ba
y(\th) &:=& \sh\frac{\th}{2} e^{\Delta(\th)}\;,\nonum
\Delta(\th) &:=& \int_0^{\infty} \frac{dt}{t} 
\frac{\ch t(1 + i \th/\pi) -1}{(1 + e^t) \sh t}\;,
\label{y1}
\ea
which enjoys the following properties
\ba
&& y(\th) = S_2(\th) y(-\th)\;,\quad y(\th + 2\pi i) = y(-\th)\,,
\nonum
&& y(\th) y(\th + i\pi) = - \frac{\pi^{3/2} e^{\Delta(0)}}%
{\Gamma\left(\frac{1}{2} - \frac{\th}{2\pi i}\right)
\Gamma\left(\frac{\th}{2\pi i}\right)}\;,\quad y(i\pi) = i\;.
\label{y2}
\ea 
We also note 
\be
\Delta(0) = 0.304637\;,\sspace 
\lim_{\th\to+\infty}\left[\Delta(\th)  - \Delta(-\th)\right]= 
\frac{i\pi}{2}\;.
\label{y3}
\ee
Taking into account the residue condition (\ref{curr2}) one 
obtains 
\be
f_{+-}(\th_2,\th_1) = - f_{-+}(\th_2,\th_1) = i \frac{y(\th_2-\th_1)}%
{\ch\frac{\th_2 -\th_1}{2}}\;.
\label{curr3}
\ee 
With the 2-particle form factor explicitly known one can 
proceed to the 4-particle form factor. The formulas now get 
more involved and we defer the details to appendix B. 
The Lorentz spin $s=1$ is readily checked to be compatible 
with $SU_{-1}(2)$ covariance via (\ref{ffspin-stat}) -- 
(\ref{cphases}). The (one-component) Noether current in the O(2) model 
is the neutral member of a hidden quantum group triplet \cite{Smir1}, 
although the charge $e=\pm 2$ partners are nonlocal in the spin field. 

For the $\Phi_s^{\alpha}(x)$ fields a natural normalization is 
\be
\bra 0|\Phi_s^{\alpha}(0)|\th,\beta\ket = f_{\beta}^{\alpha}(\th) = 
\delta_{\alpha}^{\beta}\,e^{\alpha s\,\th}\;.
\label{spin2}
\ee 
Proceeding to the 3-particle form factor we note that 
because of charge conservation and hermiticity 
there are only three independent components
\ba
&& f^+_{++-}(\th_3,\th_2,\th_1) = f_1(\th_1,\th_2,\th_3) = 
f^-_{--+}(\th_3,\th_2,\th_1)\;,\nonum 
&& f^+_{+-+}(\th_3,\th_2,\th_1) = f_2(\th_1,\th_2,\th_3) = 
f^-_{-+-}(\th_3,\th_2,\th_1)\;,\nonum 
&& f^+_{-++}(\th_3,\th_2,\th_1) = f_3(\th_1,\th_2,\th_3) = 
f^-_{+--}(\th_3,\th_2,\th_1)\;. 
\label{fpcomp}
\ea
The general 3-particle residue equation (\ref{ffres}) in the 
charge $e = \pm 1$ sectors 
\ba
\frac{i}{2} {\rm res}_{\th_{3,2} = i\pi} 
f^{\alpha}_{\alpha_3 \alpha_2\alpha_1}(\th)
\!\is \!\delta_{\alpha_3 + \gamma}\left[
\eta_{\gamma}(e)\inv \, S_{\alpha_2 \alpha_1}^{\beta \;\,\gamma}(\th_{21}) 
-\delta_{\alpha_2}^{\gamma} \delta_{\alpha_1}^{\beta} \right]
f^{\alpha}_{\beta}(\th_1)\;,
\ea
thus translates into 
\be
f_k(u,v,\th) \approx \frac{2 i}{u - v - i\pi} W_k(v - \th)\;,
\sspace u \ra v + i\pi\;,
\label{translates}
\ee
where with $\eta := \eta_+(1) = e^{2\pi i s}$ and the notation from
(\ref{S4}) 
\be
W_1(\th) = -\eta S_R(\th)\;,\quad 
W_2(\th) = 1-\eta S_T(\th)\;,\quad 
W_3(\th) = 1 - \eta^{-1} S_2(\th)\;.
\ee
These functional equations can be solved for generic $s$, the 
solution is described in appendix C. For the spin field $s=0$ fixes 
our candidate form factors. For the parafermion
field we know that $SU_{-1}(2)$ covariance (regardless of irreducibility) 
requires $s = 1/4$. (The specific construction 
used in appendix C removes the additive mod $1/2$ ambiguities). As 
explained in appendix A the condition that the parafermion field 
(and hence its form factors) transform irreducibly as a isospin $1/2$ 
doublet requires in addition 
\be
\zeta(\th_3,\th_2,\th_1) := 
f_1(\th_3,\th_2,\th_1) - f_2(\th_3,\th_2,\th_1) +
f_3(\th_3,\th_2,\th_1) \stackrel{{\displaystyle !}}{=} 0\,.
\label{irredcond}
\ee
For the solution constructed in Appendix C, $\zeta(\th_3,\th_2,\th_1)$
can be shown to be proportional to $\eta^2 +1$, so that $s = 1/4$ 
also entails the desired irreducible transformation law. 

So far we didn't say anything about the local structure 
of the parafermion field $\Phi_{1/4}^{\alpha}(x)$ supposed to 
underly the above solution of the form factor equations. 
We can address this point by employing a result by Smirnov 
\cite{Smir2} on the existence of parafermionic fields in the 
Sine-Gordon model. Smirnov's fields have well defined exchange 
relations of the form $\Psi_a(x) \Psi_b(y) = S_{ab}^{cd}(\pm \infty) 
\Psi_d(y) \Psi_c(x)$, for $\pm x^1 \! > \! \pm y^1$, where $x^1,y^1$ 
are the space components of $x,y$ in a fixed Lorentz frame. 
For generic Sine-Gordon coupling $\beta$ these fields are nonlocal
because the two limiting S-matrices are distinct. From the 
analysis of the form factors of the energy momentum tensor 
one can see that our parafermion field is the $\beta^2 \ra 
8\pi$ limit of Smirnov's field. But thanks to (\ref{Slimit}) 
it is now a {\it local} field. Indeed for the charged components 
the exchange relations assume the simple form
\ba
&& \Phi^{\pm}_{1/4}(x)\Phi^{\pm}_{1/4}(y)
= \Phi^{\pm}_{1/4}(y) \Phi^{\pm}_{1/4}(x)\,,\nonum
&& \Phi^{+}_{1/4}(x)\Phi^{-}_{1/4}(y)
= - \Phi^{-}_{1/4}(y) \Phi^{+}_{1/4}(x)\,,
\ea 
for {\it all} spacelike separated points $x,y$. Likewise the transformation
properties under the quantum group change qualitatively. 
As analyzed by several authors \cite{Smir2, SmiLeCl,FeLeCl} in general 
a dynamical quantum group symmetry of the S-matrix acts in a nonlocal 
way on the field operators. In contrast the field $\Phi_{1/4}^{\alpha}(x)$ 
transforms nicely as an irreducible $SU_{-1}(2)$ doublet. The 
situation is thus reminiscent of the Ising model, which 
can be viewed as the $n=1$ case of (\ref{LagOn}).
There both the spin and the fermion are local fields    
of Lorentz spin $0$ and $1/2$, respectively. Both 
are relatively non-local but generate equivalent
sets of scattering states. Though in the absence of a field 
theoretical construction of the parafermion field it is difficult 
to examine this point, we expect the interplay between the spin 
and the parafermion field in the XY QFT to be 
analogous. 



\newpage
\newsection{Lattice computations}
\vspace{-1cm}

\newsubsection{General setup}

For the lattice regularization we consider a square lattice 
with action
\be
S=-K\sum_{x,\mu}S(x)\cdot S(x+\hat{\mu})\,,
\label{staction}
\ee
where $S(x)\cdot S(x)=\sum_a S^a(x)S^a(x)=1$. Solving the constraint with 
$S^1(x)=\cos\varphi(x)$, $S^2(x)=\sin\varphi(x)$
the action reduces to the standard XY lattice action Eq.~(\ref{standardS}).
The correlation functions are defined as in the continuum theory except 
that the spatial integrals are replaced by discrete sums. There is 
an enormous literature, both on numerical simulations of the XY model 
\cite{XYMC} and on its high temperature expansion \cite{HTXY1,HTXY2}.
Presently the best numerical estimate of the critical
point for the standard action is \cite{Hasenbusch}
\begin{equation}
K_c=1.1199(1)\,.
\end{equation}
Previous numerical investigations mostly concentrated on the 
comparison with the KT theory. We will outline
the aspects relevant here and some refinements in Section 6.1. 
Our main goal however is to compare the continuum limit of the 
XY model with the O(2) bootstrap theory.

In the rest of this section we collect some general information 
on our simulations and continue with detailed discussions of the 
measurement of various observables in Sections 5 and 6. 
All numerical simulations were done on an $L\times T$ lattice
with periodic boundary conditions in each direction. 
During the entire investigation two independently written programs
were employed and many cross checks were made. 
Both used multi-cluster updating. The Ising spins are embedded
like in Wolff's single cluster algorithm \cite{Uli}; the 
resulting Ising model is then updated with a generalization of 
the Swendson-Wang \cite{SW} multi-cluster algorithm. 
In one application of the program each run started from a random
configuration and consisted of a large number of sweeps of which
a large initial proportion were used solely for equilibration. 
In another application after initial equilibration 
the final configuration of the run was stored
and read in for the starting configuration of the next run.
In most cases the observables were measured using 
improved (cluster) estimators. The final data sample of the many 
runs was averaged and the error was computed using the jack-knife method.

Since we aim at achieving high precision,
for many quantities to an accuracy of $<1\%$, a considerable
source of concern to us was the random number generator (RNG). 
Indeed at an initial stage of this project we found that 
results obtained by various RNGs could differ by many standard
deviations. We thus subjected the RNGs to several tests, specific 
to the model and the quantities considered here. The results are 
reported in Appendix E. 

Numerical simulations of course are restricted to work in finite 
volume. In the next two subsections we therefore discuss finite volume
effects, first in the continuum and then on the lattice.  


\newsubsection{Finite volume effects in the continuum}

If a continuum QFT is confined in a box
the physical observables will depend on the geometry
and boundary conditions. Consider first the mass $m(L)$
of the 1-particle state on a cylinder of circumference $L$.
L\"{u}scher \cite{LuVol} has shown that in a theory
without a ``three--point coupling" of this particle to itself
or to any other single--particle state, for large physical
volumes $z=ML\gg1$ 
the finite volume dependence of the mass is given by:
\be
D(z)\equiv\frac{m(L)-M}{M}=
\frac{1}{\pi}\int_0^{\infty}d t\,e^{-z\cosh t}\cosh(t)f(t)
+\dots\,,
\label{volmassdep}
\end{equation}
where $f(t)$ is the forward scattering amplitude at an off--shell
point. Thus in these models the infinite volume limit is
approached rapidly at a rate $\sim\exp(-z)$ (from above if \break
$f(0)>0$).

More explicitly in the O($n$) sigma-models the amplitude $f$ is given by
\be
f(t)=n - \frac{1}{2n}\Big[2 S_0(\th) + n (n-1) S_1(\th) 
+ (n+2)(n-1) S_2(\th) \Big]_{\th = t + i\pi/2}\,.
\ee
With the proposed S-matrix (\ref{SOn}), (\ref{S1}) of the O(2) model 
one obtains
\be
f(t)=2+\frac{2\pi^2}{t^2+\pi^2/4}
\left|\frac{\Gamma(3/4-it/2\pi)} {\Gamma(1/4-it/2\pi)}\right|^2\,,     
\end{equation}
in which case Eq.~(\ref{volmassdep}) results in 
\be
D(z)\sim 1.162475182 \,\frac{e^{-z}}{\sqrt{z}}
\left[ 1+0.36432/z+ O(1/z^2)\right]\,.
\label{volmassdepO2}
\end{equation}

Next we consider the zero-momentum coupling in a 
square box of length $L$ in each direction. 
To get a feeling of the finite volume effects consider
the expression for $\gr(L)$ in the leading order $1/n$
expansion~\cite{CPRVgR}. One obtains 
\begin{equation}
\gr(L)=\gr(\infty)[1- \sqrt{8\pi}\, z^{A_4}
{\rm e}^{-z}(1+O(1/z))+\cdots]\,,
\label{FS1n}
\end{equation}
with $A_4=1/2$. Although the $1/n$ expansion is not expected to 
be quantitatively applicable to $n=2$, one might hope that the
qualitative features are correct, in particular that the
finite volume effects are exponentially suppressed and
secondly that the exponent $A_4$ of the multiplicative power correction
is independent of $n$. Some corroboration of this might come from 
investigating the finite volume effects in a square box in an effective 
Lagrangean framework similar to \cite{LuVol}. 


\newsubsection{Finite volume effects in the lattice regularization}

The particular lattice sizes used will be specified later,
but they were generally selected to enable studies of finite size 
effects at fixed correlation length and vice versa, subject of course 
to restrictions due to the CPU power available to us.
  
For a coupling $\gr(K,L)$ depending on the
bare coupling $K$ and size $L\times L$ which tends to $\gr$
in the continuum and infinite volume limits, we adopted the 
O$(n)$ definition   
\be
\gr(K,L)= \left(\frac{L}{\xi}\right)^2
\Bigl[1 + \frac{2}{n} -\frac{\langle (\Sigma^2)^2\rangle}%
{\langle \Sigma^2\rangle^2}\Bigr]\,,
\label{lattgr}
\ee
for $n=2$, where $\Sigma^a=\sum_x S^a(x)$. In Eq.~(\ref{lattgr}) 
and throughout Sect.~6, $\xi$ is taken to be
\be
\xi=\frac{1}{2 \sin(\pi/L)} \sqrt{\frac{G(0)}{G(k_0)}-1}\,,
\label{xisecmom}
\ee
where $k_0=(2\pi/L,0)$; c.f.~ref.~\cite{FMPPT}. This correlation 
length converges in the
thermodynamic limit to the second moment correlation length $1/\Mr$. 
In this connection we would like to draw the reader's attention 
to a subtlety which is discussed at the end of Section 4 in
ref.~\cite{SigmacollabII}.

The Noether current on the lattice is defined by
\be
J_\mu(x)=K\left\{ S^1(x)\partial_\mu S^2(x)
-S^2(x)\partial_\mu S^1(x)\right\} \,,
\label{latcurr1}
\ee
where $\partial_\mu f(x)=f(x+\hat{\mu})-f(x)$.
Introducing its two point function as 
\be
J_{\mu\nu}(q|K,L)=\sum_x e^{iqx}\langle J_\mu(x)J_\nu(0)\rangle\,,
\label{latcurr2}
\ee
with $q = (q_1,q_2)$, the well-known Ward identity (for the standard action) 
reads
\be
\sum_\mu \left(1-e^{iq_\mu}\right)J_{\mu\nu}(q|K,L)=\frac{K}{2}
\left(1-e^{iq_\nu}\right)E(K,L)\,,
\label{WI}
\ee
where the energy expectation $E$ is 
\be
E(K,L)=\sum_\mu \langle S(x)\cdot S(x+\hat{\mu})\rangle\,.
\ee
Next we wish to define for a finite periodic lattice the counterpart 
of the transverse current correlation function $I(k^2)$ in (\ref{contCurr}).
$J_{\mu\nu}$ can naturally be decomposed into 3 pieces (`transverse', 
`longitudinal' and `harmonic'), as discussed in \cite{PSward}. 
The harmonic piece is concentrated at the origin in momentum space 
and has the value $J_{11}(0|K,L)$. 

In the thermodynamic limit, because of the presence of a mass gap,
$J_{\mu\nu}(q|K,L)$ will be a real analytic function;  so it cannot
contain any remnant of the harmonic piece (which would be proportional
to a $\delta$ function). The longitudinal and the transverse parts
have to go to the same limit as $q\to 0$ to avoid any non-analyticity
at zero momentum. Being a contact term the value of $J_{\mu\nu}(0|K,L)$ 
has no physical meaning. For this reason, as explained already 
in Section 2.1, we define the transverse part in such a way that it
vanishes at zero momentum in the thermodynamic limit.
This suggests the definition
\be
I((0,q_2)|K,L) :=J_{11}((0,0)|K,L)-J_{11}((0,q_2)|K,L)\,,
\label{SUB0}
\ee
to which we shall refer as the SUB definition. It ensures $I(0)=0$
at finite $L$ and $\xi$.  

Another possible definition is suggested by the Ward identity, which 
for momenta of the form $q=(0,q_2)$ reads
\be
I((0,q_2)|K,L) :=\frac{K}{2}E(K,L)-J_{11}((0,q_2)|K,L)\,.
\label{WARD}
\ee
We can use (\ref{WARD}) as an alternative definition for a lattice 
counterpart of $I(k^2)$ in (\ref{contCurr}), to which 
we shall refer to as the WARD definition. The normalization $I(0)=0$ 
is then only restored in the thermodynamic limit, but for numerical
purposes the WARD version often is advantageous. To verify that 
(\ref{WARD}) vanishes at $q=0$ in the thermodynamic limit 
note that Eq.~(\ref{WI}) also entails  
\be
J_{11}((q_1,0)|K,L)=\frac{K}{2}E(K,L),\sspace \forall \,q_1\not=0\,.
\label{identity}
\ee
Since $J_{11}((q_1,0)|K,L)$ becomes a real 
analytic function of $q_1 \in [-\pi,\pi]$ for $L \ra \infty$ 
Eq.~(\ref{identity}) remains valid also for $q_1=0$ \cite{PSward}. 
In the thermodynamic limit therefore the WARD and the SUB 
definitions are equivalent. In Section 6 we shall consider both options.

Unfortunately few rigorous results exist on finite size effects for 
$z=L/\xi(K,L)\gg1$ on the lattice. In general we expect in the 
O($n$) models, for a large class of observables ${\cal O}$ 
and for fixed bare coupling $K$, the finite size effects to be 
either of the form 
\be
{\cal O}(K,L)={\cal O}(K,\infty)+z^{A^{\cal O}}e^{-z}
\left[C_0^{\cO}(K)+C^{\cO}_1(K)\frac{1}{z}+O(1/z^2)\right]
+\dots \,,
\label{finvollattansatz}
\ee
or the same with 
$C_i^{\cO}(K)$ replaced by ${\cal O}(K,\infty)C_i^{\cO}(K)$.  
Here either the amplitudes $C^{\cO}_i(K)$ or 
${\cal O}(K,\infty)C_i^{\cO}(K)$  are hoped to be almost constant for
large 
correlation lengths. Further the dynamical assumption enters that the 
fall-off is 
$e^{-\alpha z}$ with $\alpha=1$, as expected in the absence of 
two-particle bound states. Further it seems reasonable to expect 
$A^{\cal O}$ to depend only on the form of the observable and not 
on the dynamics e.g.~not on $n$. 

As mentioned earlier one can probably, for certain correlation functions,
better justify these assumptions by performing a Feynman diagram analysis
in the framework of an effective massive lattice field theory 
analogous L\"{u}scher's in the continuum theory \cite{LuVol}. 
Here we only indicate some plausibility considerations based on 
the leading order in the lattice $1/n$ expansion (viewed as a 
summation of bubble diagrams). For the coupling this gives
\ba
&& n\gr(K,L) = 2\triangle(0)/m_0^2 + O(1/n)\,,
\nonum
&& \mbox{with} \quad \triangle(0)^{-1} =\frac{1}{L^2}\sum_p
(E_p+m_0^2)^{-2}\,,\sspace 
K =\frac{n}{L^2}\sum_p \left(E_p+m_0^2\right)^{-1}\,.
\ea
Here $\xi=1/m_0$ and the sums range over momenta $p_{\mu} = 
2\pi\, n_{\mu}/L$, with $0\le n_\mu\le L-1$, $\mu =1,2$, 
and $E_p=4\sum_\mu \sin^2 p_\mu/2$. 
For the current two point function the leading term is 
\be
J_{\mu\nu}(q|K,L)=\frac{n(n-1)}{2}\frac{1}{L^2}\sum_p
\frac{\Bigl[e^{ip_{\mu}}-e^{-iP_{\mu}}\Bigr]
\Bigl[e^{-ip_{\nu}}-e^{iP_{\nu}}\Bigr]}%
{\Bigl[E_p+m_0^2 \Bigr]\Bigl[E_P+m_0^2 \Bigr]}\,,
\ee
where $P=p+q$, and for the energy expectation it is 
\be
E(K,L)= \frac{n}{K}\frac{1}{L^2}\sum_p \frac{\sum_{\mu}\cos
p_{\mu}}{E_p+m_0^2}\,.
\ee
Starting from these formulae one can study separately the 
finite volume effects and lattice artifacts in this approximation;
the results confirm the before mentioned assumptions.  
We stress again the big qualitative difference between the finite 
volume effects and the lattice artifacts. Whereas the finite volume 
effects represent continuum physics and hence are expected to be 
structurally universal, the lattice artifacts are in general non-universal.
In particular for the spin 2-point function one finds in this framework
$A^{\cal O}= A_2=-1/2$ for the exponent in (\ref{finvollattansatz}), 
whereas for the current and $\gr$ we get $A^{\cal O}=A_4=1/2$.
(Both the current 2-point function and $\gr$ depend linearly on
spin $4$-point functions and perhaps this accounts for the same 
exponent $A_4$.) If one does not wish to adopt this framework the 
exponents can be kept as fit parameters; c.f.~Section 6.   

\newsubsection{Lattice artifacts}

After the extrapolation to infinite volume has been performed, 
the results can be regarded as corresponding to a lattice 
O$(2)$ action in infinite volume. The extrapolation to 
infinite correlation length is usually hampered by the 
lack of information about the rate of approach. 
Based on the Sine-Gordon description of the KT transition \cite{Amit} 
one of us \cite{logs} has argued that for the XY model, 
say with standard action (\ref{staction}), the leading lattice artifacts 
are {\it calculable} from the continuum Sine-Gordon theory. This applies to 
observables like the S-matrix or the two-point function of the 
Noether current, where already at finite correlation length a 
preferred normalization exists. Implicit in this proposal is a 
certain degree of action-independence of the leading lattice 
artifacts, but at present it is not clear to which class 
of actions it applies. Later on we test this proposal 
for the standard action and the two observables mentioned. 

Let $U_{\rm XY}(\xi)$ denote such an observable, where $\xi$ is the 
correlation length in lattice units and the dependence on other 
variables is suppressed. Then $U_{\rm XY}(\xi)$, computed e.g.~with 
the standard action is predicted to 
be of the form \cite{logs} 
\begin{equation}
U_{\rm XY}(\xi)=
u_0-\frac{u_1\pi^2}{4(\ln\xi+u)^2}+ O\big((\ln\xi)^{-4}\big).
\label{xiexp}
\end{equation}
Here $u_0$ is the continuum value and $u_1$ is the leading 
correction. The parameter $u$ is action-dependent but should not 
depend on the physical quantity considered. Note the 
extremely slow decay $\sim 1/(\ln\xi)^2$.  
As explained in \cite{logs} the peculiar structure of 
the KT phase diagram allows one to relate both $u_0$ and $u_1$ 
to the continuum SG theory. Namely if $U_{\rm SG}(\nu)$ 
denotes the counterpart of the physical quantity considered 
in the continuum SG theory with coupling $\nu$ close to 
$\nu =0$, then 
\begin{equation}
U_{\rm SG}(\nu)=u_0+u_1\nu^2 + O(\nu^4)\,.
\label{gammaexp}
\end{equation}
The equality of the leading $u_0$ term in (\ref{xiexp}) and (\ref{gammaexp}) 
simply reexpresses the link between the XY model and the SG QFT alluded to 
in the introduction. Remarkably the coefficient $u_1$ of the 
first correction is likewise the same in both cases.
 
Our first application of these formulae is to the scattering phase shifts. 
Recall the SG model S-matrix (\ref{SOn}) with (\ref{SSG}).
Introducing the phase shifts by $S_I(\th|\nu) = \exp (2 i \delta_I(\th|\nu))$,
$I=0,1,2$, and expanding in $\nu$ at fixed $\th$ yields
\begin{equation}
\delta_I(\th|\nu) = \delta_I(\theta)+ \nu^2 \delta'_I(\theta)+ O(\nu^4)\,.
\label{phaseshifts}
\end{equation}
By construction $\delta_I(\th)$ are the phase shifts of the 
proposed O(2) S-matrix (\ref{S1}). The $O(\nu^2)$ coefficients 
can be related to the lattice artifacts by the relations (\ref{gammaexp}),
(\ref{xiexp}). They come out as 
\begin{equation}
\delta'_0(\theta)=0,\qquad
\delta'_1(\theta)=\frac{\pi\theta}{6},\qquad
\delta'_2(\theta)=-\frac{\pi\theta}{12}.
\label{1correction}
\end{equation}
This is used to predict the leading lattice artifacts in the
phase shift analysis of Section 5.

It is also feasible to calculate the leading lattice artifacts
for the two-point function of the Noether current. In \cite{logs} 
this was done in two-loop perturbation theory. Alternatively, 
using the current form factors of the SG model, the leading artifacts can
also be calculated non-perturbatively via the form factor bootstrap.
We worked out the two-particle contribution to this correction. 
The comparison with numerical data is presented in Section 6.
\vfill
\eject

\clearpage
\newpage
\newsection{MC results for the phase shifts}

We begin by numerically investigating the S-matrix. It is the 
prime input for the bootstrap formulation and an appreciable 
discrepancy to the bootstrap result (\ref{S1}) would
immediately rule out that the O(2) bootstrap theory describes
the continuum limit of the XY model. The technique to 
numerically determine the S-matrix takes advantage of the fact 
that the large volume dependence of the spectrum in a periodic 
(spatial box) encodes information on the infinite volume S-matrix. 
For example the volume dependence of the (stable) 1--particle mass is
governed by the forward scattering amplitude \cite{LuVol},
and a determination of the low-lying two--particle spectrum gives
a measurement of the low energy two-particle phase shifts \cite{LuWo}.
Here we restrict the discussion to a field theory in $1+1$ dimensions,
i.e. the ``spatial box" in this case is just a circle of
circumference $L$. To our knowledge the first attempt to determine 
the phase shifts of the XY model was by Vohwinkel \cite{Vohwinkel}, 
and we record his results in Appendix F.  

\newsubsection{1-particle masses}

We chose to measure on the same lattices as Vohwinkel,
firstly since they are practical and secondly for having the 
advantage of being able to compare independent measurements.
These lattices are listed in Table~\ref{betal} together
with the measured 1--particle masses. 
In each case the ``time" extent of the lattice is $T=2L$ 
and periodic boundary conditions are imposed in each direction.

\begin{table}[htbp]
  \begin{center}
    \begin{tabular}{|crll|}
\hline
$K$ & $L$ & $m(L)\;$\cite{Vohwinkel} & $m(L)$ \\
\hline
0.86&  64  & 0.1711(1)  & 0.17096(4) \\
0.92& 128  & 0.09465(3) & 0.09461(6)\\
0.97& 256  & 0.04620(1) & 0.04603(14)\\
\hline
\end{tabular}
\caption{\footnotesize Values of $K$ and $L$ used in the measurements,
with the 1-particle mass $m(L)\,$\cite{Vohwinkel}
obtained by Vohwinkel \cite{Vohwinkel}
and our measurements, $m(L)$.
}
\label{betal}
\end{center}
\end{table}
\vspace{-3mm}

What is quoted are the results for the single particle masses
obtained by fitting the zero-momentum spin 2-point function
with a 2-mass formula, with the second mass constrained to be
$m_2=3m_1$. (The 1-mass fit and the unconstrained 2-mass fit
give the same values for $m_1$ within the errors.) 
The three different values of $K$ correspond to correlation 
lengths $\sim 6, 11, 22$. In all cases there is good agreement with the 
results of Vohwinkel.

All the lattices are chosen to have $m(L)L>10$, so that finite volume
effects on the 1--particle masses are expected to be very small
according to Eq.~(\ref{volmassdepO2}). We denote the resulting
1-particle mass by $m$.

\begin{table}[htbp]
\centering
\begin{tabular}{|crlll|}
\hline
$K$ & $L$ & $m(L)$ & $D$ & $D_{\rm theor}$\\
\hline
0.86&  32  & 0.17135(3) & 0.0023(4)  & 0.0022 \\
0.92&  64  & 0.09478(3) & 0.0018(10) & 0.0012 \\
0.97& 128  & 0.04622(4) & 0.0041(39) & 0.0014 \\
\hline
\end{tabular}
\caption{\footnotesize 1--particle masses and finite volume effects.}
\label{betalhalf}
\end{table}

In order to quantitatively test the latter expectation 
we measured the 1-particle masses on lattices with the same 
three bare couplings $K$ as those in Table~\ref{betal} but with 
half the previous spatial extent. Our results are listed in 
Table~\ref{betalhalf}. Also shown are the measured values of the 
finite volume mass shifts $D=(m(L)-m(2L))/m(2L)$ and $D_{\rm theor}$ 
computed from Eq.~(\ref{volmassdepO2}).
At the smallest correlation length the measured shift 
is completely consistent with our ansatz for the candidate S-matrix.
However, lattice artifacts are to be expected and unfortunately
our measurements at the larger correlation lengths are too 
imprecise to study these effects.

\newsubsection{Phase shifts}

For a numerical test of the proposed S-matrix (\ref{S1}) 
it is useful not to presuppose the symmetry enhancement. 
That is we adopt the generic O(2) invariant parameterization 
(\ref{SOn}) with $n=2$. In terms of the $S_I(\th)$ the phase
shifts are defined as 
\be
S_I(\theta)=\exp\big\{2i\delta_I(\theta)\big\}\;,
\end{equation}
and can simplified to 
\begin{subeqnarray}
\delta_0(\theta)&=&\delta_1(\theta)+\frac{\pi}{2}-
\arctan\frac{\theta}{\pi} \,,
\\
\delta_1(\theta)&=&\int_0^{\infty}\frac{{\rm d} w}{w}
\frac{\sin w\theta}{\left(1+e^{\pi w}\right)}\,,
\\
\delta_2(\theta)&=&\delta_1(\theta)-\frac{\pi}{2}\,.
\label{XYphaseshifts}
\end{subeqnarray}
The last relation is responsible for the symmetry enhancement 
discussed in section 3.1.

To measure the phase shifts on the lattice 
one sets out to determine the center-of-mass momenta
of 2-particle eigenstates of the transfer matrix, since 
due to the periodic boundary condition in the spatial direction
these momenta are quantized according to
\begin{equation}
\label{pnl}
p_n L+2\delta(\theta_n)=2\pi n \,,\,\,\,\,\,\,\,\,\,
p_n=m\sinh\frac12\theta_n\,.
\end{equation}
To accomplish this one measures correlators 
\begin{equation}
C_{xy}^{(I)}(t)=\langle O_I(x,0) O_I(y,t)\rangle\,, 
\end{equation}
where the $O^{(I)}$ are 2-spin operators with zero
total momentum in the isospin channels $I=0,1,2$:
\begin{equation}
O_I(x,t)=\frac{1}{L}\sum_{z=0}^{L-1} (P_I)_{1b_I}^{cd}\, 
S^c(z,t)S^d(z+x,t) \,,\,\,\,\,\,b_0=1\,,b_1=b_2=2\,,
\end{equation}
the $P_I$ being the projectors in Eq.~(\ref{proj}).
Taking $T$ large enough so that terms proportional to ${\rm e}^{-2mT}$ 
can be neglected, one has 
\begin{equation}
\label{cxy}
C_{xy}^{(I)}(t)=
\sum_n {\rm e}^{-E_n t} \psi_n^{(I)}(x)\psi_n^{(I)}(y) \,,
\end{equation}
where 
\begin{equation}
 \psi_n^{(I)}(x) = \langle {\rm vac}| O_I(x,0)| n \rangle \,,
\end{equation}
is the ``wave function'' of the corresponding state. 
In the $I=0$ channel the vacuum also contributes as
an intermediate state. In this case one can subtract this
contribution from the beginning, i.e.~consider the connected
correlator. 
Alternatively, one can take the full $I=0$ correlator,
keeping in mind that in this case the vacuum is the lowest energy 
intermediate state.

Now there are at least two ways to proceed.
The first is to extract the energies $E_n$ of the 2-particle 
states which dominate the correlation function Eq.~(\ref{cxy})
for sufficiently large $t$, and then compute the
corresponding center of mass momenta via
\be
\label{enpn}
E_n=2E^{(1)}(p_n)=2\sqrt{p_n^2+m^2} \,.
\ee
This was the strategy used in the pioneering paper of 
L\"{u}scher and Wolff \cite{LuWo} and adopted by
Vohwinkel in his studies \cite{Vohwinkel}.
In Eq.~(\ref{enpn}) lattice artifacts have been neglected 
and the physical volume is taken so large that the
finite volume dependence of the single particle masses is
negligible.

In ref.~\cite{SigmacollabI} an alternative was suggested
which starts from the observation that
the relative momentum $2p_n$ of the two particles is also
encoded in the wave function: in the symmetric channels ($I=0,2$)
one should have
\begin{equation} \label{psin}
\psi_n(x)=A \cos p_n(x-L/2), \text{ for } R < x < L-R \,,
\end{equation}
and similarly with $\sin p_n(x-L/2)$ for the $I=1$ channel.
Here $R$ is the ``interaction range'' characterized by the requirement
that for a relative distance $x > R$ the two particles propagate 
essentially freely.
Note that Eq.~(\ref{pnl}) assumes that the box is large enough to
accommodate the two particles without ``squeezing'' them, i.e. 
$L/2 > R$.

The rank $N$ of the matrix $C(t)$ in Eq.~(\ref{cxy}) is
$L/2$, $L/2-1$ and $L/2+1$ in the $I=0,1,2$ channels, 
respectively. (This is when the connected correlation 
function is considered in the $I=0$ channel, otherwise $N=L/2+1$ 
also in this channel.)
We assume that for $t\ge t_0$ (with some $t_0$)
no more than $N$ states contribute to $C^{(I)}_{xy}(t)$, i.e. that 
the contribution from the states $n>N$ can be neglected completely.

L\"uscher and Wolff \cite{LuWo} suggested to determine the energies
$E_n$ from the generalized eigenvalue problem\footnote{This equation
was considered already earlier by Michael \cite{Michael}, in
connection with a variational approach evaluating the static potential
in lattice gauge theory.}
\begin{equation} \label{genev}
C(t)v_n = \lambda_n(t,t_0) C(t_0)v_n\,.
\end{equation}
Provided the sum in Eq.~(\ref{cxy}) is restricted to $N$ terms,
$1 \le n \le N$,
the eigenvalues of Eq.~(\ref{genev}) are given {\em exactly} by 
\begin{equation} \label{lambdan}
\lambda_n(t,t_0) = {\rm e}^{-E_n (t-t_0)} \,.
\end{equation}
It is easy to show that (apart from the normalization)
\begin{equation} \label{psiv}
\psi_n(x)=\sum_y C_{xy}(t_0) v_n(y) \,.
\end{equation}
A problematic feature of the generalized eigenvalue equation (\ref{genev})
is that its solutions become unstable if $C(t_0)$ has very small
eigenvalues. This can be seen by observing that 
$\lambda_n(t,t_0)$ are the eigenvalues of the ordinary eigenvalue 
equation for the matrix $C(t_0)^{-1/2} C(t) C(t_0)^{-1/2}$.
Of course, the exact correlation matrix $C(t_0)$ is positive 
definite, but the statistical noise will spoil this property,
and the measured matrix can have even negative eigenvalues,
especially for larger values of $t_0$ and for large number of 
operators $N$. 
For this reason in ref.~\cite{LuWo} $N \sim L/4$ operators 
were used (actually, in momentum space rather then in $x$-space)
and the values of $t_0$ were restricted to 0 and 1.

To avoid the instability, we restrict first the correlation matrix 
to an $M$-dimensional subspace ($M<N$) spanned 
by the first $M$ eigenvectors of $C(t_0)$ with the largest 
eigenvalues (still stable against the statistical fluctuations)
\cite{SigmacollabI,glueballs}.
The generalized eigenvalue problem is then written for the 
new correlation matrix $\overline{C}(t)$ in this reduced basis.
Of course, to read off the momenta, the wave functions have to
be transformed back into the original basis labeled by the relative 
distance $x$.

Following refs.~\cite{Michael,LuWo} one can obtain $E_n$ 
from the plateau of the ``effective energy''
\begin{equation} \label{Eeff}
E_n^{\rm eff}(t)= \ln \frac{\lambda_n(t,t_0)}{\lambda_n(t+1,t_0)} \,,
\end{equation}
and determine the corresponding momentum $p_n$ from Eq.~(\ref{enpn}).

The alternative way is to fit the wave function $\psi_n(x)$
by the ansatz in Eq.~(\ref{psin}) for $x_0 \le x \le L/2$.
We have verified that it is safe to take $x_0 \ge 3/m$.
There is also a large window where the results are not sensitive
to the variation of $M$, for different choices of $t_0$.
For the largest correlation length $\xi_{\rm exp} \approx 21.6$ 
we could take $t_0$ as large as 10, which would be impossible 
without the preceding truncation.

In~\cite{SigmacollabI} we concluded that the wave function
method has somewhat smaller errors and is more stable.
In particular the smallest momentum obtained from the 2-particle
energy is quite sensitive to the error in the single particle mass,
while in the wave function method the value of this mass is not used at all.
Although we measured the phase shifts by both methods
and checked their consistency, only our results from the wave 
function method will be presented here. Vohwinkel's results
on the energy levels are recorded in Table~\ref{clausdata}.

\begin{figure}[thb]
\begin{center}
\vskip -3mm
\leavevmode
\epsfxsize=98mm
\epsfysize=70mm
\epsfbox{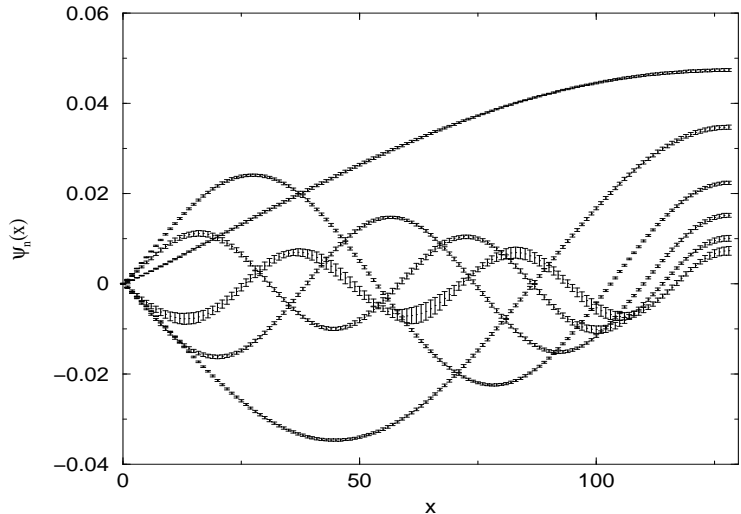}
\vskip -8mm
\end{center}
\caption{\footnotesize The first six wave functions
$\langle {\rm vac} | O_I(x) | n \rangle$ in the $I=0$ channel
for $K=0.97$, $L=256$.}
\label{fig:psi_I0a}
\end{figure}

\begin{figure}[bht]
\begin{center}
\vskip +3mm
\leavevmode
\epsfxsize=100mm
\epsfysize=70mm
\epsfbox{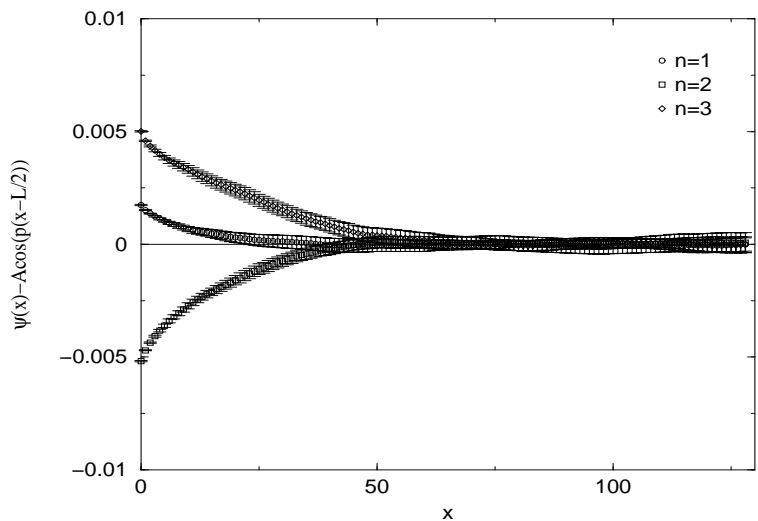}
\vskip -8mm
\end{center}
\caption{\footnotesize Deviations of the first 3 wave functions from the 
ansatz $A\cos p(x-L/2)$.}
\label{fig:psi_diff_I0}
\vskip 5mm
\end{figure}

Fig.~\ref{fig:psi_I0a} shows the wave functions of the first six
2-particle states in the $I=0$ channel obtained using 
Eqs.~(\ref{genev}), (\ref{psiv}). Fig.~\ref{fig:psi_diff_I0} displays 
the deviations of the first three wave functions
from the corresponding free one, $A\cos p(x-L/2)$ in the $I=0$ channel.
As expected, the true wave functions deviate
from the free ones only for small relative distances of O($\xi_{\rm exp}$).
The plots illustrate that the momentum $p$ can be determined quite 
precisely by fitting the wave function in some properly chosen 
range $x_0 \le x \le L/2$. Note, however, that in Eq.~(\ref{pnl}) 
the momentum $p_n$ is multiplied by $L$, hence it has to be 
determined to good precision in order to yield a reasonable error 
for the phase shift. In our simulations therefore only the phase shifts 
at the first 3--4 momenta could be determined with a reasonable error.

\begin{figure}[thb]
\begin{center}
\vskip -3mm
\leavevmode
\epsfxsize=70mm
\epsfbox{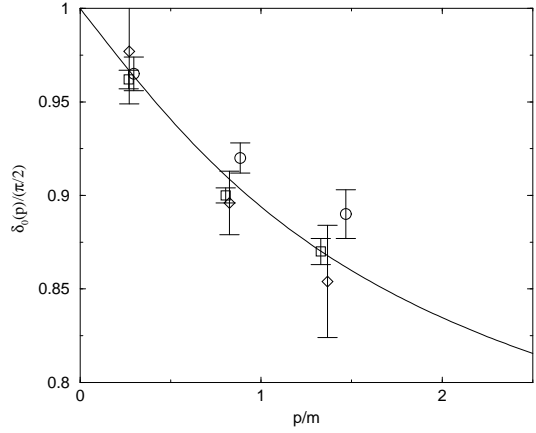}
\vskip -8mm
\end{center}
\caption{\footnotesize $I=0$ phase shifts. Points denoted by circle, square
and diamond correspond to correlation length $\approx$ 6, 11 and 22,
respectively.
The solid line is the continuum result, Eq.~(\ref{XYphaseshifts}a).
}
\label{fig:I0_phase}
\end{figure}

\begin{figure}[bht]
\begin{center}
\leavevmode
\epsfxsize=70mm
\epsfbox{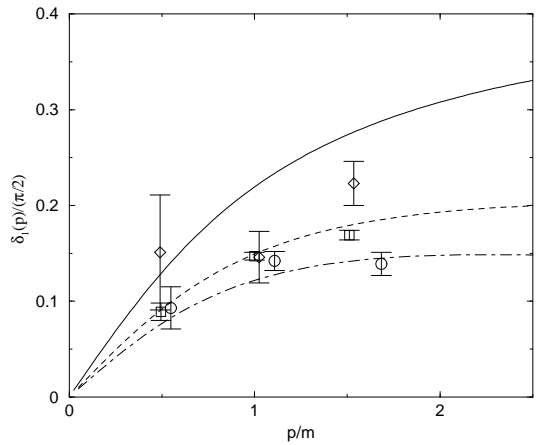}
\vskip -8mm
\end{center}
\caption{\footnotesize $I=1$ phase shifts. 
The notations for data points are the same as in Fig.~\ref{fig:I0_phase}.
The solid line is the continuum result, the other two lines
include leading corrections due to finite $\xi_{\rm exp}$ of 
Eq.~(\ref{1correction}) with $u=1.46$ (see \cite{Hasenbusch}): the dashed
line for
$\xi_{\rm exp}\approx 22$, the dot-dashed line for 
$\xi_{\rm exp}\approx 11$.
}
\label{fig:I1_phase}
\end{figure}

\begin{figure}[htb]
\begin{center}
\leavevmode
\epsfxsize=70mm
\epsfbox{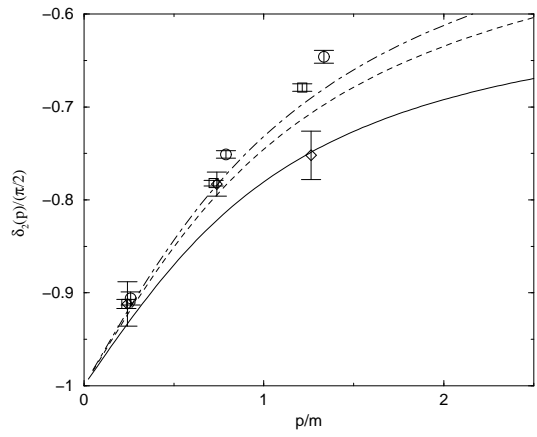}
\vskip -10mm
\end{center}
\caption{\footnotesize $I=2$ phase shifts.
The notations are the same as in Fig.~\ref{fig:I1_phase}.}
\label{fig:I2_phase}
\end{figure}

\newpage
{}\bigskip \bigskip
A summary of our measurements of the phase shifts is given in 
Table~\ref{delta_res}. The Figures \ref{fig:I0_phase} to 
\ref{fig:I2_phase} compare these results with the theoretical
curves of Eqs.~(\ref{XYphaseshifts}a-c). The leading lattice artifacts
according to Eq.~(\ref{1correction}) for the largest correlation length 
$\xi_{\rm exp}=21.645(5)$ are also shown. The present overall 
results are certainly consistent with the theoretical expectations.
\newcommand{\m}{\phantom{$-$}}
\begin{table}[htbp]
  \begin{center}
    \begin{tabular}{|crcclclc|}
\hline
$K$ & $L$ & $I$ & $n$ & \quad$p/m$  &  $[p/m]_{\rm ex}$ & 
        \quad$\delta^{(I)}(p)$ & \quad$[\delta^{(I)}(p)]_{\rm ex}$ \\
\hline
0.86    &  64 & 0 & 1 & 0.297(3) & 0.2976  &  \m1.517(15) & \m1.5138 \\
        &     &   & 2 & 0.885(2) & 0.8895  &  \m1.446(12) & \m1.4186 \\
        &     &   & 3 & 1.468(4) & 1.4760  &  \m1.397(20) & \m1.3529 \\
0.92    & 128 & 0 & 1 & 0.269(1) & 0.2677  &  \m1.512(8)  & \m1.5193 \\
        &     &   & 2 & 0.804(1) & 0.8007  &  \m1.413(6)  & \m1.4307 \\
        &     &   & 3 & 1.330(2) & 1.3297  &  \m1.366(11) & \m1.3669 \\
0.97    & 256 & 0 & 1 & 0.272(7) & 0.2745  &  \m1.535(43) & \m1.5180 \\
        &     &   & 2 & 0.825(5) & 0.8210  &  \m1.408(27) & \m1.4279 \\
        &     &   & 3 & 1.367(8) & 1.3632  &  \m1.342(47) & \m1.3180 \\
\hline
0.86    &  64 & 1 & 1 & 0.548(6) & 0.5349  &  \m0.146(34) & \m0.2161 \\
        &     &   & 2 & 1.108(3) & 1.0828  &  \m0.223(16) & \m0.3616 \\
        &     &   & 3 & 1.684(4) & 1.6416  &  \m0.218(19) & \m0.4474 \\
0.92    & 128 & 1 & 1 & 0.495(2) & 0.4855  &  \m0.140(15) & \m0.1992 \\
        &     &   & 2 & 0.999(1) & 0.9806  &  \m0.231(7)  & \m0.3405 \\
        &     &   & 3 & 1.511(1) & 1.4846  &  \m0.265(7)  & \m0.4280 \\
0.97    & 256 & 1 & 1 & 0.491(16)& 0.4969  &  \m0.237(94) & \m0.2031 \\
        &     &   & 2 & 1.024(7) & 1.0041  &  \m0.230(42) & \m0.3455 \\
        &     &   & 3 & 1.535(6) & 1.5205  &  \m0.350(36) & \m0.4328 \\
\hline
0.86    &  64 & 2 & 1 & 0.260(2) & 0.2663  & $-1.423(11)$ & $-1.4562$ \\
        &     &   & 2 & 0.790(1) & 0.8072  & $-1.179(6) $ & $-1.2726$ \\
        &     &   & 3 & 1.334(2) & 1.3612  & $-1.014(10)$ & $-1.1607$ \\
0.92    & 128 & 2 & 1 & 0.236(1) & 0.2419  & $-1.433(7) $ & $-1.4660$ \\
        &     &   & 2 & 0.721(1) & 0.7318  & $-1.228(5) $ & $-1.2930$ \\
        &     &   & 3 & 1.213(1) & 1.2317  & $-1.067(7) $ & $-1.1812$ \\
0.97    & 256 & 2 & 1 & 0.242(6) & 0.2475  & $-1.432(38)$ & $-1.4638$ \\
        &     &   & 2 & 0.739(4) & 0.7491  & $-1.230(21)$ & $-1.2881$ \\
        &     &   & 3 & 1.262(7) & 1.2614  & $-1.181(41)$ & $-1.1763$ \\
\hline
    \end{tabular}
    \caption{\footnotesize 
    The phase shifts $\delta_I(p)$ for the first 3 states 
   ($n=1,2,3$) in different isospin channels, with the values
   of $p/m$ determined from the wave function.
   The continuum results $[p/m]_{\rm ex}$ and $[\delta_I(p)]_{\rm ex}$
   calculated from the S-matrix are also given.}
    \label{delta_res}
  \end{center}
\end{table}

\vfill
\eject

\clearpage
\newpage
\newsection{MC results for correlation functions and 4-point coupling}

Here we describe our MC results for the intrinsic 4-point coupling 
$\gr$ and for the current and spin 2-point correlation functions in 
Fourier space. In addition we reconsider some aspects of the 
Kosterlitz-Thouless (KT) theory. 

A list of the lattices considered together with some of the MC
results is given in Table~\ref{lattices}. We group the lattices 
into families labeled 1 to 12. Members of a given family correspond
to the same coupling $K$ but different size $L$.
Throughout this section $\xi$ denotes the second moment correlation length 
\eqref{xisecmom}, which is a function of $K$ and $L$. We denote the
`apparent' physical size of the lattice by $z'=L/\xi(K,L)$.
For most lattices we made 200k measurements where we measured all 
the physical quantities above supplemented by an additional
2M measurements for the \lq\lq bulk" quantities 
($\xi$, $\chi$ and $\gr$) only.
We also took data on `thermodynamic' lattices ($L/\xi\approx14$). These
are reported in Appendix E. We did not use them in our analysis because
there were some not completely understood inconsistencies between
data taken on different machines, with different random number
generators and different programs. 

\newsubsection{KT theory}

We begin by studying the dependence of the correlation length $\xi$ and 
the susceptibility $\chi$ on the coupling $K$ and compare our MC results 
to the predictions of the KT theory. For this purpose we first need
to extrapolate the measured values of $\xi$ and $\chi$ to infinite
volume. This is done by a finite size scaling analysis.

For fixed coupling the size dependence of the correlation length is 
assumed to be given by
\begin{equation}
\ln\xi(K,L)=\ln\xi(K,\infty)+z^{A_2}e^{-z}\left[
C^\xi_0(K)+C^\xi_1(K)\frac{1}{z}+\ldots\right].
\label{FSxi}
\end{equation}
Here $z:=L/\xi_{\rm exp}(K,\infty)$ where $\xi_{\rm exp}$ is the true
(``exponential'') correlation length. $\xi_{\rm exp}$ is expected to 
be very close to the second moment correlation length: the form factor 
bootstrap predicts $\xi_{\rm exp}/\xi=\sqrt{\gamma_2/\delta_2}=1.00089$
and the lattice models are certainly not very far from the form factor
construction; therefore at our $\approx 0.001$ accuracy the distinction
between the two correlation lengths can be neglected.
For the exponent in \eqref{FSxi} we expect $A^{\xi}= A_2 =-1/2$ on account 
of the considerations outlined in Section 4.3, but we also tried
different values of that exponent in order to see if the data indeed
support this expectation.

\begin{table}[h]
\centering
\begin{tabular}[t]{|l|d{4}|d{0}|d{6}|d{6}|d{6}|d{1}|}
 \hline
label & \multicolumn{1}{c|}{$K$} &\multicolumn{1}{c|}{$L$} &
\multicolumn{1}{c|}{$\xi$} & \multicolumn{1}{c|}{$\chi$} &
\multicolumn{1}{c|}{$\gr$} & \multicolumn{1}{c|}{$z'$} \\
\hline \hline
$\phantom{1} 1$ & 0.86 & 24 & 5.728(1) & 57.07(11) & 7.533(3) & 4.2 \\
\hline
$\phantom{1} 1$ & 0.86 & 29 & 5.795(1) & 58.82(12) & 8.147(4) & 5.0 \\
\hline
$\phantom{1}1 $ & 0.86 & 32 & 5.815(2) & 59.33(2)  & 8.357(8) & 5.5 \\
\hline
$\phantom{1}1 $ & 0.86 & 40 & 5.833(1) & 59.814(9) & 8.672(7) & 6.9 \\
\hline
$\phantom{1}1 $ & 0.86 & 64 & 5.839(1) & 59.956(7) & 8.774(14)& 11.0 \\
\hline
\hline
$\phantom{1} 2$ & 0.92 & 42 & 10.314(2)& 153.76(4) & 7.484(3) &4.1    \\
\hline
$\phantom{1} 2$ & 0.92 & 52 & 10.466(2) & 159.58(4)& 8.201(5) & 5.0 \\
\hline
$\phantom{1} 2$ & 0.92 & 68 & 10.536(2) & 162.28(4) & 8.690(8) & 6.5  \\
\hline
$\phantom{1} 2$ & 0.92 & 94 & 10.548(5) & 162.79(7) & 8.855(27)& 8.9 \\
\hline
\hline
$\phantom{1} 3$ & 0.93 & 64 & 11.861(4) & 198.05(7) & 8.436(10)&  5.4  \\
\hline
$\phantom{1} 3$ & 0.93 & 80 & 11.905(2) & 199.95(3)& 8.755(7)  & 6.7 \\
\hline
\hline
$\phantom{1} 4$ & 0.97 & 86 &  21.146(4) & 525.11(11) & 7.539(2) & 4.1   \\
\hline
$\phantom{1} 4$ & 0.97 & 108 & 21.467(6) & 545.85(17) & 8.313(6) & 5.0  \\
\hline
$\phantom{1} 4$ & 0.97 & 136 & 21.589(9) & 553.88(28) & 8.745(21) & 6.3\\
\hline
$\phantom{1} 4$ & 0.97 & 194 & 21.633(11) & 556.33(29) & 8.934(26) & 9.0\\
\hline
\hline
$\phantom{1} 5$ & 0.975 & 128 & 23.546(6) &641.53(19)  & 8.514(9)  & 5.4\\
\hline
\hline
$\phantom{1} 6$ & 1.00 & 160 & 39.219(8) & 1522.2(4)& 7.585(3) & 4.1 \\
\hline
$\phantom{1} 6$ & 1.00 & 200 & 39.801(7) &1581.2(4) & 8.345(4)  &5.0\\
\hline
$\phantom{1} 6$ & 1.00 & 256 & 40.025(11) & 1605.0(5) & 8.798(10) & 6.4 \\
\hline
$\phantom{1} 6$ & 1.00 & 360 & 40.104(15) & 1611.2(7) & 8.955(22) & 9.0 \\
\hline
\hline
$\phantom{1} 7$ & 1.005 & 256 & 45.247(13) &1980.7(6) & 8.618(9)& 5.7 \\
\hline
\hline
$\phantom{1} 8$ & 1.0174 & 360 & 63.889(22) &3596.4(1.2) & 8.631(10)& 5.6\\
\hline
$\phantom{1} 8$ & 1.0174 & 500 & 64.314(91) &3639.2(6.2) & 9.05(12)&
7.8\\
\hline
\hline
$\phantom{1} 9$ & 1.02 & 276 & 67.934(45) & 3936.7(3.1) & 7.587(7)&  4.1\\
\hline
$\phantom{1} 9$ & 1.02 & 344 & 68.942(47) &4091.1(3.0) & 8.375(13)&  5.0\\
\hline
$\phantom{1} 9$ & 1.02 & 560 & 69.500(25) &4170.6(1.8) & 8.997(18)&  8.1\\
\hline
\hline
$\phantom{1} 10$ & 1.04 & 578 & 142.09(15)& 14164.(20.)& 7.600(15)& 4.1 \\
\hline
$\phantom{1} 10$ & 1.04 & 726 & 144.43(12) &14743.(15.)& 8.404(18)& 5.0\\
\hline
\hline
$\phantom{1} 11$ & 1.05 & 930 & 230.10(29) & 32589.(46.)& 7.586(19) & 4.0\\
\hline
$\phantom{1} 11$ & 1.05 & 1160 & 232.83(22) &33800.(37.)& 8.392(19) & 5.0\\
\hline
\hline
$\phantom{1} 12$ & 1.06 & 2100 & 418.0(1.6) &93724.(428.)& 8.410(79) & 5.0\\
\hline
\hline
\end{tabular}
\caption{{}\footnotesize Lattice parameters and results.}
\label{lattices}
\end{table}
\clearpage

We produce a global fit of the finite size effects by fitting all the 
data in Table \ref{lattices} to the form \eqref{FSxi} truncating after
the $C^\xi_0$ term and taking $C^\xi_0$ to be independent of $K$.
The values of $\xi(K,\infty)$ are fit parameters, one for each of the
twelve families.
Since initially we do not know the values of $L/\xi(K,\infty)$ we
first replace it by $L/\xi(K,L)$; this leads to a first estimate of
$\xi(K,\infty)$ which is then used in the fit ansatz. Iterating this
procedure about 5 times leads to convergence of our extrapolated
values of $\xi(K,\infty)$. It turns out that this type of fit favors a 
value of $A_2$ near $-1/2$ in agreement with expectation; for this value
of the exponent we obtain ${\rm chi^2}/{\rm dof}=0.9$ with 19 degrees
of freedom.

Furthermore, including a $C^\xi_1$ term makes the fit very
insensitive to the value of the exponent $A_2$, both as far as the
fit quality and the extrapolated values of $\xi(K,\infty)$ are
concerned. For $A_2=-1/2$ the coefficient $C^\xi_1$ comes out
consistent with zero. Therefore in Table \ref{xichi_TD} we only report
the values obtained with the simplest fit with $A_2=-1/2$ and only
one finite size correction term.

\begin{figure}[htb]
\begin{center}
\leavevmode
 \epsfxsize=80mm
\epsfbox{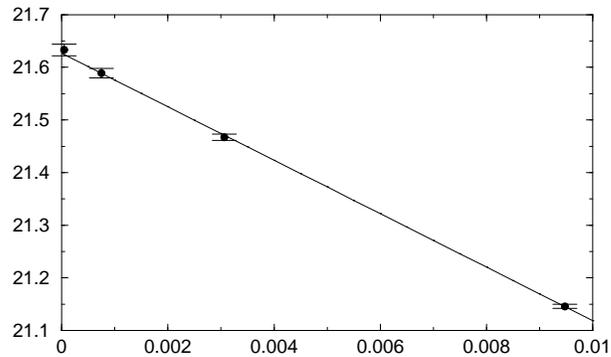}
\vskip -6mm
\end{center}
\caption{{}\footnotesize
$\xi(K,L)$ vs. $\exp(-z)/\sqrt{z}$ at $K=0.97$.}
\label{xi_b097}
\end{figure}

Figure \ref{xi_b097} illustrates the $z$ dependence of the correlation 
length for $K=0.97$.

The FS analysis of the susceptibility is analogous, but we don't need
any iteration, since we start already with the right $z$ values. Again 
the data favor a value near $-1/2$ if we truncate with the $C^\chi_0$
term, and if we include the next term, the fit becomes insensitive to 
$A_2$. In Table \ref{xichi_TD} we present the results from the simplest 
fit assuming $A_2=-1/2$, which has a ${\rm chi^2}/{\rm dof}$ of 0.7 
with 19 degrees of freedom.

\begin{figure}[htb]
\begin{center}
\vskip 5mm
\leavevmode
\epsfxsize=0.45\linewidth
\epsfbox{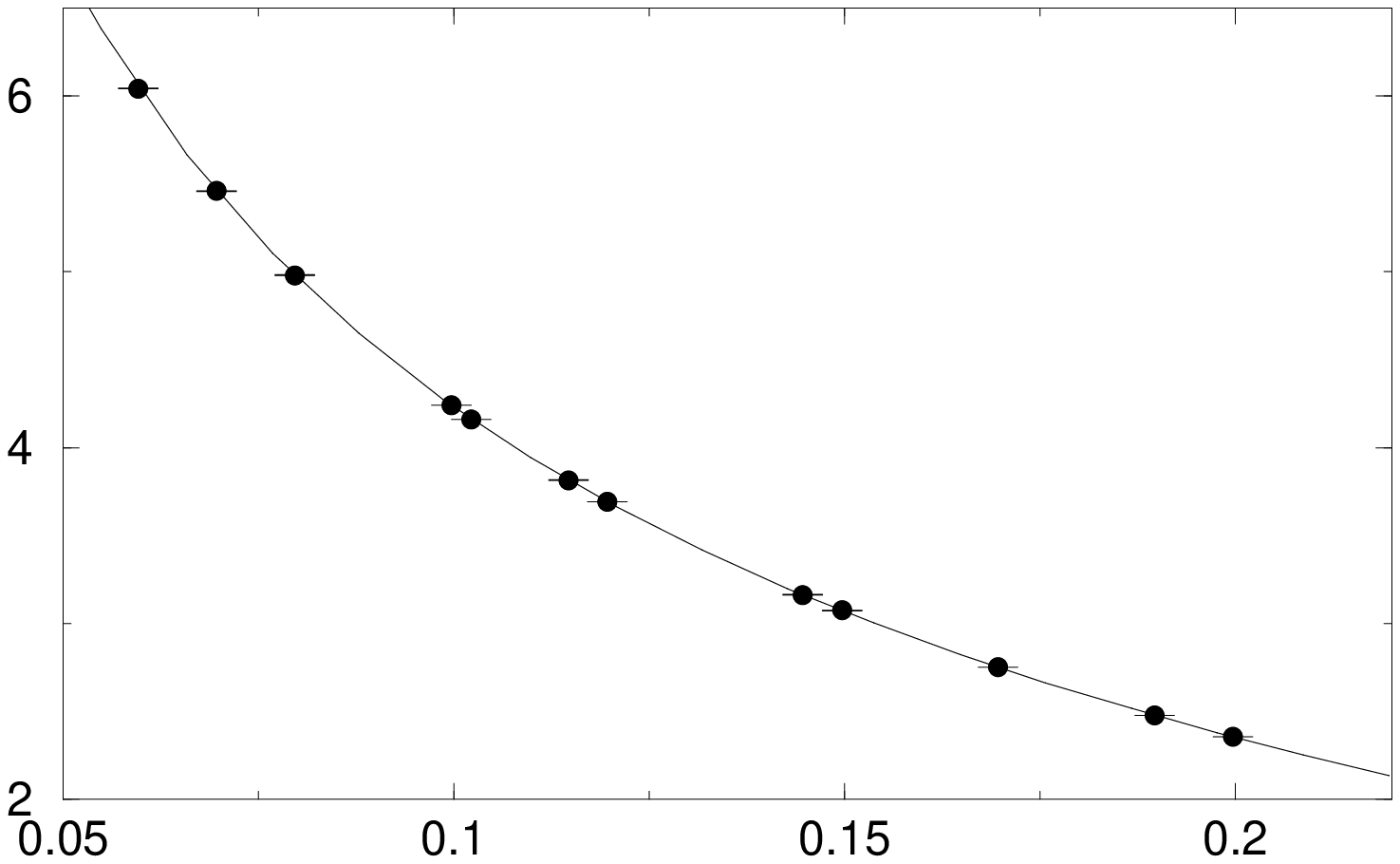}
\hspace{5mm} 
\epsfxsize=0.45\linewidth
\epsfbox{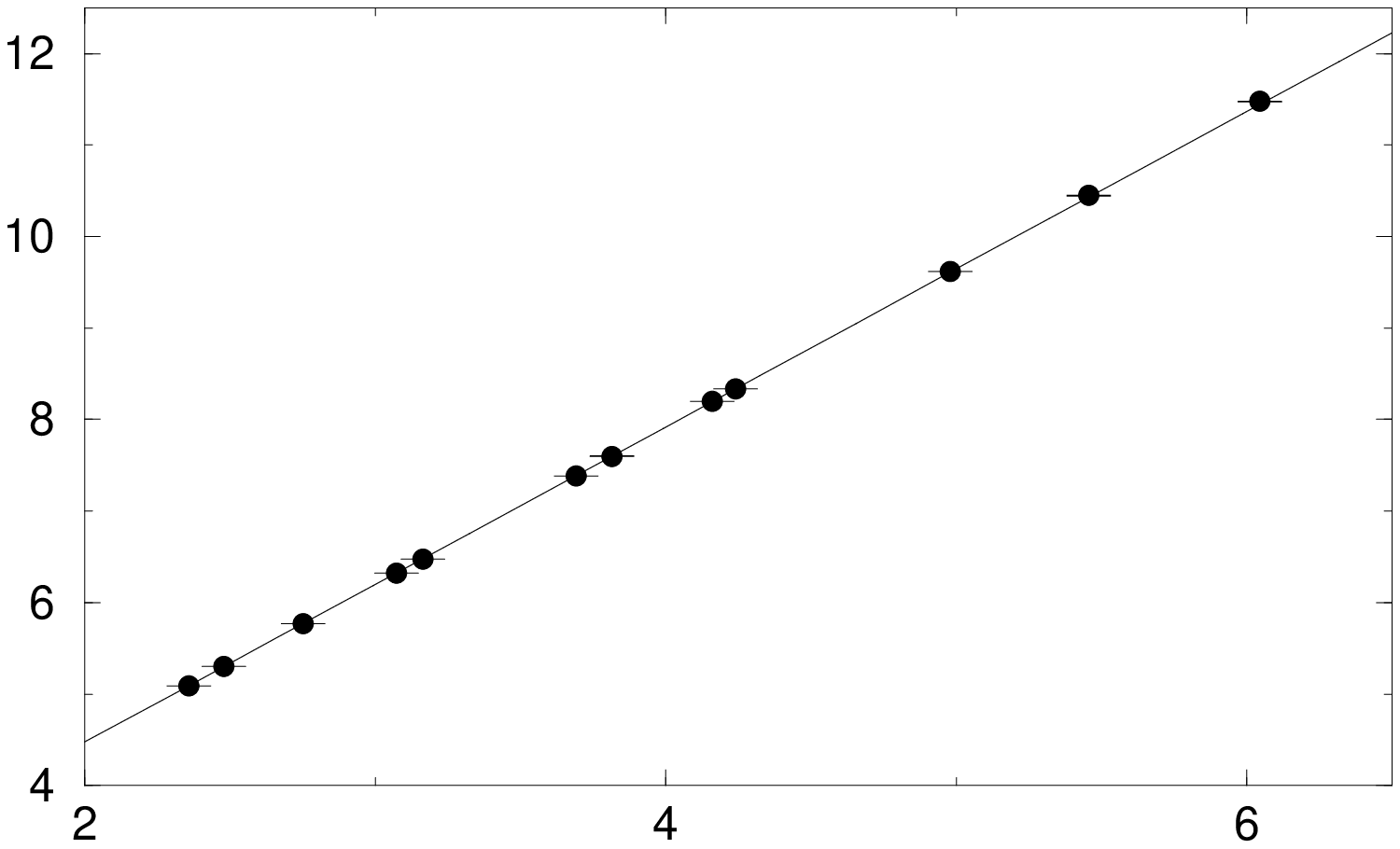}
\vskip -4mm
\end{center}
\caption{{}\footnotesize
Illustration of KT scenario: Left, $\ln\xi$ versus $\tau$ and a fit to 
\eqref{KTxi}. Right, $\ln\chi$ versus $\ln\xi$ and a linear fit.
}
\label{pllogxi}
\end{figure}

Next we study the $K$-dependence of the infinite volume quantities.
One of the best known predictions of KT theory \cite{KT} is 
the unusual coupling dependence of the correlation length. 
Close to the critical point $K_c$, $\xi$ is predicted to diverge as
$\xi\sim\exp(b/\sqrt{\tau})$, where $\tau=K_c-K$ is
the reduced coupling and $b$ is a non-universal constant.
In more detail, the Sine-Gordon description of the KT transition
entails (using eqs. (8.13) and (8.14) of \cite{Amit}
\footnote{note that (8.15) contains a misprint.})
\begin{equation}
\ln\xi(K,\infty)=\frac{b}{\sqrt{\tau}}-u+c\sqrt{\tau}+\dots,
\label{KTxi}
\end{equation}
where $u$ and $c$ are again non-universal constants and the dots
stand for higher powers in $\tau$. We fitted the $\xi(K,\infty)$ data 
to the form \eqref{KTxi} without higher terms, leaving out different	
numbers $n_{\rm skip}$	 
of the low $\xi$ families to see how stable the fit parameters are. 
The results are presented in Table \ref{ktfit}. Note that some of the
fits have an unacceptable ${\rm chi^2}$. In any case, the determination
of $u$ is not very stable. This situation could be a sign of the
inappropriateness of the ansatz \eqref{KTxi}, or of some problem with
our data or it could mean that we are still too far from the critical
point, so that asymptotic formulae cannot yet be applied reliably.
In further fits below that do involve $u$, we use the value
appropriate to the number of discarded families.
A visual illustration of the $n_{\rm skip}=1$ fit is shown 
in the left part of Fig.~\ref{pllogxi}.

We should mention that Hasenbusch and Pinn \cite{Hasenbusch} have   
determined the constants $b$ and $u$ by their method of matching to the
exactly solvable BCSOS model; their values are $u=1.46(1)$ and
$b=1.879(4)$, in rough agreement at least with some of our estimates.
It should be noted, however, that their method avoids the problem of
controlling subleading corrections like the third term in eq.
\eqref{KTxi}.

\begin{table}[h]
\centering
\begin{tabular}[t]{|c||c|c|c|c|c|c|}
\hline
$K$   & 0.86     & 0.92     & 0.93      & 0.97      &  0.975    & 1.00  \\
\hline 
$\xi(K,\infty)$ 
      & 5.8391(5) & 10.549(1) &11.918(2)& 21.627(4) & 23.655(6) &
      40.096(5) \\
\hline
$\chi(K,\infty)$
      &59.962(5) & 162.86(26) &200.52(28) & 556.32(11)& 649.15(19)
      & 1611.4(3)\\
\hline\hline
$K$   & 1.005 & 1.0174    & 1.02     &  1.04    &  1.05    & 1.06     \\
\hline 
$\xi(K,\infty)$
      &45.410(13)& 64.137(22) & 69.505(20) & 145.424(95) & 234.93(18) &
      421.1(1.6)\\
\hline
$\chi(K,\infty)$
      & 1999.1(6)&3631.1(1.2) & 4173.0(1.4)& 15017.(13.) & 34512.(30.) &
      95502.(436.)\\
\hline
\end{tabular}
\caption{{}\footnotesize Correlation length and susceptibility extrapolated
to infinite volume using a global fit.}
\label{xichi_TD}
\end{table}

\begin{table}[h]
\centering
\begin{tabular}[t]{|c|l|l|c|}
\hline
$n_{\rm skip}$  & $b$    & $u$     &   ${\rm chi^2}/{\rm dof}$     \\
\hline \hline
1 & 1.865(1) & 1.280(8) & 27/8 \\
\hline
2 & 1.866(2) & 1.286(9) & 27/7 \\
\hline
3 & 1.875(3) & 1.340(16)& 11/6 \\
\hline
4 & 1.873(4) & 1.332(21)& 11/5 \\
\hline
5 & 1.886(1) & 1.414(40)& 4.9/4\\
\hline
6 & 1.885(9) & 1.406(58)& 4.9/3\\
\hline
7 & 1.86(2)  & 1.22(15) & 3.3/2\\
\hline
8 & 1.86(3)  & 1.21(18) & 2.2/1\\
\hline
\end{tabular}
\caption{{}\footnotesize Fit of the infinite volume correlation length
to the ansatz \eqref{KTxi}; $n_{\rm skip}$ denotes the number of low $\xi$
families discarded.}
\label{ktfit}
\end{table}

KT theory also predicts the asymptotics of $\chi(\xi)$ for large $\xi$ 
(at least up to possible logarithmic corrections),
\begin{equation}
\chi\sim\xi^{2-\eta},\quad \mbox{with}\quad \eta = 1/4\,.
\label{chixi1}
\end{equation}
A linear fit of $\ln \chi$ versus $\ln \xi$ is shown in
Fig.~\ref{pllogxi}.
The slope is 1.73, which is very close to the expected result. 
The fit is visually good, but even if we omit families 1,2 and 3 we get
a huge value of ${\rm chi}^2/{\rm dof}\approx 288/7$,
indicating the presence of non-negligible subleading terms.
Irving and Kenna \cite{IK}, following Butera and Comi \cite{BuCom}
(see also \cite{Janke}) argued that the Kosterlitz-Thouless 
theory implies the following refinement of \eqref{chixi1}:
\begin{equation}
\chi\sim\xi^{2-\eta}(\ln \xi)^{-2r}\,,
\label{chixi2}
\end{equation}
with $r=-1/16$.
This would in particular mean that $\chi$ grows faster than
$\xi^{1.75}$, whereas the data on the contrary indicate a slower increase.

On the other hand one of us \cite{logs} has argued that
the Kosterlitz-Thouless theory implies $r=0$, with a specific 
additive correction to \eqref{chixi1}: the renormalization group 
invariant quantity $Q$ was introduced, which for large $\xi$ behaves as
\begin{equation}
Q=\frac{\pi^2}{2(\ln\xi+u)^2}+O\left\{(\ln\xi)^{-5}\right\}\;.
\label{Qdef}
\end{equation}
It was then argued that the correct asymptotic formula, instead of 
\eqref{chixi2}, is
\begin{equation}
\ln\chi\sim \frac{7}{4} \ln\xi+O(Q).
\label{chixiQ}
\end{equation}
Taking $u$ from Table \ref{ktfit} the relation \eqref{chixiQ} can again
be tested against the data. We fitted  $\ln \chi -1.75 \ln \xi$ against 
$Q$, discarding successively more and more low $\xi$ families; in the 
fits we used the best value of $u$ corresponding to the same number of 
discarded families. The results are given in Table \ref{Qfit}.

\begin{table}[h]
\centering
\begin{tabular}[t]{|c||c|c|c|c|c|c|c|c|}
\hline
$n_{\rm skip}$  & 1 & 2 & 3 &  4 & 5 & 6 & 7 & 8     \\
\hline 
${\rm chi^2}/{\rm dof}$  & 44/9   & 32/8    &8.2/7     & 5.4/6
  &  2.9/5   & 0.6/4    & 0.5/3     & 0.5/2     \\
\hline 
\end{tabular}
\caption{{}\footnotesize Fits of $\ln\chi-1.75\ln\xi$ to linear function
of $Q$;  $n_{\rm skip}$ denotes the number of low $\xi$ families discarded.}
\label{Qfit}
\end{table}

Starting with $n_{\rm skip}=3$ the fits are acceptable, but of course it
should be remembered that we are not quite sure which value of $u$
should be used.  We do not list the fit parameters, but they are quite 
stable. So one can say the data support the prediction 
\eqref{chixiQ}.

\begin{figure}[htb]
\begin{center}
\leavevmode
\epsfxsize=80mm
\epsfbox{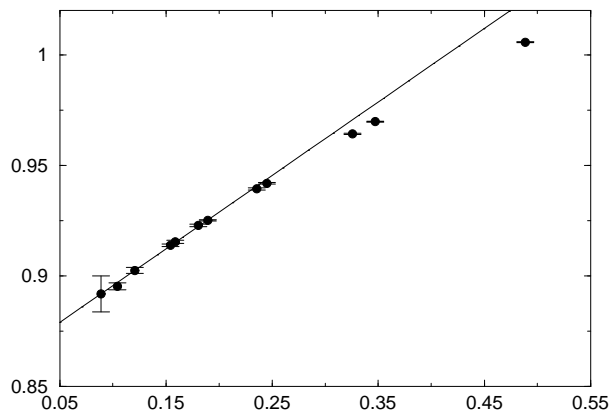}
\vskip -7mm
\end{center}
\caption{{}\footnotesize
$\ln \chi - 1.75 \ln \xi$ versus $Q$ and a linear 
fit on the data from the $\xi \geq 39$ lattices. 
}
\label{SQ}
\end{figure}

We will use the variable $Q$ as the quantity characterizing the distance 
from the continuum limit. 

\newsubsection{Determination of $\gr$}

Lattice determinations of $\gr$ can be based either on the high temperature 
expansion or on numerical simulations. The results obtained via
the high temperature expansion and the standard action \eqref{staction} 
are 
\begin{equation}
\gr = 9.15(10) \;\;\cite{HTXY1} \sspace 
\gr = 9.10(5) \;\;\cite{HTXY2}\,.
\label{grHT}
\end{equation}

In the numerical simulations we again aimed at achieving a precision 
of better than one percent. 
For the necessary extrapolation to the infinite volume we use
the procedure outlined before, i.e. we fit the data to the
ansatz
\begin{equation}
\ln\gr(K,L)=\ln\gr(K,\infty)+z^{A_4}e^{-z}\left[
C^{\gr}_0(K)+C^{\gr}_1(K)\frac{1}{z}+\ldots\right]  \,,
\label{FSgr}
\end{equation}
where, as described in Section 4.3, there are arguments suggesting
$A^{\gr}=A_4=1/2$. As opposed to the situation with $\xi$ and $\chi$,
with this value, the leading finite size correction alone does not 
properly describe the FS dependence. If we instead use the optimized value
$A_4=0.8$,  a subleading finite size correction
is not needed. It is gratifying that the extrapolated values are
almost independent of which of the two options we choose.
We report the infinite volume values obtained with $A_4=1/2$ and two
FS correction terms as well as those with only the leading FS correction
and 
$A_4=0.8$ in Table \ref{gr_global}. Both fits have a ${\rm chi}^2/{\rm
dof}$
around 1. 

\begin{table}[h]
\centering
\begin{tabular}[t]{|c||c|c|c|c|c|c|}
\hline
$K$   & 0.86     & 0.92     & 0.93      & 0.97      &  0.975    & 1.00  \\
\hline 
$\gr(K,\infty)$
      & 8.790(6)& 8.877(6) & 8.900(7) & 8.952(6)& 8.967(11) &   8.989(6) \\
\hline 
$\gr(K,\infty)$
    & 8.794(3)& 8.882(4) & 8.906(6) & 8.958(4)& 8.973(10) &   8.995(4) \\
\hline\hline
$K$   & 1.005 & 1.0174    & 1.02     &  1.04    &  1.05    & 1.06     \\
\hline 
$\gr(K,\infty)$
   & 8.993(10)& 9.015(11) & 9.026(8) & 9.039(14)& 9.053(16)& 9.062(9) \\
\hline 
$\gr(K,\infty)$  
   & 8.999(9)& 9.022(10) &9.031(7)  & 9.044(13)& 9.058(15) & 9.067(9)  \\
\hline
\end{tabular}
\caption{{}\footnotesize Fits to $\gr$ with $A_4=1/2$ using
2 finite size parameters (upper line) and with $A_4=0.8$ using only
the leading finite size parameter (lower line).}
\label{gr_global}
\end{table}

Finally we turn to the continuum limit of $\gr$. One of us 
\cite{logs} has argued that the leading lattice artifacts are 
proportional to the quantity $Q$, which in turn depends on the 
parameter $u$ extracted from the fits in table \ref{ktfit}. 
We present in Table \ref{grcont} the results of fits of the infinite
volume $\gr$ values to a linear function of $Q$; we are reporting
the results obtained by discarding different numbers of low $\xi$
values, obtained with the corresponding $u$ value. The fits are 
generally of good quality, but the resulting continuum values of
$\gr$ depend noticeably on the number of skipped values.

\begin{table}[h]
\centering
\begin{tabular}[t]{|c||c|c||c|c|}
\hline
$n_{\rm skip}$  & $\gr(K_c,\infty)$ & ${\rm chi^2}/{\rm dof}$ &
$\gr(K_c,\infty)$ &  ${\rm chi^2}/{\rm dof}$ \\
\hline \hline
1 & 9.124(8) & 2.2/9 & 9.129(7) & 4.4/9 \\	
\hline
2 & 9.120(10)& 2.4/8 & 9.124(9) & 3.5/8 \\
\hline
3 & 9.127(15)& 2.1/7 & 9.132(12)& 3.0/7 \\
\hline
4 & 9.125(20)& 2.1/6 & 9.131(18)& 3.0/6 \\
\hline
5 & 9.140(25)& 1.4/5 & 9.148(22)& 1.6/5 \\
\hline
6 & 9.134(33)& 1.3/4 & 9.138(31)& 1.4/4 \\
\hline
7 & 9.108(40)& 0.4/3 & 9.112(38)& 0.3/3 \\
\hline
8 & 9.099(43)& 0.8/2 & 9.105(41)& 0.9/2 \\
\hline
\end{tabular}
\caption{{}\footnotesize Fits of $\gr(K,\infty)$ to a linear function
of $Q$;  $n_{\rm skip}$ denotes the number of low $\xi$ families discarded.
The two columns correspond to the two rows of Table \ref{gr_global}.}
\label{grcont}
\end{table}
Our MC results are to be compared with the result of the form factor
computation
from Appendix D
\be
\gr = 9.10(4).
\label{gr_ff}
\ee

\newsubsection{The current correlation function}

Here we compare the bootstrap result for the current two-point
function (\ref{contCurr}) with the lattice measurements. The 
extrapolation of the lattice data to the continuum is done 
by means of a two-step procedure, which is a variant of the
method used for $\gr$. We first perform a FS analysis for those
relatively short correlation lengths for which we could afford to measure
the correlation function on lattices of large physical size. Using
the FS scaling coefficients determined this way we are able to
extrapolate the results of our measurements, corresponding to
moderately large physical size, to infinite size. In the second step
we take the continuum limit by extrapolating our results for
infinite correlation length.

For the first step we adopt an additive form of the FS scaling 
hypothesis:
\begin{equation}
I(q;K,L)=I(q;K,\infty)+z^{A^I}e^{-z}\left[
C^I_0(q;K)+C^I_1(q;K)\frac{1}{z}+\cdots\right].
\label{FScurr}
\end{equation}
We note that for the WARD case the subtracted correlation function
is a linear combination of a 2-point function and a 4-point function;
therefore by the arguments used before one expects $A^I=A_4=1/2$.

Compared to the case of $\gr$, the analysis here is complicated by  
momentum dependence and also the fact that we have used two alternative
definitions (SUB and WARD) of the current correlation function.
To be able to compare the results with each other and with the bootstrap
calculation we used the dimensionless momentum variable $q:=p/M_R$,
where $M_R$ is the second moment mass. We interpolated our lattice results
to integer $q$ values $q=1,2,\dots,50$ by fitting the measured values
with the 8-parameter formula
\begin{equation}
I(q)=\sum_{k=3}^{10}\,m_k\,\frac{q^2}{k^2+q^2}.
\end{equation}
This formula is motivated by the spectral representation and with the
8 parameters $\{m_k\}$ it gives excellent representation of the current
correlation function for all our lattices in the range $q<50$.

\begin{figure}[htb]
\begin{center}
\vskip 1mm
\leavevmode
\epsfxsize=80mm
\epsfbox{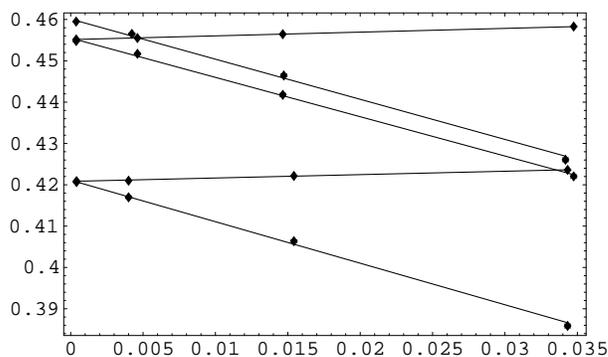}
\vskip -6mm
\end{center}
\caption{{}\footnotesize Test of FS scaling: 
$I(25)$ in SUB and WARD definition versus $\sqrt{z} e^{-z}$.
The three (approximately) parallel lines with negative slope correspond 
to the SUB data;
increasing values correspond to families 2, 4 and 6 respectively.
The two (approximately) parallel lines with positive slope are the WARD
data for families 2 and 4. There are no WARD data available for family 6.
}
\label{T25}
\end{figure}

We fitted the leading coefficient 
$C^I_0(q)$ (ignoring its $K$ dependence and any subleading terms) using
the
four data points of families 2, 4 and 6 for the SUB case and families
2 and 4 for the WARD case. 
(Unfortunately the
WARD data are not available for family~6, except for the lattice 6 
with $L=200$.)
The results of these fits for $q=25$ are shown in Figure \ref{T25}.
The fact that the linear fits are nearly parallel shows that our assumption
of FS scaling works here. 
More importantly one sees that the difference between the two definitions
disappears at large $z$, as it should\footnote{Note that in this analysis 
we replaced $z$ by $z^\prime$.}.
We also note that FS 
corrections have opposite signs for the SUB and WARD cases and that 
the latter are much smaller. These qualitative features remain valid also 
for other $q$-values, although for small momenta the FS corrections for 
WARD data are no longer minute compared to the SUB ones at the same $q$.
On the other hand for larger momenta the difference is even more pronounced.

\begin{figure}[htb]
\begin{center}
\vskip -3mm
\leavevmode
\epsfxsize=80mm
\epsfbox{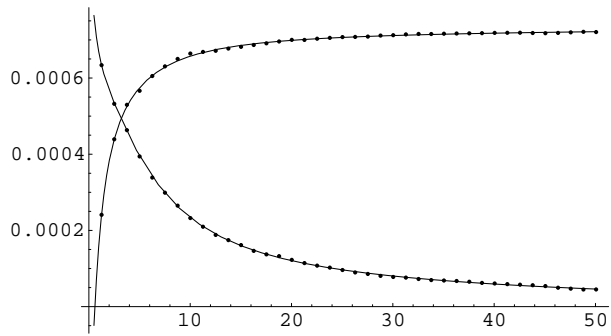}
\vskip -6mm
\end{center}
\caption{{}\footnotesize
Absolute statistical errors for SUB (increasing) and WARD (decreasing)
current data as a function of $q$, for $\xi = 421,\,L=2100$. The
continuous lines are fitted for convenience.
}
\label{sherrsh1err}
\end{figure}

Before turning to the infinite volume extrapolation let us briefly 
digress on the 
relative size of the statistical errors in the WARD and SUB data. 
In Figure \ref{sherrsh1err} these errors are shown as a function
of the momentum $q$ for both methods, for lattice 12. 
One sees that, with the exception of the first few points,
the WARD data have much smaller statistical errors, for large
momenta by about an order of magnitude! This remains true
for all other lattices. The explanation is that before
subtraction, the zero momentum component of the Fourier transform
of the current correlation function has the largest fluctuation and its
fluctuations decrease with momentum. Already at $q\sim$~4-5 the 
fluctuations are much smaller. If we subtract the zero momentum component, 
all the SUB points inherit its large error and beyond $q>5$ 
the errors are practically constant. On the other hand, since the 
action density is known with a very good precision the errors of the 
WARD data points are almost the same as the unsubtracted ones and 
rapidly decrease with increasing momentum. For $q$ less than $3\sim4$,
the SUB data have smaller errors since the fluctuations of $I(q)$
and $I(0)$ cancel due to their strong correlation.
Because of these two advantages of the WARD method (for not too low
momenta), from now on we will use the WARD data exclusively.

Let us now address the extrapolation to infinite lattice size. 
This is done by determining the FS coefficients in (\ref{FScurr})  
from a small reference lattice and then use them to do the extrapolation 
for all other lattices. We used lattice family 4 to determine 
the coefficients, which is the family with four sizes and the 
largest coupling. Using family 2 instead leads to slightly different 
results that allow one to estimate the systematic error in the 
extrapolation procedure. For illustration let us quote the 
(absolute) statistical error ${\rm stat}(q)$ and the 
(absolute) systematic error ${\rm sys}(q)$ 
obtained thereby at $q=5,15,25$: ${\rm stat}(5)=0.0002$, ${\rm sys}(5)= 
0.0006$,  ${\rm stat}(15)=0.0001$, ${\rm sys}(15)= 0.0003$,   
${\rm stat}(25)=0.0001$, ${\rm sys}(25)= 0.0001$. One sees that the 
errors are small and the infinite volume extrapolation is under control.
 
\begin{figure}[htb]
\begin{center}
\vskip 3mm
\leavevmode
\epsfxsize=100mm
\epsfbox{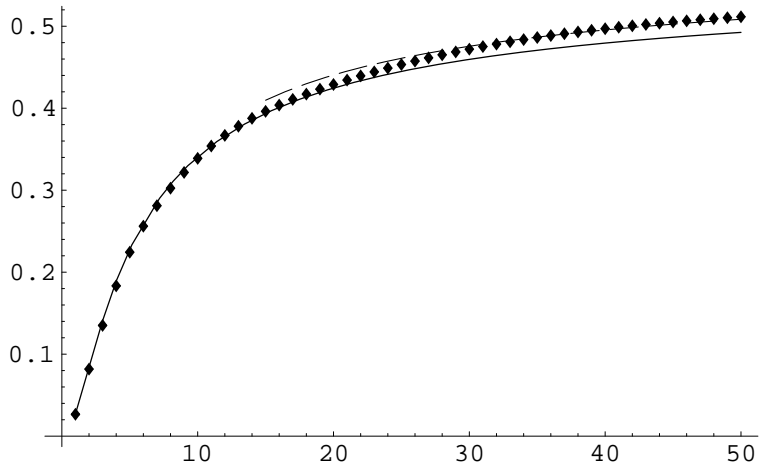}
\vskip -8mm
\end{center}
\caption{{}\footnotesize
Thermodynamic values of the current two-point function at $\xi =418$ 
versus $q$. The solid line is the $2+4$ particle form factor result 
and the dashed line is the absolute upper limit.
}
\label{best}
\end{figure}

In the final step the extrapolation to infinite correlation length
has to be performed. Since the largest lattice 12 already corresponds
to a correlation length of $\xi =418$ one might be tempted to 
regard this as superfluous. However if one were to take these data
as representing the continuum limit one would have to conclude that the 
XY QFT does not coincide with the O$(2)$ bootstrap theory! This 
is because, in contrast to the $\gr$ measurement, the statistical 
errors here are very small and the data differ significantly from the 
bootstrap result. Moreover, as explained below, the truncation error 
in the bootstrap computation is under good control. The situation 
is illustrated in Figure \ref{best}. 

In general it is difficult to strictly control the systematic error 
in a form factor computation caused by the truncation in the number 
of intermediate particles. One only knows that the truncated result provides 
a strict lower bound on the exact answer, since (for a two-point
function) all multi-particle contributions are positive.    
In the O$(2)$ model however we also have a strict upper bound.
This is because the exact $I(q)$ is known to be increasing and to 
approach the value $2/\pi \approx 0.637$ at infinite momentum; 
c.f.~(\ref{2overpi}).  
On the other hand the 2+4 particle approximation is likewise increasing 
and approaches 0.621 at infinity; the difference $0.016$ is an upper 
bound on the error made by the truncation, because also the higher
particle contributions are monotonically increasing. The true value of the 
bootstrap function $I(q)$ is somewhere between the 2+4
approximation and this approximation + 0.016, shown as a 
dashed line in Fig.~\ref{best}. For the relatively low
momentum range we are interested in, the true value is probably closer
to the 2+4 value than to the upper limit.  

\begin{figure}[htb]
\begin{center}
\leavevmode
\epsfxsize=70mm
\epsfbox{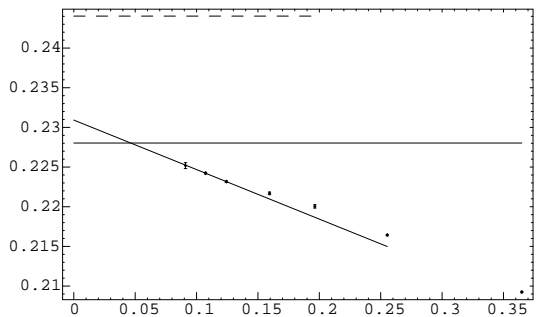}
\vskip -6mm
\end{center}
\caption{{}\footnotesize
Thermodynamic values of the current correlation function for 
lattice families 2,4,6,9,10,11,12
for $q=5$ versus $Q$. 
The solid horizontal line shows the $2+4$ form factor result 
and the dashed horizontal line is the absolute upper limit.
The linear fit is based on the three biggest lattices, corresponding
to families 10, 11 and 12.
}
\label{ywdlog5}
\end{figure}

\begin{figure}[htb]
\begin{center}
\vskip -3mm
\leavevmode
\epsfxsize=0.45\linewidth
\epsfbox{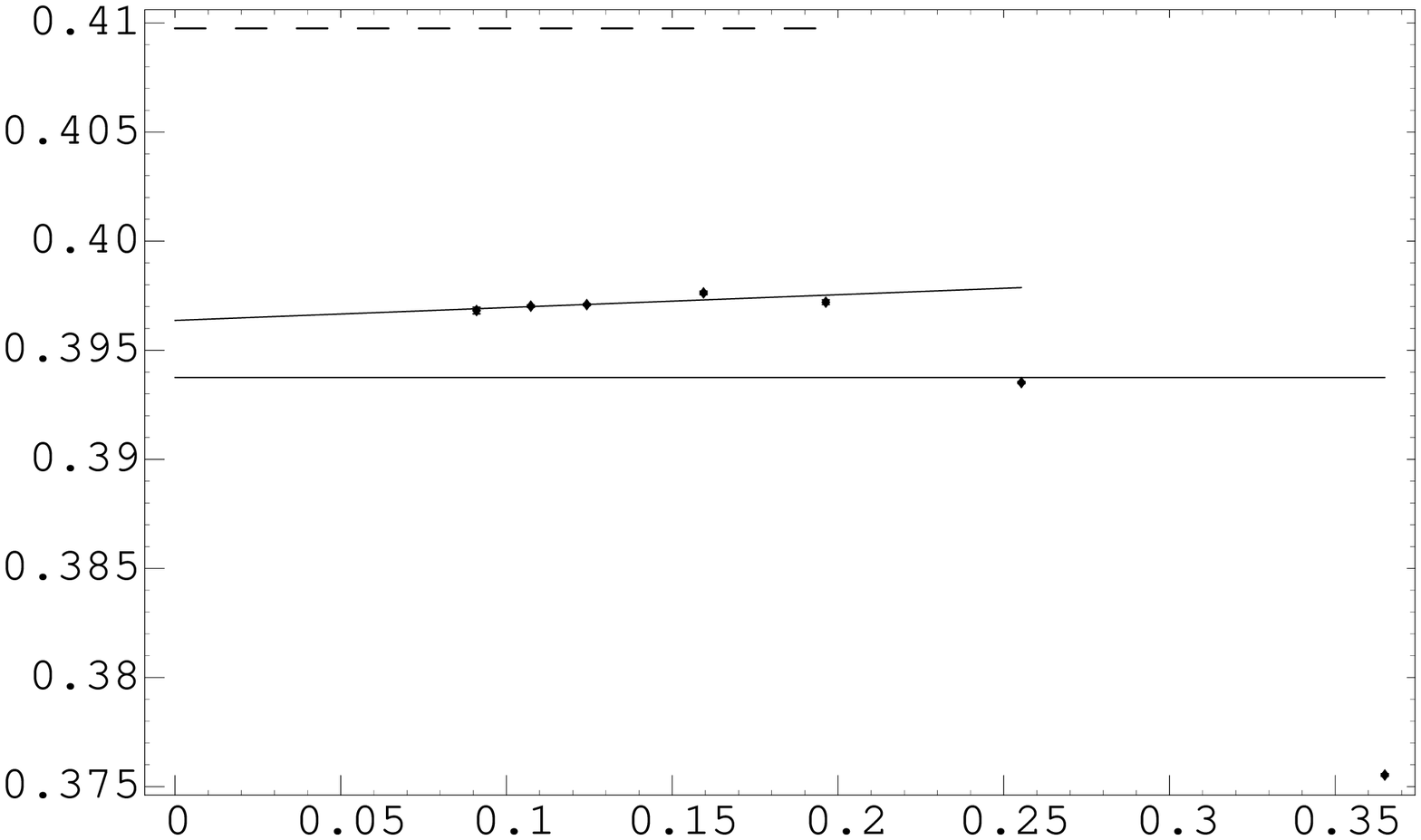}
\hspace{5mm} 
\epsfxsize=0.45\linewidth
\epsfbox{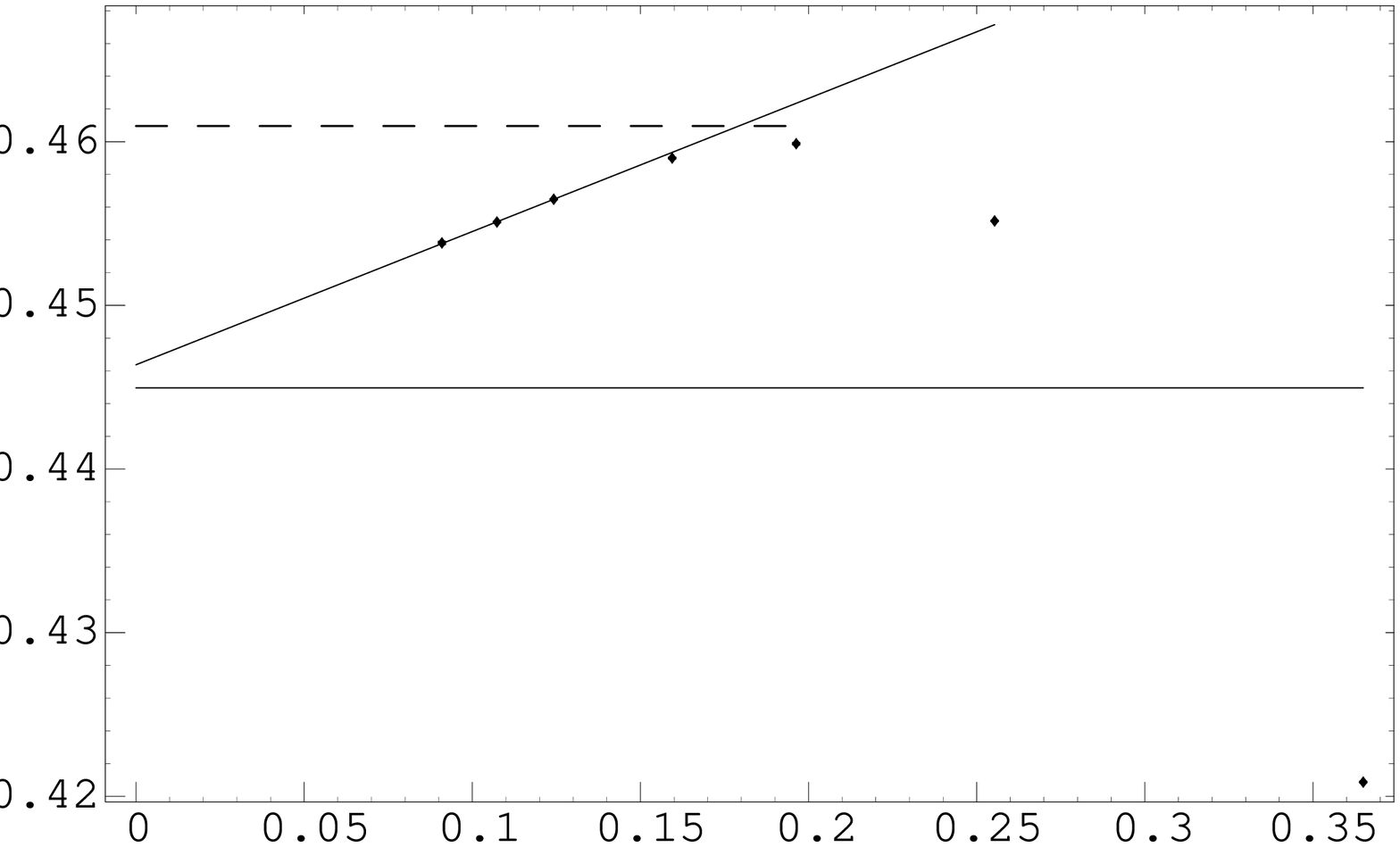}
\vskip -4mm
\end{center}
\caption{{}\footnotesize
Left: same as Fig.~\ref{ywdlog5} for $q=15$. Right: 
same as Fig.~\ref{ywdlog5} for $q=25$. 
}
\label{ywdlog15_25}
\end{figure}

If we use all our data to perform an extrapolation to infinite 
correlation length, the situation changes drastically. We assume
a cutoff dependence asymptotically linear in $Q$ 
\begin{equation}
I(q;K,\infty)=I(q)+I'(q)\, Q + O(Q^2)\,,
\label{linQ}
\end{equation}
and extrapolate our
measurements to infinite correlation length by means of a linear fit to
the data points corresponding to families 10, 11 and 12.
Our thermodynamic data (for families 2,4,6,9,10,11,12) 
together with this fit is
shown in Figures \ref{ywdlog5} and \ref{ywdlog15_25} for $q=5,\, 15, \,25$. 
For concreteness we used $u=1.33$, corresponding to $n_{\rm skip}=4$
in Table 6,
in all our fits. Varying $u$ around this value we observed that our 
results are rather insensitive to the precise choice of $u$ in the 
range 1.2 - 1.5.

\begin{figure}[htb]
\begin{center}
\leavevmode
\epsfxsize=100mm
\epsfbox{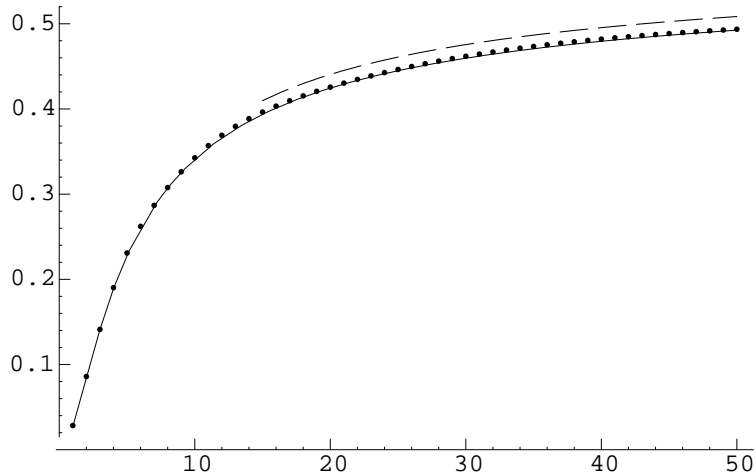}
\vskip -7mm
\end{center}
\caption{{}\footnotesize
Thermodynamic values of the current two-point function extrapolated 
to the continuum limit versus $q$. The solid line is the $2$+$4$ 
particle form factor result and the dashed line is the absolute upper limit.
}
\label{extralog}
\end{figure}

From these figures one infers that the approach to the continuum
limit, for $q>10$, is non-monotonic. While this is not unusual in itself,
it is most remarkable that for the momentum range $q>20$, 
the turning point is around $\xi\sim40$,
a rather large value. Beyond the turning point the data behave as
expected and follow a curve that is approximately linear in $Q$.
If we extrapolate our measurements to the continuum limit using the
linear fit, the extrapolated points agree reasonably well with the 
form factor calculation. This is shown in Figure \ref{extralog}, which is 
one of the main results of this paper. According to this figure the
current two point function of  
the XY QFT is very close to that of the O$(2)$ bootstrap theory. The
absolute
difference between the extrapolated points and the 2+4 FF result is in the 
range [0.0003,0.0032]. This difference is positive and much less than
0.016, consistent with the hypothesis that it is due to higher particle
contributions.

\begin{figure}[htb]
\begin{center}
\vskip -3mm
\leavevmode
\epsfxsize=80mm
\epsfbox{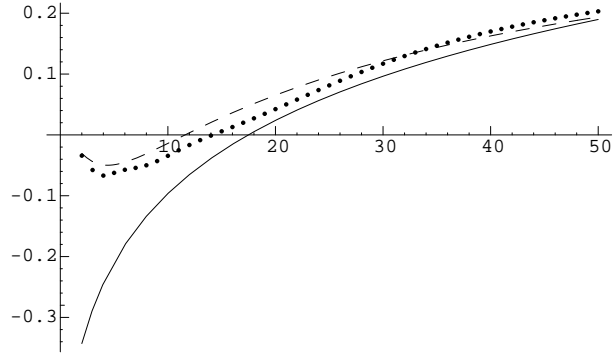}
\vskip -6mm
\end{center}
\caption{{}\footnotesize
Fitted values of the coefficients $I'(q)$ versus $q$. The solid
curve is the perturbative result of \cite{logs} and the dashed curve
is the leading (2-particle) form factor contribution.
}
\label{plCORRlog}
\end{figure}

Not only the extrapolated continuum values $I(q)$ but also the 
coefficient $I'(q)$ governing the rate of approach can be compared to theory.
In the theoretical framework presented in ref.\cite{logs} (and recalled
in Section 4) this coefficient can be
calculated from the continuum SG theory. In \cite{logs} this was done 
in asymptotically free perturbation theory at two-loop order. 
This is expected to be valid for large momenta subject to the 
additional constraint~\cite{logs} $\log q \ll \sqrt{3/2Q}$. Taking
$Q\sim0.1$ our data for $10<q<50$ are just in the window of validity.
Here we calculated the leading 2-particle contribution to the
first correction coefficients using the known \cite{KarWe} 2-particle 
form factor of the Noether current in the SG theory.
In the lattice theory we define $I'(q)$ as the slope of the 
linear fit on the data for the 3 largest lattice families, 10,11,12. 
(Recall that all our results correspond to the choice $u=1.33$.)
The comparison with the theoretical results is shown in 
Figure \ref{plCORRlog}. The agreement is quite remarkable (for the value
of $u$ chosen). 
In particular both analytical computations predict a change of 
sign in $I'(q)$ between $q=10$ and $q=20$, which is indeed observed 
for the fitted numbers obtained from the MC measurements. 

\newsubsection{The spin correlation function}

Finally we consider the two-point correlation function of the spin
operator in Fourier space. While the current correlator had to be
subtracted, the spin correlator has to be normalized
(multiplicatively) before it can be compared to analytical calculations.
We define
\begin{equation}
G^{(\mu)}(q)=\frac{G_0(q)}{G_0(\mu)},
\label{Ga}
\end{equation}
where $G_0(q)$ is the bare spin correlator. Traditionally one takes
$\mu=0$, which amounts to normalizing the spin correlator with
the susceptibility $\chi$. A problematic feature of this definition is that, 
just like with the current defined by the SUB method, the statistical 
errors are big, dominated by the large error of $\chi$. 
We thus introduced $\hat G^{(\mu)}(q):=(q/\mu)^2G^{(\mu)}(q)$, whose 
values are closer to unity, and decided to take $\mu=5$. For $q>5$ this
leads 
to much smaller errors, in many cases smaller by a factor 3-4. 
We adopted this unusual choice of normalization because  
in this paper we focus our attention to the range $q>5$.
The reason for concentrating on this range is that it is here the cutoff
effects show interesting non-monotonic behaviour.
(For the low momentum range $q<5$ it would be better to take the
normalization at $\mu=0$, but this is not investigated here.)

\begin{figure}[htb]
\begin{center}
\vskip +3mm
\leavevmode
\epsfxsize=80mm
\epsfbox{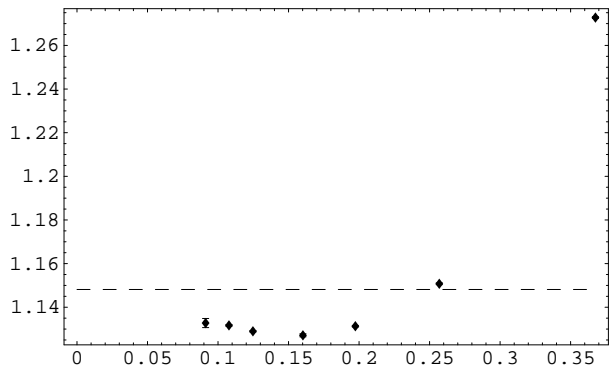}
\vskip -5mm
\end{center}
\caption{{}\footnotesize
Thermodynamic values of the spin correlation function $\hat G^{(5)}(15)$
for lattice families 2, 4, 6, 9, 10, 11 and 12 versus $Q$. 
The dashed line shows the $1+3$ form factor result.
}
\label{spYD515}
\end{figure}

\begin{figure}[bht]
\begin{center}
\vskip +8mm
\leavevmode
\epsfxsize=80mm
\epsfbox{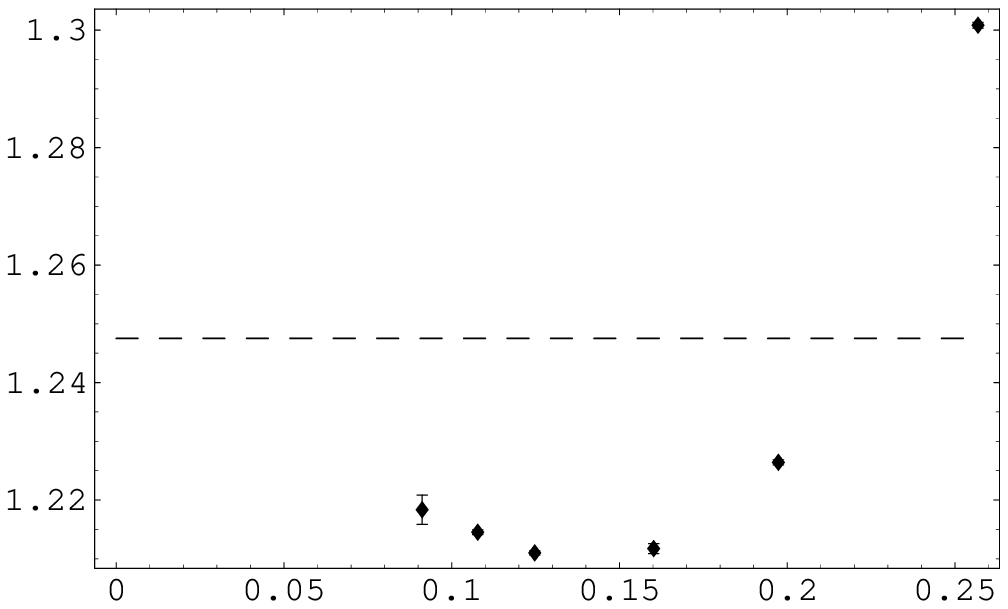}
\vskip -5mm
\end{center}
\caption{{}\footnotesize
Thermodynamic values of the spin correlation function $\hat G^{(5)}(25)$
for lattice families 4, 6, 9, 10, 11 and 12 versus $Q$. 
The dashed line shows the $1+3$ form factor result.
}
\label{sphiYD525}
\end{figure}

\begin{figure}[bht]
\begin{center}
\vskip -3mm
\leavevmode
\epsfxsize=80mm
\epsfbox{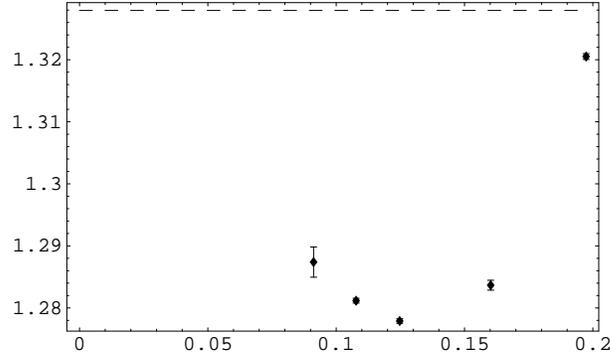}
\vskip -5mm
\end{center}
\caption{{}\footnotesize
Thermodynamic values of the spin correlation function $\hat G^{(5)}(35)$
for lattice families 6, 9, 10, 11 and 12 versus $Q$. 
The dashed line shows the $1+3$ form factor result.
}
\label{sphihiYD535}
\end{figure}

Adopting the normalization at $q=5$ the FS analysis is 
analogous to the current case. The FS scaling hypothesis is applicable
and allows one to extrapolate $\hat G^{(5)}(q)$ to thermodynamic lattices.
The results are shown in Figures \ref{spYD515},
\ref{sphiYD525} and \ref{sphihiYD535} for $q=15$, 25 and 35, respectively.
Again, similarly to the current case, one sees a non-monotonic approach
to the continuum limit, with a turning point which is for $q>20$
around $\xi\sim100$. The actual points are still significantly away
from the analytical prediction, but beyond the
turning point they move into the right direction. We did not attempt
to fit a linear function to the data but our results
for the spin correlator are not inconsistent with the form factor
bootstrap.

\vfill

{}
\pagebreak[1]
{}
\newsection{Conclusion}

Since we already surveyed our motivation and some of the theoretical 
issues involved in the introduction, let us return here to the 
question raised in the title. Screening the comparison between 
bootstrap and lattice theory for the various quantities considered,
we would tend to answer the question in the affirmative.
Probably the strongest Pro argument stems from the intrinsic 
coupling $\gr$. The final results in both approaches have an 
estimated (systematic) error of less than one percent, so that the
good agreement is remarkable. For the two-point function 
of the Noether-current the prediction \cite{logs} for the  
lattice artifacts could be tested. After, and only after, the 
lattice artifacts are taken into account a good and non-trivial 
agreement with the form factor result emerges. A final decision
whether the remaining small differences are due to the neglected higher 
particle contributions or signify in fact a true difference of the
two constructions cannot be reached at this stage.
Quantitatively the least convincing are the phase shift
results. However in the lattice framework measurements of the phase
shifts are technically difficult and here, as in other models, 
mainly the qualitative features at low energies can be probed. 
But the latter do agree with the bootstrap prediction. 
Each comparison considered separately certainly leaves room 
for doubt, but collectively they do suggest
that the continuum limit of the XY model and the O$(2)$ bootstrap
theory are the same QFT.   

Concerning future work, a more detailed exploration of the superselection
structure should be interesting. The new parafermionic superselection
sector found here is probably accompanied by a third (`disorder-like') 
sector. Their interplay e.g.~on the level of the operator product
expansion as well as an explicit field theoretical construction remains
to be found. Finally, as a test case for other sigma-models, it 
would be important to understand which quantities in the XY QFT 
can be understood, qualitatively or quantitatively, in terms of a 
perturbed conformal field theory description. 
\vspace{1cm}

{\tt Acknowledgements:} We wish to thank M.~Karowski and K.H.~Rehren 
for helpful discussions as well as P.~Butera and M.~Hasenbusch 
for useful correspondences. 
This investigation was supported in part by the Hungarian National 
Science Fund OTKA (under T030099, T029802 and T034299) and by
Schweizerischer Nationalfonds.

\clearpage
\newpage
\setcounter{section}{0}
\newappendix{Quantum group covariant form factors}

Here we derive the necessary and sufficient conditions (\ref{cphases})
on the statistics phases that ensure the quantum group covariance 
of the form factors. The additional conditions like (\ref{irredcond})
required for multiplets transforming irreducibly are also detailed. 

To fix our conventions we begin by recalling a few definitions for 
the action of $\cU_q(su(2))$ on some irreducible representation.
Background material on quantum groups in 2-dimensional physics can be 
found in the book \cite{GSRbook}. The Hopf algebra $\cU_q(su(2))$ 
is generated by $X_{\pm},\,H$ that act for generic $q$ on an 
irreducible spin $j$ module according to 
\ba
&& X_{\pm} |j,m\ket = \sqrt{ [j\mp m]_q[j \pm m + 1]_q}
\,|j, m \pm 1\ket\,,\nonum
&& H |j,m\ket = 2 m \,|j,m\ket\,. 
\ea
Here $|j,m\ket$, $m\in \{ -j,-j+1,\ldots, j-1,j\}$ denotes a 
basis of the $(2j+1)$-dimensional irreducible module ${\bf 2 j +1}$,
and $[n]_q = (q^n - q^{-n})/(q - q^{-1})$. 
For $q = - e^{-i\pi/ p}$, $p \geq 3$, an upper bound on the allowed 
isospins $j$ exists. It reads $j \leq p/2 -1$ and is related to 
an enlarged center; see e.g.~\cite{Arn}. To the best of our 
knowledge the case $q =-1$ has not been studied explicitly 
in the literature, but it is not hard to work out the aspects
needed here. First, as a Lie algebra $\cU_{-1}(su(2))$ is isomorphic 
to $su(2)$, but the co-multiplication differs by signs. Guided by 
the sample computations presented below and the formal $p \ra \infty$ 
limit of the above relation, we expect that for $q=-1$ no trunction 
of the allowed isospins occurs. For definiteness we fix the 
roots $q^{1/2} = i,\,q^{-1/2} = -i$. 

For generic $q$ the comultiplication of $\cU_q(su(2))$ is 
\ba
\Delta X_{\pm} \is X_{\pm} \otimes q^{H/2} + q^{-H/2} \otimes X_{\pm}\,,
\nonum
\Delta H \is H \otimes \1 + \1 \otimes H\,,
\ea
for $q \ra -1$ we define it with the above choice of roots. 
For $j=1/2$ we write $|\pm\ket = |1/2, \pm 1/2\ket$ for the basis of the 
defining representation ${\bf 2}$. In the $n$-fold tensor product 
${\bf 2}^{\otimes n}$ we write $|\alpha_n,\ldots ,\alpha_1\ket :=
|\alpha_n\ket \otimes \ldots \otimes |\alpha_1\ket$, $\alpha_j \in \{\pm\}$,
for the natural basis. The `charged' components of a form 
factor are introduced as the coefficients with respect to this basis, i.e.
\be
|f\ket = \sum_{\alpha_n,\ldots ,\alpha_1} f_{\alpha_n\ldots \alpha_1} 
|\alpha_n\ldots \alpha_1\ket\;.
\label{fb}
\ee 
By construction the quantum group generator $H^{\otimes n}$ 
is diagonal on this basis and its eigenvalues 
$e: = \alpha_n + \ldots + \alpha_1$ are the U(1) charges used 
in section 2. The raising and lowering operators $\Delta^{(n)} 
X_{\pm}$ act as a $2^n \times 2^n$ matrix $\Sigma_{\pm}$ 
on the basis $|\alpha_n, \ldots ,\alpha_1\ket$. We choose a 
lexicographical ordering of the basis vectors  
that is symmetric under the flip $\alpha_j \ra -\alpha_j$. 
Then $\Sigma_- = \Sigma_+^T$. Further there is an induced action 
on the coefficients $f_{\alpha_n\ldots \alpha_1}$ in (\ref{fb})
which is implemented by $\Sigma_-$ for $\Delta^{(n)} X_+$ and 
by $\Sigma_+$ for $\Delta^{(n)} X_-$. The  $2^n \times 2^n$ 
matrices $\Sigma_{\pm}$ are triangular and `sparse' 
with only a few blocks different from zero. The block structure 
arises because evidently $\Delta^{(n)} X_{\pm}$ maps the charge 
$e$ sector into the charge $e \pm 2$ sector. Explicitly the 
matrix $\Sigma_-$ acts on the form factor components as
\be
\Sigma_- : \quad f_{\alpha_n \ldots \alpha_1} \rra
e^{i\frac{\pi}{2}(e-1)} 
\sum_{j =1}^n \Big(\frac{1 + \alpha_j}{2} \Big)\,(-1)^{n-j}
f_{\alpha_n \ldots -\alpha_j \ldots \alpha_1}\;,
\label{xf}
\ee
and similarly for $\Sigma_+$. The mapping (\ref{untwist}) could
be used to `untwist' the $SU_{-1}(2)$ co-multiplication, i.e.~to
remove the phases in (\ref{xf}). We refrain from doing so 
because the `untwisting' does not induce a physically interesting
correspondence between the form factor sequences of the 
$SU(2)$ and the $SU_{-1}(2)$ bootstrap theories. 

As usual the n-particle form factors carry a representation 
of the permutation group $S_n$. Its representation matrices 
are $2^n \times 2^n$ matrices $L_s(\th),\,s \in S_n$.
The quantum group invariance of the S-matrix (\ref{S5})
generalizes to 
\be
\Sigma_{\pm}\, L_s(\th) =  L_s(\th)\, \Sigma_{\pm} \,,\sspace
\forall \,s \in S_n\,.
\label{inv1}
\ee  
For completeness let us note the explicit definition. One 
sets
\be
L_{s_j}(\th)_{\alpha_n \ldots \alpha_1}^{\beta_n \ldots \beta_1} 
= \d_{\alpha_n}^{\beta_n} \ldots
S_{\alpha_{j+1}\alpha_j}^{\beta_j \;\beta_{j+1}}(\th_{j+1,j}) 
\ldots  \d_{\alpha_1}^{\beta_1}\;,\quad j =1,\ldots n-1\;,
\ee
for the generators $s_1,\ldots, s_{n-1}$, acting by 
$s_j(\th_n,\ldots,\th_1) = (\th_n, \ldots ,\th_j,\th_{j+1} ,
\ldots,\th_1)$ on the rapidities $\th := (\th_n,\ldots,\th_1)$.
By means of $L_{ss'}(\th) = L_s(\th) L_{s'}(s\inv \th)$ 
this extends to all $s,s' \in S_n$. The invariance (\ref{inv1}) 
clearly entails that the form factor Eq.~(\ref{ff3}a) (and its 
generalization to generic $s \in S_n$) are covariant under the 
quantum group action.

It is natural to ask whether the same can be achieved for the 
cyclic form factor equations (\ref{ff3}b). In that case the 
cyclic Eqs in the charge $e$ sector 
\be
f_{\alpha_n\ldots \alpha_1}(\th_n + 2\pi i,\th_{n-1},\ldots,\th_1) = 
\eta_{\alpha_n}(e) \,f_{\alpha_{n-1}\ldots \alpha_1\alpha_n}
(\th_{n-1},\ldots,\th_1,\th_n)\,,
\label{inv2}
\ee
with $e =\alpha_n + \ldots + \alpha_1$ and those in the charge $e \pm 2$ 
sector would again be compatible with the quantum group symmetry. 
Explicitly this means that starting from (\ref{inv2}) and performing 
e.g.~the substitutions (\ref{xf}) the result 
should be an identity by virtue of the cyclic Eq.~in 
the charge $e \pm 2$ sectors. This condition gives rise to
a set of overdetermined relations among the phases 
$\eta_{\alpha}(e)$, -- which turn out to be self-consistent. 
Using (\ref{xf}) the solution (\ref{cphases}) 
can be verified.

Finally we turn to the residue equations. Consistency 
requires that the inverse of the matrix $\Gamma_{\alpha}^{\beta}$
in (\ref{ff3}b) appears on the right hand side, irrespective of 
its concrete form. In the charged basis and for $n \geq 3$ one 
has:
\ba
&\nspace & \frac{i}{2} {\rm res}_{\th_n =\th_{n-1} + i\pi} 
f_{\alpha_n \ldots \alpha_1}(\th_n,\ldots ,\th_1) \\
&\nspace & = \delta_{\alpha_n + \gamma} \left[
\eta_{\gamma}(e)\inv L_{s_{n-2} \ldots s_1}(\th)%
_{\alpha_{n-1} \alpha_{n-2} \ldots \;\alpha_2 \alpha_1}%
^{\beta_{n-2} \beta_{n-3} \ldots \;\,\beta_1 \gamma} - 
\d_{\alpha_{n-1}}^{\gamma} \d_{\alpha_{n-2}}^{\beta_{n-2}}
\ldots \d_{\alpha_1}^{\beta_1}
\right]f_{\beta_{n-2} \ldots \beta_1}(\th_n\!-\!2,\ldots ,\th_1)\;.
\nonumber
\label{ffres}
\ea
Here $e = \alpha_n + \ldots + \alpha_1 = \beta_{n-2} + \ldots + \beta_1$ 
refers to the charge sector. For the specific choice of phases 
(\ref{cphases}) these Eqs can be seen to be likewise quantum group 
covariant.

In summary there exists a preferred (and up a normalization 
uniquely determined) choice of the statistics phases 
$\eta_{\alpha}(e)$ for which the form factor equations
(\ref{ff3}), (\ref{ffres}) are covariant with respect to the 
quantum group $\cU_{-1}(su(2))$. This means its solutions 
can be grouped into multiplets that transform covariantly 
under the symmetry group, and one can restrict attention to 
those transforming irreducibly. Technically the irreducibility 
condition can be encoded into a parameterization of the form 
factors that is adapted to the embedding of the irreducible 
spin $j$ module ${\bf 2 j +1}$ into ${\bf 2}^{\otimes n}$,
including multiplicities. Essentially it amounts to 
determining the generalized Clebsch-Gordon coefficients.   

To faciliate the comparison with the familiar $\cU_1(su(2)) = 
su(2)$ case we first consider the decomposition for 
generic $q$ and specialize to $q = -1$ only at the end. 
Thus let ${\bf 2}$ again denote the defining two-dimensional
representation of $\cU_q(su(2))$ and consider the decomposition 
of ${\bf 2}^{\otimes n}$ into irreducible representations.
It assumes the familiar form 
\be
{\bf 2}^{\otimes n} = \bigoplus_{j_0 \leq j \leq n/2} 
m_j(n) {\bf (2 j+1)}\;,
\label{Q1}
\ee
where $m_j(n)$ is the multiplicity with which  ${\bf 2 j+1}$
occurs and $j_0 = 0,1/2$ for $n$ even,odd, respectively. For 
generic $q$ these multiplicities are the same as for $su(2)$,
only the Clebsch Gordon coefficients differ. As outlined before 
we expect the limit $q \ra -1$ to be regular in the sense that 
no trunctation of the isospins occurs and that the multiplicities
$m_j(n)$ are the same as in the undeformed case. The multiplicities
then are conveniently computed from a generalized `Pascal triangle' 
described by the recursion relations
\ba
&& m_{n/2}(n) =1\;,\;\;\;m_j(n) =0\;,\;\; j<0\;,\nonum
&& m_j(n) = m_{j-1/2}(n-1) + m_{j +1/2}(n-1)\;, \;\;\;
j = j_0, \ldots, n/2\;.
\label{Q2}
\ea
The highest weight conditions $\Delta^{(n)} X_+ v =0,
\,v \in {\bf 2}^{\otimes n}$, are readily solved and yield 
$m_j(n)$ linearly independent solutions on each of which a spin $j$ 
multiplet can be based.   

Below we list for $n=2,3,4$ a basis for the spin $j$ sector in 
${\bf 2}^{\otimes n}$. The 
multiplicites are taken into account by displaying a $m_j(n)$-dimensional
family of highest weight vectors.  

$n=2:$
\bas
j=0: && q^{1/2} |+-\ket - q^{-1/2} |-+\ket\;,\nonum
j=1: && |++\ket\;,\quad 
\Sigma_- |++\ket \sim q^{1/2} |-+\ket + q^{-1/2} |+-\ket\;,
\quad \Sigma_-^2 |++\ket \sim |--\ket\;.
\eas 

$n=3:$
\bas
j=1/2: && v_{1/2} = \lb_1 |-++\ket + \lb_2 |+-+\ket + \lb_3 |++-\ket 
\;,\quad q^2\lb_1 + q \lb_2 + \lb_3 =0\,, \nonum
&& \Sigma_- v_{1/2} = (\lb_2 + q\inv \lb_1  ) |+--\ket + 
(\lb_1 + \lb_3)|-+-\ket + (q \lb_1 +\lb_2) |--+\ket\;,\nonum 
j=3/2: &&  \Sigma^k_- |+++\ket\;, \quad k =0,1,2,3\,.
\eas 

$n=4:$
\bas
j=0: && v_0 = |--++\ket -(q - q^{-1}) |-+-+\ket - |-++-\ket 
- |+--+\ket + q^2 |++--\ket\;,\nonum
&& v'_0 = q^{-1} |-+-+\ket + |-++-\ket + |+--+\ket 
- q |+-+-\ket\;,\nonum
j=1: && v_1 = \lb_1 |-+++\ket + \lb_2 |+-++\ket + \lb_3 |++-+\ket + 
\lb_4 |+++-\ket\;, \nonum
&& q^3 \lb_1 + q^2 \lb_2 + q \lb_3 + \lb_4 =0\;,
\nonum
&& \Sigma_- v_1  \sim q(q \lb_1 + \lb_2)|--++\ket + 
q(\lb_1 + \lb_3) |-+-+\ket + (\lb_3 + q \lb_4) |-++-\ket \nonum
&& \sspace + (q \lb_2 + \lb_3)|+--+\ket + (\lb_2 + \lb_4) |+-+-\ket
+ (\lb_1 + q\inv \lb_4) |++--\ket\;.
\nonum
&& \Sigma_-^2 v_1  = q^2 \lb_1 |+---\ket + 
q^2[\lb_2 + (q - q\inv) \lb_1]|-+--\ket \nonum
&& \sspace + [\lb_3 -(q - q\inv) \lb_4]|--+-\ket + \lb_4 |---+\ket\;, 
\nonum
j=2: &&  \Sigma^k_- |++++\ket\;, \quad k =0,1,2,3,4\,.
\eas 

To illustrate the use of this table let us look at the $n=3,\, j=1/2$ 
entry. The form factor components $f_1,f_2,f_3$ in (\ref{fpcomp}) play 
the role of the $\lb$'s and for $q =-1$ one obtains (\ref{irredcond}).

\newappendix{3-particle Form Factors}

Here we construct the 3-particle form factors of the 
spin and the parafermion field by an adaptation of the 
technique of \cite{Karowskietal}. From the SG viewpoint
these fields are nonlocal which is why they have not been 
considered in \cite{Karowskietal}. All even particle form 
factors of the SG fields were constructed by Smirnov 
\cite{Smir1} where also the Bethe ansatz technique instrumental 
in \cite{Karowskietal} is implicit; see also \cite{TarVar}
for related results in the mathematical literature.     

Adapting theorem 4.1 of \cite{Karowskietal} the 3-particle FF
can be represented as the contour integral
\begin{equation}
\begin{split}
f^\alpha_{\epsilon_3\epsilon_2\epsilon_1}(\theta_3,\theta_2,\theta_1)&=
Y(\theta_3,\theta_2,\theta_1)\int_{{\cal C}}du\,
\gamma^\alpha_{\alpha_1\alpha_2}(\theta_3,\theta_2,\theta_1;u)
v_{\beta_3\beta_2\beta_1}\\
& \quad \;\,S_{\epsilon_1\alpha_1}^{\beta_1\gamma_1}(\theta_1-u)  
S_{\epsilon_2\gamma_1}^{\beta_2\gamma_2}(\theta_2-u)  
S_{\epsilon_3\gamma_2}^{\beta_3\alpha_2}(\theta_3-u)\,.  
\end{split}
\label{BA}
\end{equation}
Here
\begin{equation}
Y(\theta_3,\theta_2,\theta_1)=y(\theta_3-\theta_2)\,
y(\theta_3-\theta_1)\,y(\theta_2-\theta_1)\,,
\label{Y}
\end{equation}
and the only nonvanishing component of $\gamma^\alpha_{\alpha_1\alpha_2}$ is
\begin{equation}
\gamma^+_{+-}(\theta_3,\theta_2,\theta_1;u)={\cal N}
e^{s(\theta_1+\theta_2+\theta_3-2u)}
\prod_{m=1}^3\,\frac{\phi(\theta_m-u)}{S_2(\theta_m-u)}\,,
\label{gamma++-}
\end{equation}
where
\be
{\cal N}=\frac{i}{4\pi^{11/2}}\,e^{-\Delta(0)}e^{-i\pi s}\;,\sspace
\phi(\theta):=\Gamma\left(\frac{1}{2}+\frac{x}{2\pi i}\right)
\Gamma\left(-\frac{x}{2\pi i}\right)\,.
\label{phi}
\ee
Finally the \lq\lq pseudo-vacuum" vector is
\begin{equation}
v_{\beta_3\beta_2\beta_1}=
\delta_{\beta_3+}\delta_{\beta_2+}\delta_{\beta_1+}\,.
\label{vjjj}
\end{equation}
The integration contour ${\cal C}$ consists of several
pieces. First, there are three small clockwise circles
around the three points $\theta_1,\theta_2$ and $\theta_3$.
In addition ${\cal C}$ also contains a line integral parallel
to the real axis such
that the integration path goes between $\theta_m-i\pi$
and $\theta_m-2i\pi$ for all the three $\theta_m$.

Since ${\cal C}$ is defined relative to the arguments
$\theta_m$, when we analytically continue (\ref{BA})
it is best to deform the contour parallel to the arguments.
This way it is trivial to see that the solution satisfies
(\ref{ffspin}). It is also relatively easy to see that
the Bethe Ansatz like construction (\ref{BA})
ensures that (\ref{ff3}a) is also satisfied, independently
of the contour ${\cal C}$. 
(To show this one has to use the Yang-Baxter equation 
satisfied by the S-matrix.) 
It is more difficult to verify (\ref{ff3}b)
because here one has to take into account that
the contours are different on the two sides of the
equation. Similarly for the residue equations (\ref{translates})
the contour is different from the original ${\cal C}$.

Inserting the S-matrix (\ref{S1}) Eq.~(\ref{BA}) can be rewritten as 
\begin{equation}
f_m(\theta_3,\theta_2,\theta_1)={\cal N}
Y(\theta_3,\theta_2,\theta_1)\int_{{\cal C}}du\,
e^{s(\theta_1+\theta_2+\theta_3-2u)}
\left[\prod_{k=1}^3\phi(\theta_k-u)\right]
t_m(\theta_3,\theta_2,\theta_1;u)\,,
\label{fm}
\end{equation}
where
\begin{eqnarray}
t_1(\theta_3,\theta_2,\theta_1;u)&=&
\frac{\theta_3-u}{i\pi-\theta_3+u}\,
\frac{\theta_2-u}{i\pi-\theta_2+u}\,
\frac{i\pi}{i\pi-\theta_1+u}\,,\nonumber\\
t_2(\theta_3,\theta_2,\theta_1;u)&=&
\frac{\theta_3-u}{i\pi-\theta_3+u}\,
\frac{i\pi}{i\pi-\theta_2+u}\,,\label{tm}\\
t_3(\theta_3,\theta_2,\theta_1;u)&=&
\frac{i\pi}{i\pi-\theta_3+u}\,.\nonumber
\end{eqnarray}

Next we examine whether the solution (\ref{tm}) is compatible 
with SU$_{-1}(2)$ symmetry, which requires the vanishing of the linear
combination (\ref{irredcond}). This can be written as
\be
\zeta(\theta_3,\theta_2,\theta_1)={\cal N}Y(\theta_3,\theta_2,\theta_1)
\left[z_1(\theta_3,\theta_2,\theta_1)+z_2(\theta_3,\theta_2,\theta_1)
\right]\,,
\label{zeta1}
\end{equation}
where
\ba
z_1(\theta_3,\theta_2,\theta_1)&=&\int_{{\cal C}}e^{s(\theta_1+\theta_2+
\theta_3-2u)}\left[\prod_{k=1}^3\frac{\theta_k-u}
{i\pi-\theta_k+u}\,\phi(\theta_k-u)\right],\label{z1}\\
z_2(\theta_3,\theta_2,\theta_1)&=&\int_{{\cal C}}e^{s(\theta_1+\theta_2+
\theta_3-2u)}\left[\prod_{k=1}^3\phi(\theta_k-u)\right].
\label{z2}
\end{eqnarray}
Using the identity
\be
\frac{z}{i\pi-z}\,\phi(z)=\phi(z-2\pi i)
\end{equation}
(\ref{z1}) can be rewritten as
\be
z_1(\theta_3,\theta_2,\theta_1)=e^{4\pi is}
\int_{{\cal C}+}e^{s(\theta_1+\theta_2+
\theta_3-2u)}\left[\prod_{k=1}^3\phi(\theta_k-u)\right],
\label{shiftz1}
\end{equation}
where the contour ${\cal C}+$ is shifted by $2\pi i$, i.e. it consists
of three small circles around $\theta_m+2\pi i$ and the line integral
goes between $\theta_m+i\pi$ and $\theta_m$. From (\ref{phi}) we can see
that the small circles do not contribute here since the integrand is regular
there. The only relevant singularities are those at $\theta_m$ and it is
easy to see that the contribution of the shifted line integral is
precisely the same as that of ${\cal C}$. We thus have
\be
\zeta(\theta_3,\theta_2,\theta_1)=(\eta^2+1)
{\cal N}Y(\theta_3,\theta_2,\theta_1)z_2(\theta_3,\theta_2,\theta_1)\,.
\label{zeta2}
\end{equation}
This proves the assertion after (\ref{irredcond}). Remarkably the 
quantum group invariance is not visible on the level
of the integrand in (\ref{BA}) but only after the integral
has been performed. In addition this fixes the value 
of the spin to be $s = \pm 1/4$, without mod(1/2) 
ambiguities. This is because the integral 
in (\ref{BA}) exists for $|s| < 3/4$ only. 

\newappendix{4-particle Form Factors of the Noether Current}

The form factors of the Noether current can be found in 
Smirnov's book \cite{Smir1}.
We have adapted this result to our notations and conventions for the
4-particle case.

Let us introduce the reduced form factors $g$ that are defined by 
\be
f_{\epsilon_1\epsilon_2\epsilon_3\epsilon_4}
(\theta_1,\theta_2,\theta_3,\theta_4)=
Y(\theta_1,\theta_2,\theta_3,\theta_4)
g_{\epsilon_1\epsilon_2\epsilon_3\epsilon_4}
(\theta_1,\theta_2,\theta_3,\theta_4)\,,
\end{equation}
where
\be
Y(\theta_1,\theta_2,\theta_3,\theta_4)=
\prod_{i<j}y(\theta_i-\theta_j)\,.
\end{equation}

Using the O$(2)$ symmetry and charge conjugation, we need only the following
components:
\ba
g_{++--}(\theta_1,\theta_2,\theta_3,\theta_4)
&=&-g_{--++}(\theta_1,\theta_2,\theta_3,\theta_4)\,,\nonumber\\
g_{+--+}(\theta_1,\theta_2,\theta_3,\theta_4)
&=&-g_{-++-}(\theta_1,\theta_2,\theta_3,\theta_4)\,,\\
g_{+-+-}(\theta_1,\theta_2,\theta_3,\theta_4)
&=&-g_{-+-+}(\theta_1,\theta_2,\theta_3,\theta_4)\,.\nonumber
\ea

Further, using the axioms, we can express everything in terms of a
single function \break
$A(\theta_1,\theta_2,\theta_3,\theta_4)$ as follows.
\ba
g_{++--}(\theta_1,\theta_2,\theta_3,\theta_4)
&=&A(\theta_1,\theta_2,\theta_3,\theta_4)\,,\nonum
g_{+--+}(\theta_1,\theta_2,\theta_3,\theta_4)
&=&A(\theta_4+2\pi i,\theta_1,\theta_2,\theta_3)
\ea
and
\be
\begin{split}
g_{+-+-}(\theta_1,\theta_2,\theta_3,\theta_4)
=\frac{i\pi+\theta_3-\theta_4}{\theta_4-\theta_3}&\Big[
A(\theta_1,\theta_3+2\pi i,\theta_4,\theta_2)\\
&-\frac{i\pi}{i\pi-\theta_4+\theta_3}\,
A(\theta_1,\theta_4+2\pi i,\theta_3,\theta_2)\Big]\,.
\end{split}
\label{pmpm}
\end{equation}

This function is given by
\be
A(\beta_1,\beta_2,\beta_3,\beta_4)=
\frac{2ie^{-2\Delta(0)}}{\pi^4}
\frac{e^{-\frac{1}{2}\big(\sum_{j=1}^4\beta_j\big)}
}{\big(\sum_{j=1}^4e^{-\beta_j}\big)}p(\beta_1,\beta_2,\beta_3,\beta_4)
I(\beta_1,\beta_2,\beta_3,\beta_4)\,,
\label{Aexplicit}
\end{equation}
where
\be
p(\beta_1,\beta_2,\beta_3,\beta_4)=(\beta_1+\beta_2-\beta_3-\beta_4-2\pi i)
\Big[\prod_{i=1}^2\prod_{j=3}^4\frac{1}{\beta_i-\beta_j-i\pi}\Big]
\label{pdef}
\end{equation}
and
\be
\begin{split}
I(\beta_1,\beta_2,\beta_3,\beta_4)=
\int_{-\infty}^\infty d\alpha &e^\alpha
\big\{q(\alpha,\beta_1)q(\alpha,\beta_2)x(\alpha,\beta_3)x(\alpha,\beta_4)\\
&+x(\alpha,\beta_1)x(\alpha,\beta_2)q(\alpha,\beta_3+2\pi i)
q(\alpha,\beta_4+2\pi i)\big\}\,.
\end{split}
\label{Idef}
\end{equation}
Here we defined
\ba
x(\alpha,\beta)&=&\Gamma\Big(\frac{1}{4}-\frac{\alpha-\beta}{2\pi i}\Big)
\,\Gamma\Big(\frac{1}{4}+\frac{\alpha-\beta}{2\pi i}\Big)=
\phi(\beta-\alpha-\frac{i\pi}{2})\,,\nonum
q(\alpha,\beta)&=&\Gamma\Big(\frac{1}{4}-\frac{\alpha-\beta}{2\pi i}\Big)
\,\Gamma\Big(\frac{5}{4}+\frac{\alpha-\beta}{2\pi i}\Big)\,.
\ea

\newpage
\newappendix{Calculation of the subleading correction to $\gr$}

The leading contribution to the intrinsic coupling $\gr$ in the O$(2)$ model
was calculated in \cite{SigmacollabII}. Here we present the calculation
of the most important subleading term reducing the theoretical uncertainty
in this quantity to a few per mille. This appendix relies heavily on 
\cite{SigmacollabII}, especially Sections 3, 6 and Appendix C.
We use the notations and conventions of that paper.

The first few contributions to $\gamma_4$ for the Ising
model and for the O$(3)$ model were also calculated in \cite{SigmacollabII}.
Let us compare the results.

\begin{table}[htbp]
\begin{center}
\begin{tabular}{|c|r|r|}
\hline
contribution & Ising model &  O$(3)$ model \\
\hline
121 & $-4.99343$ & $-4.16835$\\
\hline
123/1 & $-0.01348$ & $-0.01351$\\
\hline
123/2 & 0.10610 & 0.11901\\
\hline
123/3 & 0.00000 &$-0.00200$\\
\hline
141 & $-0.00265$ &$-0.00407$\\
\hline
\end{tabular}
\caption{\footnotesize
The first few contributions to $\gamma_4$. 123/i stand for
the contribution of the integrals $V^{(i)}$ for i=1, 2 and 3.}
    \label{data}
  \end{center}
\end{table}
The pattern is strikingly similar for the two models. For the
XY model we so far only have the $121$ contribution, which is somewhere
inbetween the corresponding values for the Ising and the O(3) model. 
If we assume that the corrections follow the same pattern also for 
the O$(2)$ model, already the calculation of the 123/2 term  
yields $\gamma_4$ with a precision better than one percent. 
We will see that we have all the ingredients necessary for this 
calculation.

Using the FF axioms and some of the expressions in Appendix C of
\cite{SigmacollabII} we have%
\ba
g^{(3)}(\beta,\alpha_1,\alpha_2)
&=&{\cal G}^1_{1bb}(i\pi,\beta,-\beta)\,
{\cal G}^1_{bx_1x_2}(i\pi-\beta,\alpha_1,\alpha_2)\,
{\cal G}^{*1}_{bx_1x_2}(\beta,\alpha_1,\alpha_2)\label{g3}\\
&-&{\cal G}^1_{1bb}(i\pi,-\beta,\beta)\,
{\cal G}^1_{bx_1x_2}(i\pi-\beta,\alpha_1,\alpha_2)\,
{\cal G}^{*1}_{x_1x_2b}(\alpha_1,\alpha_2,\beta)\,.\nonumber\\
\nonumber
\ea
Now it is easy to see that $g^{(3)}(0,\alpha,-\alpha)=0$ so
\be
V^{(4)}=0
\label{V4}
\ee
in general. Further using (\ref{g3}) in $(C.43)$ of \cite{SigmacollabII}
we get
\ba
g^{(5)}(\beta)
&=&{\cal G}^1_{1bb}(i\pi,\beta,-\beta)\,\Big\{
{\cal G}^{*1}_{bb1}(\beta,-\beta,0)+
{\cal G}^{*1}_{b1b}(\beta,0,-\beta)\Big\}\label{g5}\nonum
&-&{\cal G}^1_{1bb}(i\pi,-\beta,\beta)\,\Big\{
{\cal G}^{*1}_{b1b}(-\beta,0,\beta)+
{\cal G}^{*1}_{1bb}(0,-\beta,\beta)\Big\}\,.
\ea 
We see that
\be
g^{(5)}(\beta)+g^{(5)}(-\beta)=0\,,
\ee
so also for generic $n$
\be
V^{(5)}=0\,.
\label{V5}
\ee
(\ref{V4}) and (\ref{V5}) together imply that the conjecture $(C.57)$ 
of \cite{SigmacollabII} is true. 

Similarly simplifying $(C.37)$ of \cite{SigmacollabII} gives
\be
g^{(2)}(\beta)=
{\cal G}^1_{1bb}(i\pi,\beta,-\beta)\,
{\cal G}^{*1}_{bb1}(\beta,-\beta,0)
\ee
and using this in $(C.36)$ of \cite{SigmacollabII} we have
\be
{\cal F}^1_{1bb}(i\pi,v,-v)\,
{\cal F}^1_{bb1}(-v,v,0)\,.
\label{V2}
\ee
It is easy to see that we already have everything that is necessary
to compute (\ref{V2}) since
\be
{\cal F}^1_{xy1}(-v,v,0)=S_{y1;qp}(v)\, f^1_{pqx}(i\pi+v)\,.
\ee
Putting everything together we have
\be
V^{(2)}= \frac{1}{64\pi}\int_0^\infty dv\, \frac{\sinh^2v}{\cosh^4v}\,
e^{H(v)}\,{\cal M}(v)\,,
\label{V21}
\ee
where
\be
H(v)=\Delta(i\pi+v)+\Delta(i\pi-v)+\Delta(2v)+\Delta(-2v)+
\Delta(v)+\Delta(-v)-2\Delta(0)
\ee
and
\ba
{\cal M}(v)&=& \frac{1}{(i\pi-v)\big[(n-2)v-2\pi i\big]}\Big\{
K(v)K(i\pi+v)\,\big[2\pi^2+n(n-2)v(i\pi-v)\big]\nonumber\\
&+&2K(v)L(i\pi+v)\,\big[(n+1)\pi^2+(n-2)v(i\pi-v)\big]\\
&+&2L(v)K(i\pi+v)\,\big[2\pi^2+(n-2)v(i\pi-v)\big]\nonumber\\
&+&4L(v)L(i\pi+v)\,\big[2\pi^2-i\pi v+(2-n)v^2\big]\Big\}\,.\nonumber\\
\nonumber
\ea

To summarize, using the results of \cite{SigmacollabII}
(\ref{V21}) in the $n=2$ case can be represented as
\ba
V^{(2)}&=&
\frac{1}{3600\pi^7}\int_0^\infty d\theta 
\frac{\sinh^3\theta}{\cosh^4\theta}
\Bigg\{e^{H+D_3+D_1}\sqrt{(4\theta^2+25\pi^2)
(4\theta^2+49\pi^2)}
\Big[\frac{A_1}{\theta}+ \frac{\pi A_2}{2(\theta^2+\pi^2)}\Big]
\nonumber\\
&&\qquad\qquad\qquad\qquad\qquad+\,\,
e^{H+D_5+D_1}(4\theta^2+9\pi^2)
\Big[\frac{A_1}{\theta}+ \frac{3\pi A_2}{2(\theta^2+\pi^2)}\Big]
\Bigg\}\,,\label{V22}\\
\nonumber
\ea
where
\begin{subeqnarray}
H(\theta)&=&2\int_0^\infty \frac{d\omega}{\omega}\,
\frac{\cos\omega\theta+
\cosh\pi\omega(\cos2\omega\theta+\cos\omega\theta-1)-2}%
{\sinh\pi\omega(1+e^{\pi\omega})}\,,\\
D_1(\theta)&=&\int_0^\infty \frac{d\omega}{\omega}\,
\frac{\cos \frac{\omega\theta}{2} -1}{2\sinh
\frac{\pi\omega}{2}}\,k(\omega)\,,\\
D_3(\theta)&=&\int_0^\infty \frac{d\omega}{\omega}\,
\frac{\cosh\pi\omega \cos\frac{\omega\theta}{2}-
\cosh \frac{\pi\omega}{2} }%
{\sinh\pi\omega}\,k(\omega)\,,\\
D_5(\theta)&=&\int_0^\infty \frac{d\omega}{\omega}\,
\frac{\cos \frac{\omega\theta}{2} -\cosh \frac{\pi\omega}{2}}%
{\sinh\pi\omega}\,k(\omega)\,. 
\end{subeqnarray}
Further
\be
k(\omega)=-e^{-\frac{5}{4} \pi\omega}-e^{-\frac{7}{4}\pi\omega}\,,
\ee
and finally
\be
A_1+iA_2=i(i\pi-2\theta)(3i\pi+2\theta)(5i\pi+2\theta)
e^{iD_2(\theta)}\,,
\ee
where
\be
D_2(\theta)=-\int_0^\infty \frac{d\omega}{\omega}\,
\frac{\sin \frac{\omega\theta}{2}}{2\cosh
\frac{\pi\omega}{2}}\,k(\omega)\,.
\ee

Numerically we find
\be
V^{(2)}=0.00724518.
\ee

Now we are in a position to be able to fill in some O$(2)$ entries
in the previous table.

\begin{table}[htbp]
  \begin{center}
\begin{tabular}{|c|r|c|r|}
\hline
contribution & Ising model & O$(2)$ model &  O$(3)$ model \\
\hline
121 &$-4.99343$ & $-4.65718$ & $-4.16835$\\
\hline
123/1 & $-0.01348$ &$*$& $-0.01351$\\
\hline
123/2 & 0.10610 & ${\phantom{-}}0.11592$ & 0.11901\\
\hline
123/3 & 0.00000 &$*$& $-0.00200$\\
\hline
141 & $-0.00265$ &$*$& $-0.00407$\\
\hline
\end{tabular}
\caption{\footnotesize
The first few contributions to $\gamma_4$. 123/i stand for
the contribution of the integrals $V^{(i)}$ for i=1, 2 and 3.}
    \label{dataplus}
  \end{center}
\end{table}

As expected, the available O$(2)$ data follow the same pattern as
before. Furthermore the O$(2)$ numbers are in between the O$(1)$
and O$(3)$ ones. So (with some confidence) we can predict the uncalculated 
$(*)$ contributions to be close to the avarage of the corresponding
Ising and O$(3)$ entries. In this way we get
\be
(*)=-0.01786 \pm 0.00893\,,
\ee
where (generously) we allowed for 50\% error here.
This gives a total $k+l+m=6$ contribution of
\be
0.09806 \pm 0.00893\,.
\label{klm6}
\ee
Estimating the $k+l+m\geq8$ contributions, as usual, to be less than
10\% of (\ref{klm6}) our final estimate is
\be
\gamma_4=-4.559\pm0.019\,.
\label{gamma4}
\ee

We also need the product $\gamma_2\delta_2$. We have computed 
numerically the 3-particle contribution to both $\gamma_2$
and $\delta_2$ using the 3-particle form factors
constructed in Appendix B. We found 
\be
\gamma_{2;3}=0.001813(1)\qquad\qquad
{\rm and}\qquad\qquad
\delta_{2;3}=0.00003016(2)\,.
\label{gammadelta2}
\end{equation}
Thus $\gamma_2\delta_2=1.00184$ and finally we get $\gr$ with an 
error of a few permille: 
\be
\gr=9.10(4)\,.
\ee

\newpage
\newappendix{Test of random number generators}

Since we are interested in achieving numerical precision
for many quantities to an accuracy of $<1\%$, a considerable
source of concern to us was the random number generator (RNG). 
Indeed at an initial stage of this project we found
large standard deviations between results obtained by
various generators. 

Our first test concerned comparison of computations using various RNG's, 
with exact results on small lattices. 
The (practically) exact result on a $3\times 3$ lattice is obtained 
by discretizing the spins, taking ${\rm O}(2) \to {\rm Z}(N)$ 
and summing over all $N^{V-1}$ terms\footnote{Due to the global symmetry
one spin can be fixed.} in the partition function.
The convergence to the ${\rm O}(2)$ case is extremely (exponentially)
fast. As illustration in Table~\ref{exact_chi} we give the values 
of the susceptibility for $K=0.25$, $L=3$ and $N=6,\dots,10$.
Some generators already failed this test. The exact numbers were also 
useful to check our programs.

\begin{table}[ht]
\begin{center}
\begin{tabular}{|c|c|}
\hline
$N$ &  $\chi$ \\
\hline
 6 & 1.7619848372 \\
 7 & 1.7619804581 \\
 8 & 1.7619803546 \\
 9 & 1.7619803525 \\
10 & 1.7619803524 \\
\hline
\end{tabular}
\caption[]{{}\label{exact_chi}
The susceptibility of the ${\rm Z}(N)$ model on a $3\times 3$
lattice at $K=0.25$.
}
\end{center}
\end{table}

As a next step we compared results obtained by different RNG's
on larger lattices, see e.g. Table~\ref{rng} where we tabulated
our results for the susceptibility at $K=1.0$, $L=256$.
Here {\tt rand} is the RNG listed in {\it Language Reference}
XL Fortran for AIX (Version 3 Release 2)
and {\tt SGI} is the RNG provided by Silicon Graphics for the
SGI 2000 machine.
The {\tt nag} (the {\tt g05caf} RNG by Numerical Algorithms Group)
and {\tt ranlux} ~\cite{ranlux} are portable RNG's.
The latter has a single- and a double precision version,
both with an extra choice, a ``luxury parameter''.
The notations ``rlxs\_0'', ``rlxd\_1'' and ``rlxd\_2''
refer to the precision and the value of the luxury level parameter.

\begin{table}[ht]
\begin{center}
\begin{tabular}{|c|c|c|}\hline
\rule[-0.5ex]{0ex}{3.2ex}
program
&RNG
&$\chi$
\\
\hline
$1$ & {\tt rand}      &  $1609.5(9)$ \\
$1$ & {\tt SGI}   &  $1607.3(9)$ \\
$1$ & {\tt nag}       &  $1604.7(8)$ \\
\hline
$2$ & {\tt nag}       &  $1605.5(9)$ \\
$2$ & {\tt rlxd\_2}   &  $1604.1(9)$ \\
\hline
\end{tabular}
\caption[]{{}\label{rng}
The susceptibility at $K=1.0$ and $L=256$ using 
different RNG's and two different programs 
}
\end{center}
\end{table}

To our knowledge {\tt ranlux} is the only generator known with proven 
randomness qualities. 
Unfortunately, for ``historical reasons'' we used it only in the later 
stages of the project, while most of other runs were using {\tt nag}.
Reassuringly we found in all our tests, that the {\tt nag} generator 
produced results consistent with {\tt ranlux}. The combined 
{\tt nag} result (same RNG but different programs) in Table~\ref{rng}
is $1605.1(6)$, which is only 1-sigma away from rlxd\_2.
Note however, the {\tt rand} result is 4.2-sigma away from {\tt rlxd\_2}
while the {\tt SGI} result is 2.5-sigma away.

Although the latter deviation is still not too serious, 
the {\tt SGI} RNG gave also  suspicious results
on lattices with very large physical size, $z\approx 14$:
the data obtained by this RNG were several sigma too high
above the FS scaling lines.
Since the $z > 10$ data were not needed in the extrapolations
anyhow, we simply omitted these data points from Table~\ref{lattices}
and did not use them in our fits.
Nevertheless, to make sure that the discrepancy was indeed due to
the failure of the RNG we remeasured a few of these points with
{\tt ranlux}.  Table~\ref{z14} shows the results
for these two RNG's, together with the fits of Table \ref{xichi_TD}.
Note that the {\tt SGI} results show large deviations
from both {\tt ranlux} and the FS fit, especially for 
the susceptibility, which is always too high, by 5.6 to 7.6
standard deviations.

\begin{table}[ht]
\begin{center}
\begin{tabular}[t]{|d{2}|d{0}|d{8}|d{7}|d{7}|c|}
\hline
 \multicolumn{1}{|c|}{$K$} & \multicolumn{1}{c|}{$L$} &
 \multicolumn{1}{c|}{$\xi$} & \multicolumn{1}{c|}{$\chi$} & 
 \multicolumn{1}{c|}{$\gr$} & RNG \\
\hline \hline
0.92 & 150 & 10.559(3)  & 163.110(36) & 8.879(25) & {\tt rand} \\
     &     & 10.5508(9) & 162.843(13) & 8.864(8)  & {\tt nag} \\
     &     & 10.5499(7) & 162.829(10) & 8.869(6)  & {\tt rlxs\_0} \\
     &     & 10.5507(9) & 162.835(12) & 8.871(8)  & {\tt rlxd\_1} \\
     &     & 10.5510(11)& 162.840(16) & 8.881(10) & {\tt rlxd\_2} \\
     &     & 10.549(1)  & 162.86(26)  & 8.877(6)  & FS fit        \\
\hline
0.97 & 300 & 21.659(6)  & 557.41(15)  & 8.958(25) & {\tt SGI} \\
     &     & 21.627(4)  & 556.32(11)  & 8.952(6)  & FS fit        \\
\hline
1.0  & 560 & 40.168(13) & 1615.68(48) & 9.037(29) & {\tt SGI} \\
     &     & 40.107(12) & 1611.78(43) & 9.010(28) & {\tt rlxd\_2} \\
     &     & 40.096(5)  & 1611.4(3)   & 8.989(6)  & FS fits       \\
\hline
1.02 &1000 & 69.647(23) & 4184.5(1.2) & 9.066(34) & {\tt SGI} \\
     &     & 69.533(24) & 4174.5(1.3) & 8.962(33) & {\tt rlxd\_2} \\
     &     & 69.505(20) & 4173.0(1.4) & 9.026(8)  & FS fits       \\
\hline
\end{tabular}
\caption[]{{}\label{z14}
The correlation length and susceptibility for $z\approx 14$
for different RNG's and the corresponding value obtained 
by finite size (FS) extrapolation taken from Table \ref{xichi_TD}.
}
\end{center}
\end{table}

The only {\tt SGI} data present in Table~\ref{lattices}
are the 4 points at $K=1.04$ and $1.05$ for $z=4$ and $5$. 
These data, in contrast to the $z=14$ points, agree with the
FS fits. We have also rechecked the $K=1.04$, $L=578$ 
point with {\tt rlxs\_0} and got 
$\xi=142.10(9)$, $\chi=14142(12)$, $\gr =7.604(9)$, 
in good agreement with the {\tt SGI}
results\footnote{Note, however, that at these very large correlation 
length our relative errors are much larger than those at 
smaller $\xi$.} in Table~\ref{lattices}
therefore we did not repeat all these measurements with
{\tt ranlux}.

In one of the programs we measured the quantities
with the standard estimator along with the improved one,
and checked that they agree within the errors.
The other program used a Ward identity for checking.
Note, however, that the agreement in these quantities does
not guarantee yet the correctness of the results:
the error of the standard estimator is usually much larger
than that of the improved estimator,
while even the bad $z=14$ results passed the WI test.

\vfill
\eject

\newappendix{Vohwinkel's results for the 2-particle energies}

We here reproduce the original data table of Vohwinkel,
giving the 2-particle energies obtained nearly 10 years ago 
\cite{Vohwinkel}. We do not know the reason why
Vohwinkel did not publish his results, but it could be that he just
confused the assignment of quantum numbers, and so could not
match his results with the proposed S-matrix. With the 
correct identification his data are listed in Table~\ref{clausdata}.

As mentioned in Section 5 his values for the single particle
masses on the lattices with $z>\sim10$ are in good agreement with ours.
It is only on the lattice with $K=0.97$ 
and $L=128$ that our values differ by $\sim4$ standard deviations.
 
\begin{table}[h]
\centering
\begin{tabular}[t]{|c||l|l|l|l|}
\hline
$K$&  0.97     & 0.97       & 0.92       & 0.86      \\
$L$   &   128      &  256       & 128        & 64        \\ \hline
$m$   & 0.04633(2) & 0.04620(1) & 0.09465(3) & 0.1711(1) \\ \hline
$2$ & 0.1019(2)  & 0.0956(2)  & 0.1949(3)  & 0.353(1)  \\
      & 0.1612(3)  & 0.1149(3)  & 0.2334(4)  & 0.435(2)  \\ 
      &            & 0.1481(6)  & 0.2982(7)  & 0.566(3)  \\ 
      &            & 0.1872(8)  & 0.3760(10) &           \\ \hline
$1$   & 0.1298(1)  & 0.1038(4)  & 0.2116(4)  & 0.389(1)  \\ 
      &            & 0.1324(4)  & 0.2676(7)  & 0.510(2)  \\ 
      &            & 0.1705(6)  & 0.3440(20) &           \\ \hline
$0$ & 0.1081(15) & 0.0967(3)  & 0.1966(10) &           \\ 
      &            & 0.1220(8)  & 0.2431(20) &           \\ 
      &            & 0.1614(14) & 0.3143(40) &           \\ \hline
\end{tabular}
\caption{\footnotesize 
Masses and energies obtained by Claus Vohwinkel in 1992}
\label{clausdata}
\end{table}

\vfill
\eject


\newpage
\begin{thebibliography}{99}

\bibitem{Amit}
{D. J. Amit, Y. Y. Goldschmidt and G. Grinstein, J. Phys. {\bf A13} 
(1980) 585.}

\bibitem{Arn} D. Arnaudon, Commun. Math. Phys. {\bf 159} (1994) 175.

\bibitem{Karowskietal}
H.~Babujian, A.~Fring, M.~Karowski and A.~Zapletal,
Nucl. Phys. B538 (1999) 535. 


\bibitem{logs}
J. Balog, J. Phys {\bf A34} (2001) 5237.

\bibitem{BH}
J. Balog and \'A. Heged\H us, J. Phys. {\bf A33} (2000) 6543.

\bibitem{SigmacollabI}
J. Balog, M. Niedermaier, F. Niedermayer,
A. Patrascioiu, E. Seiler and P. Weisz,
Phys. Rev. {\bf D60} (1999) 094508.


\bibitem{SigmacollabII}
J. Balog, M. Niedermaier, F. Niedermayer,
A. Patrascioiu, E. Seiler and P. Weisz,
Nucl. Phys. {\bf B583} (2000) 614.

\bibitem{Banksetal}
T. Banks, D. Horn, and H. Neuberger, Nucl. Phys. {\bf B108} (1976) 119.

\bibitem{pew}
B. Berg and P. Weisz, Nucl. Phys. {\bf B146} (1979) 205;\\
E. Abdalla, B. Berg and P. Weisz, Nucl. Phys. {\bf B157} (1979) 387.

\bibitem{HTXY1}
P.~Butera and M.~Comi, Phys. Rev. {\bf B54} (1996) 15828. 

\bibitem{BuCom} P.~Butera and M.~Comi, 
Phys. Rev. {\bf B47}(1993)11969. 

\bibitem{CPRVgR}
M.~Campostrini, A.~Pelissetto, P.~Rossi and E.~Vicari,
\NP{B459} (1996) 207; Nucl. Phys. Proc. Suppl. {\bf 47} (1996) 751.

\bibitem{Coleman}
S. Coleman, Phys. Rev. {\bf D11} (1975) 2088.

\bibitem{Destri}
C. Destri, Phys. Lett. {\bf 210B} (1988) 173; 
Erratum--{\it ibid} {\bf 213B} (1988) 565.

\bibitem{FMPPT}
M.~Falcioni, G.~Martinelli, M.~L.~Paciello, G.~Parisi and
B.~Taglienti, \NP{B225} (1983) 313.

\bibitem{FeLeCl} G. Felder and A. LeClair, Int. J. Mod. Phys. 
{\bf A7} (1992) 239. 

\bibitem{FrohSpen}
J. Fr\"ohlich and T. H. Spencer, Comm. Math. Phys. {\bf 81} (1981) 527.

\bibitem{GSRbook} C. Gomez, G. Sierra, M. Ruiz-Altaba, 
{\it Quantum groups in two-dimensional physics}, Cambridge University
Press, 1996.   

\bibitem{XYMC}
R. Gupta, J. De Lapp, G. G. Batrouni, G. C. Fox, C. F. Baillie 
and J. Apostolakis, Phys. Rev. Lett. {\bf91} (1988) 1996; \\ 
R. Gupta and  C. F. Baillie, Nucl. Phys. B (Proc. Suppl.) 
{\bf 20} (1991) 669; \\
 R. Gupta and C. F. Baillie, Phys. Rev. {\bf B45} (1992) 2883; \\
U. Wolff, Nucl. Phys.{\bf B322} (1989) 759; \\
R. G. Edwards, J. Goodman, A. D. Sokal, Nucl. Phys. {\bf B354} (1991) 289; \\
W. Janke and  K. Nather, Phys. Lett. {\bf 157A} (1991) 11. 

\bibitem{HMN}
P.~Hasenfratz, M.~Maggiore and F.~Niedermayer,
Phys. Lett. {\bf B245} (1990) 522;\\
P.~Hasenfratz and F.~Niedermayer,
Phys. Lett. {\bf B245} (1990) 529.

\bibitem{Hasenbusch}
M. Hasenbusch, M. Marcu and K. Pinn, Physica {\bf A208} (1994) 124.\\
M. Hasenbusch and K. Pinn, J. Phys. {\bf A30} (1997) 63.

\bibitem{Janke}
W. Janke, Phys.~Rev.~{\bf B55} (1997) 3580. 

\bibitem{Jose}
J. V. Jos\'e, L. Kadanoff, S. Kirkpatrick, and D. Nelson, 
Phys. Rev. {\bf B16} (1977) 1217.


\bibitem{KarWe}
M. Karowski and P. Weisz,
Nucl. Phys. {\bf B139} (1978) 455.

\bibitem{IK}
R. Kenna and A. C. Irving, Nucl. Phys. {\bf B485} (1997) 583;
Phys. Lett. {\bf B351} (1995) 273.

\bibitem{KlaMelz} T. Klassen and E. Melzer, Int. J. Mod. Phys.
{\bf A8} (1993) 4131. 

\bibitem{LeClair}
R. M. Konik and A. LeClair,
Nucl. Phys. {\bf B479} (1996) 619. 

\bibitem{KT}
J.M.~Kosterlitz and D.J.~Thouless,
J. Phys. {\bf C6} (1973) 1181;\\ 
J.M.~Kosterlitz,
J. Phys. {\bf C7} (1974) 1046.

\bibitem{Lash} M. Lashkevich, {\it Sectors of mutually local fields 
in integrable models of QFT}, [hep-th/9406118]. 

\bibitem{SmiLeCl} A. LeClair and F. Smirnov, Int. J. Mod. Phys. 
{\bf A7} (1992) 2997. 

\bibitem{LuWo}
M. L\"uscher und U. Wolff,
Nucl. Phys. {\bf B339} (1990) 222.

\bibitem{LuVol} 
M. L\"uscher,
Commun. Math. Phys. {\bf 104} (1986) 177.\\
M. L\"uscher, {\it On a relation between finite size effects 
and elastic scattering processes}, in: Cargese Summer Inst.~1983,
Plenum Press, New York, 1984. 

\bibitem{Lusch}
M.~L\"uscher, Nucl. Phys. {\bf B135} (1978) 1.

\bibitem{ranlux}
M. L\"uscher, Comp. Phys. Commun. {\bf 79} (1994) 100. 

\bibitem{Michael}
C. Michael, Nucl. Phys. {\bf B259} (1985) 58.

\bibitem{MNcycl} M. Niedermaier, Commun. Math. Phys. {\bf 196} 
(1998) 411.

\bibitem{glueballs}
F. Niedermayer,  Ph. R\"ufenacht and U. Wenger,
Nucl. Phys. {\bf B597} (2001) 413.

\bibitem{perc} A. Patrascioiu and E. Seiler, {\it Absence of 
Asymptotic Freedom in nonabelian models} [hep-th/0002153];
{\it Percolation and the existence of a soft phase in the classical
Heisenberg model}, [hep-th/0011199].

\bibitem{PSward} A. Patrascioiu and E. Seiler, Phys. Lett. {\bf B417}
(1998) 123; Phys. Rev. {\bf E57} (1998) 111.

\bibitem{HTXY2} 
A.~Pelissetto and E.~Vicari, \NP{B519} (1998) 626; \NP{575} (2000) 579.

\bibitem{Poly}
A. Polyakov,
Phys. Lett. {\bf B72} (1977) 224.

\bibitem{RS} N. Reshetikhin and F. Smirnov, Commun. Math. Phys. 
{\bf 131} (1990) 157.


\bibitem{Smir1} F. Smirnov, {\it Form factors in completely integrable 
models of quantum field theory}, World Scientific, 1992. 

\bibitem{Smir2} F. Smirnov, Commun. Math. Phys. {\bf 132} (1990) 415.

\bibitem{Sym}
K. Symanzik, Nucl. Phys. {\bf B226} (1983) 187.

\bibitem{SW} H. Swendsen and J. Wang, Phys. Rev. Lett. {\bf 58} 
(1987) 86. 


\bibitem{TarVar}
V. Tarasov and A. Varchenko, Amer.~Math.~Transl.~{\bf 174} 
(1996) 235. 

\bibitem{Villain}
J. Villain, J. Physique {\bf 36} (1975) 581.

\bibitem{Vohwinkel} 
C. Vohwinkel, private communication (1992).

\bibitem{Uli} U. Wolff, Phys. Rev. Lett. {\bf 62} (1989) 361.

\bibitem{Woo} C.H.~Woo, Phys. Rev. {\bf D20} (1979) 1880. 

\bibitem{ZZ}
A.B. and Al.B.~Zamolodchikov,
Ann. Phys. {\bf 120} (1979) 253; Nucl. Phys. {\bf B133} (1978) 525.

\bibitem{Zamolodchikov}
{Al. B. Zamolodchikov, Int. J. of Mod. Phys. {\bf A10} (1995) 1125.}

\bibitem{ZinnJ}
J. Zinn-Justin, {\it Quantum Field Theory and Critical 
Phenomena}, Oxford, 1989.

\end{thebibliography}
\end{document}